\documentclass[trackchanges,twocolumn]{aastex701}
\usepackage{amsmath}
\usepackage[T1]{fontenc}
\usepackage{upgreek}
\usepackage{comment}
\usepackage{xspace}

\newcommand{\targetx}{HD\,126053\xspace}
\newcommand{\targety}{HD\,168009\xspace}
\newcommand{\targetz}{HD\,10780\xspace}
\newcommand{\planetx}{HD\,126053\,b\xspace}
\newcommand{\planety}{HD\,168009\,b\xspace}
\newcommand{\planetz}{HD\,168009\,c\xspace}

\newcommand{\nlowsnx}{3\xspace} 
\newcommand{\nlowsny}{1\xspace} 

\newcommand{\ndriftx}{5\xspace} 
\newcommand{\ndriftz}{13\xspace} 

\newcommand{\nobsx}{159\xspace}
\newcommand{\nobsy}{142\xspace}
\newcommand{\nobsz}{173\xspace}

\newcommand{\snx}{289\xspace}
\newcommand{\sny}{315\xspace}
\newcommand{\snz}{296\xspace}

\newcommand{\CaII}{\ion{Ca}{2}\xspace}

\newcommand{\rvsearch}{\texttt{RVSearch}\xspace}

\shorttitle{NETS VI: \targetx, \targety, and \targetz}
\shortauthors{Gupta et al}

\newcommand{\PSUAA}{Department of Astronomy \& Astrophysics, 525 Davey Laboratory, 251 Pollock Road, Penn State, University Park, PA, 16802, USA}
\newcommand{\PSUCEHW}{Center for Exoplanets and Habitable Worlds, 525 Davey Laboratory, 251 Pollock Road, Penn State, University Park, PA, 16802, USA}
\newcommand{\PSUARC}{Astrobiology Research Center, 525 Davey Laboratory, 251 Pollock Road, Penn State, University Park, PA, 16802, USA}
\newcommand{\PSETI}{Penn State Extraterrestrial Intelligence Center, 525 Davey Laboratory, 251 Pollock Road, Penn State, University Park, PA, 16802, USA}
\newcommand{\PSUICDS}{Institute for Computational and Data Sciences, Penn State, University Park, PA, 16802, USA}
\newcommand{\PSUCASt}{Center for Astrostatistics, 525 Davey Laboratory, 251 Pollock Road, Penn State, University Park, PA, 16802, USA}
\newcommand{\UA}{Steward Observatory, University of Arizona, 933 N.\ Cherry Ave, Tucson, AZ 85721, USA}

\newcommand{\Caltech}{Department of Astronomy, California Institute of Technology, 1200 E California Blvd, Pasadena, CA 91125, USA}

\newcommand{\NOIRLab}{U.S. National Science Foundation National Optical-Infrared Astronomy Research Laboratory, 950 N.\ Cherry Ave., Tucson, AZ 85719, USA}

\newcommand{\Macquarie}{School of Mathematical and Physical Sciences, Macquarie University, Balaclava Road, North Ryde, NSW 2109, Australia}

\newcommand{\JPL}{Jet Propulsion Laboratory, California Institute of Technology, 4800 Oak Grove Drive, Pasadena, California 91109}
\newcommand{\MIT}{Kavli Institute for Astrophysics and Space Research, Massachusetts Institute of Technology, Cambridge, MA, USA}
\newcommand{\UCI}{Department of Physics \& Astronomy, The University of California, Irvine, Irvine, CA 92697, USA}
\newcommand{\Carleton}{Carleton College, One North College St., Northfield, MN 55057, USA}

\newcommand{\FlatironCCA}{Center for Computational Astrophysics, Flatiron Institute, 162 Fifth Avenue, New York, NY 10010, USA}

\newcommand{\TIFR}{Department of Astronomy and Astrophysics, Tata Institute of Fundamental Research, Homi Bhabha Road, Colaba, Mumbai 400005, India}

\newcommand{\UAm}{Anton Pannekoek Institute for Astronomy, 904 Science Park, University of Amsterdam, Amsterdam, 1098 XH}

\newcommand{\Amherst}{Department of Physics and Astronomy, Amherst College, 25 East Drive, Amherst, MA 01002, USA}
\newcommand{\SchmidtAstro}{Astrophysics \& Space Center, Schmidt Sciences, New York, NY 10011, USA}

\submitjournal{AJ}

\begin{document}

\title{The NEID Earth Twin Survey. VI. Discovery of a Low-Mass Planet Orbiting HD 126053 and RV Signals Near the Rotation Periods of HD 168009 and HD 10780}

\author[0000-0002-5463-9980]{Arvind F.\ Gupta}
\affil{\UA}
\affil{\NOIRLab}
\affil{NASA Sagan Fellow}
\email{afgupta@arizona.edu}


\author[0000-0002-9632-9382]{Sarah E.\ Logsdon}
\affiliation{\NOIRLab}
\email{sarah.logsdon@noirlab.edu}

\author[0000-0002-4927-9925]{Jacob K. Luhn}
\affiliation{\JPL}
\email{jacob.luhn@jpl.nasa.gov}

\author[0000-0003-0199-9699]{Evan Fitzmaurice}
\affil{\PSUAA}
\affil{\PSUCEHW}
\affil{\PSUICDS}
\email{exf5296@psu.edu}

\author[0000-0001-9596-7983]{Suvrath Mahadevan}
\affiliation{\PSUAA}
\affiliation{\PSUCEHW}
\affiliation{\PSUARC}   
\email{suvrath@astro.psu.edu}

\author[0000-0003-0149-9678]{Paul Robertson}
\affiliation{\UCI}
\email{paul.robertson@uci.edu}

\author[0000-0001-6545-639X]{Eric B.\ Ford}
\affiliation{\PSUAA}
\affiliation{\PSUCEHW}
\affiliation{\PSUICDS}
\affiliation{\PSUCASt}
\email{ebf11@psu.edu}


\author[0000-0003-4384-7220]{Chad F.\ Bender}
\affiliation{\UA}
\email{cbender@arizona.edu}

\author[0000-0002-0078-5288]{Mark R.~Giovinazzi}
\affiliation{\Amherst}
\email{mgiovinazzi@amherst.edu}

\author[0000-0003-1312-9391]{Samuel Halverson}
\affiliation{\JPL}
\email{samuel.halverson@jpl.nasa.gov}

\author[0000-0002-7127-7643]{Te Han}
\affiliation{\MIT}
\email{tehan@mit.edu}

\author[0000-0001-9626-0613]{Daniel M.\ Krolikowski}
\affiliation{\UA}
\email{krolikowski@arizona.edu}

\author[0000-0002-9082-6337]{Andrea S.J.\ Lin}
\affiliation{\Caltech}
\email{asjlin@caltech.edu}

\author[0000-0001-8720-5612]{Joe P.\ Ninan}
\affiliation{\TIFR}
\email{indiajoe@gmail.com}

\author[0000-0002-4677-8796]{Michael L. Palumbo III}
\affiliation{\FlatironCCA}
\email{mpalumbo@flatironinstitute.org}

\author[0000-0003-1324-0495]{Leonardo A.\ Paredes}
\affiliation{\UA}
\email{}

\author[0000-0002-2488-7123]{Jayadev Rajagopal}
\affiliation{\NOIRLab}
\email{}

\author[0000-0001-8127-5775]{Arpita Roy}
\affiliation{\SchmidtAstro}
\email{arpita308@gmail.com}

\author[0000-0002-4046-987X]{Christian Schwab}
\affiliation{\Macquarie}
\email{mail.chris.schwab@gmail.com}

\author[0000-0001-7409-5688]{Gudmundur Stefansson}
\affiliation{\SchmidtAstro}
\affiliation{\UAm}
\email{gstefansson@schmidtsciences.org}

\author[0000-0002-4788-8858]{Ryan C. Terrien}
\affiliation{\Carleton}
\email{rterrien@carleton.edu}

\author[0000-0001-6160-5888]{Jason T.\ Wright}
\affiliation{\PSUAA}
\affiliation{\PSUCEHW}
\affiliation{\PSETI}
\email{astrowright@gmail.com}

\begin{abstract}

Extreme precision radial velocity (EPRV) spectrographs are making strides in extending the limits of radial velocity exoplanet detection. Perhaps the best evidence for this is how readily EPRV instruments are facilitating the discovery and detailed characterization of small-amplitude planetary signals around stars that have already been intensively monitored for decades with less stable instruments. In this paper, we use data from the ongoing NEID Earth Twin Survey (NETS) to detect and confirm the planet HD\,126053\,b ($P=10.91$ d), which is among the least massive planets ($m \sin i = 3.60\,{\rm M}_\oplus$) around solar-type stars within 25 pc. We then jointly characterize the known planet HD\,168009\,b ($P=15.14$ d; $m \sin i = 8.64\,{\rm M}_\oplus$) orbiting a solar analog along with a previously unreported signal at 27.8 d which we tentatively attribute to rotational modulation of stellar activity ($P_{\rm rot} = 30.1$ d). We also analyze and discuss periodic RV variations in HD\,10780 arising from a combination of rotationally-modulated activity and a long-term activity cycle. We discuss HD\,126053\,b and HD\,168009\,b in the context of the broader small planet population and  our sensitivity to long-period planets in these systems, and we comment on our capacity to disambiguate planet-induced and activity-induced radial velocity signals. HD\,126053 is exceptionally RV-quiet, making it a promising target for continued RV monitoring in search of low-mass exoplanets in the habitable zone.
\end{abstract}


\section{Introduction}\label{sec:intro}

Astronomers have been monitoring the radial velocity (RV) perturbations of nearby stars for decades.
These surveys led to some of the earliest exoplanet discoveries \citep{Mayor1995} and they have continued to make invaluable contributions to the census of known exoplanets \citep[e.g., Proxima Cen b;][]{Anglada-Escude2016}.
Yet even with this decades-long heritage, the weak RV signals induced by the planetary companions of our stellar neighbors have proven challenging to detect due to barriers imposed by intrinsic stellar RV variations and by the internal stability limits of our spectrographs \citep{Burt2025}.
As instrumentation advances and as stellar variability mitigation techniques improve, we expect that RV surveys will continue to lead to new discoveries.
Indeed, the newest extreme precision radial velocity (EPRV) spectrographs, which are capable of maintaining sub-m~s$^{-1}$ night-to-night stability, have already unveiled new planetary signals around nearby stars \citep{GonzalezHernandez2024,Nari2025,Basant2025,NETSI}.

The NEID Earth Twin Survey \citep[NETS;][]{Gupta2021} is an ongoing EPRV search for low-mass, long-period exoplanets around a sample of bright, nearby stars using the NEID spectrograph \citep{Schwab2016,Robertson2019} on the WIYN 3.5\,m telescope at Kitt Peak National Observatory.\footnote{The WIYN Observatory is a joint facility of the NSF’s National Optical-Infrared Astronomy Research Laboratory, Indiana University, the University of Wisconsin-Madison, Pennsylvania State University, and Princeton University.}
\citet{Gupta2025} analyzed the first three years of NETS data, from the start of the survey in 2021 through mid-2024, to assess the sensitivity of the data set to known planets orbiting NETS target stars. In addition to recovering most of the known planets with a periodogram grid search, \citet{Gupta2025} also identified a number of new candidate signals.  In this work, we undertake a more detailed exploration of the RVs for \targetx, \targety, and \targetz to investigate and evaluate three of these signals. In addition to the NEID measurements presented in \citet{Gupta2025}, our analysis incorporates more recent NETS observations as well as archival data from other spectrographs.

For \targetx, we focus on a 10.9 d signal which was detected at a high significance by \citet{Gupta2025} but is notably absent in previous RV data sets for this star.
For \targety, we first characterize a 15.1 d signal which is consistent with a candidate identified by \citet{Hirsch2021} and later classified as the planet \planety by \citet{Rosenthal2021}, noting that \citet{Rosenthal2021} acknowledge that the planet classification is not rigorously justified therein. We then analyze and comment on the nature of an additional 28 d signal that emerges in the residuals, as well as two long-period signals that appear in archival RVs but are absent from the NEID data.
For \targetz, we jointly characterize a 25 d signal first identified by \citet{Gupta2025} along with an emerging long-term RV signal, noting that both of these variability timescales are comparable to previous measurements of the stellar rotation period and activity cycle \citep[e.g.,][]{Baliunas1996,Olspert2018}.

The RV measurements used in this work are described in Section \ref{sec:data}. In Section \ref{sec:target_stars}, we discuss the properties of the target stars. We describe the NEID RV calculations in Section \ref{sec:neidrv} and our analysis of the stellar activity indicators and RV data in Section \ref{sec:analysis}, and we discuss the results and prospects for future planet detection in Section \ref{sec:discussion}. Our results and their implications are summarized in Section \ref{sec:summary}.

\section{Radial Velocity Data}\label{sec:data}

\subsection{NEID Observations}\label{sec:neid}

\targetx, \targety, and \targetz are currently being monitored with NEID as part of the NETS program. The observing strategies and cadence for each of these stars adhere to the survey scheme described by \citet{Gupta2025}. In brief, we obtain a single exposure on each night that a star is observed, where the exposure time is set by a fixed signal-to-noise ratio (S/N) trigger corresponding to an expected measurement precision of 30 cm~s$^{-1}$ \citep[comparable to the predicted internal stability of the instrument as given in][]{Halverson2016}. All observations were taken in the high resolution mode ($R\sim115,000$) during standard NEID queue operations.
Here, we make use of data collected from September 2021 through July 2026.

The raw spectra were all processed with version $1.5$ of the NEID Data Reduction Pipeline (DRP) and retrieved from the NEID data archive\footnote{\url{https://neid.ipac.caltech.edu/}}. We discard \ndriftx measurements of \targetx and \ndriftz measurements of \targetz; some of these data were taken while NEID was recovering from a thermal instability in December 2025 and the remainder were flagged by the pipeline due to a sub-standard instrument drift solution or wavelength calibration\footnote{We only retain spectra with the standard (\texttt{dailymodel0}) drift correction and with wavelength calibration flags of \texttt{LFCplusThAr} or \texttt{LFCplusThArModified}.}. We also exclude \nlowsnx observations of \targetx and \nlowsny observation of \targety for which the extracted S/N at 5500 \AA\ was $\leq60\%$ of the desired threshold. The resulting data set consists of \nobsx measurements for \targetx at a median S/N of \snx, \nobsy measurements for \targety at a median S/N of \sny, and \nobsz measurements for \targetz at a median S/N of \snz. We divide these into separate runs corresponding to NEID's RV eras as defined in the DRP documentation\footnote{\url{https://neid.ipac.caltech.edu/docs/NEID-DRP/rveras.html}} to account for RV zero point shifts introduced by changes to the spectrograph or calibration system. Our data span four distinct eras: Run 1 (2021 September 1 -- 2022 June 16)\footnote{Our adopted start date for Run 1 is later than that given in the DRP documentation, as we choose to acknowledge an additional candidate era identified by \citet{Gupta2025}. This choice is inconsequential for the present study, as none of our targets were observed before September 2021.}, Run 2 (2022 October 18 -- 2024 August 19), Run 3 (2025 August 24 -- 2025 December 8), and Run 4 (2025 December 9 -- 2026 July 26).

The NEID DRP provides wavelength-calibrated 1-D spectra and RVs calculated using the cross-correlation function \citep[CCF;][]{Baranne1996} method. We make use of the 1-D spectra in this work, but we do not adopt the DRP RVs due to an error that was identified in the CCF calculation. In Section \ref{sec:neidrv}, we describe three separate methods we use to calculate the NEID RVs (including a CCF approach in which the DRP error is corrected). The adopted RVs are shown in \autoref{fig:all_rvs}, and we compare the results of the different methods in Figures \ref{fig:HD126053_rvs}, \ref{fig:HD168009_rvs}, and \ref{fig:HD10780_rvs}. In addition to processed spectra and RVs, the NEID DRP supplies accompanying diagnostics such as CCF shape metrics and stellar activity indicators. In this work, we use the \ion{Ca}{2} H \& K emission line strengths to assess stellar activity signals in our data. These are calibrated to the standard Mount Wilson S-index scale using the linear correction derived in \citet{Gupta2025}.

\begin{deluxetable*}{llrrr}
\tablecaption{Summary of Radial Velocity Measurements for \targetx, \targety, and \targetz \label{tab:rv_stats}}
\tablehead{HD No. & Instrument& \colhead{$N_{\rm RV}$} & \colhead{First} & \colhead{Last}} 
\startdata
126053 & Hamilton & 82 & 1987-06-11 & 2008-03-23 \\
&HIRES (pre)$^\dagger$ & 46 & 1996-07-11 & 2004-03-07\\
&HIRES (post) & 92 & 2004-08-23 & 2022-09-11 \\
&APF & 277 & 2013-12-05 & 2020-02-08 \\
&NEID Run 1 & 25 & 2021-12-18 & 2022-06-07 \\
&NEID Run 2 & 75 & 2022-12-21 & 2024-07-13\\
&NEID Run 3 & 31 & 2024-12-17 & 2025-07-10 \\
&NEID Run 4 & 28 & 2026-01-18 & 2026-07-24 \\
\hline
168009 & Hamilton & 53 & 1998-07-10 & 2008-08-19 \\
&HIRES (post) & 129 & 2008-06-24 & 2022-05-16\\
&APF & 441 & 2013-10-31 & 2019-07-26 \\
&NEID Run 1 & 31 & 2021-09-18 & 2022-06-12 \\
&NEID Run 2 & 59 & 2023-02-08 & 2024-06-28 \\
&NEID Run 3 & 35 & 2024-09-06 & 2025-11-11 \\
&NEID Run 4 & 17 & 2026-01-21 & 2026-06-29 \\
\hline
10780 & Hamilton & 22 & 1998-12-05 & 2001-08-26 \\
&HIRES (post) & 25 & 2004-08-21 & 2022-07-20 \\
&APF & 59 & 2013-11-01 & 2018-10-07\\
&NEID Run 1 & 32 & 2021-09-20 & 2022-06-13\\
&NEID Run 2 & 60 & 2022-12-01 & 2024-06-18 \\
&NEID Run 3 & 68 & 2024-09-05 & 2025-12-08 \\
&NEID Run 4 & 13 & 2025-12-31 & 2026-06-19 \\
\hline
\enddata
\tablenotetext{}{$^\dagger$We split the HIRES data for \targetx due to a zero point offset introduced by a detector upgrade in 2004 \citep{Butler2017,Tal-Or2019}.}
\end{deluxetable*}

\subsection{Archival Measurements}\label{sec:archival}

\begin{figure}
    \centering
    \includegraphics[width=\linewidth]{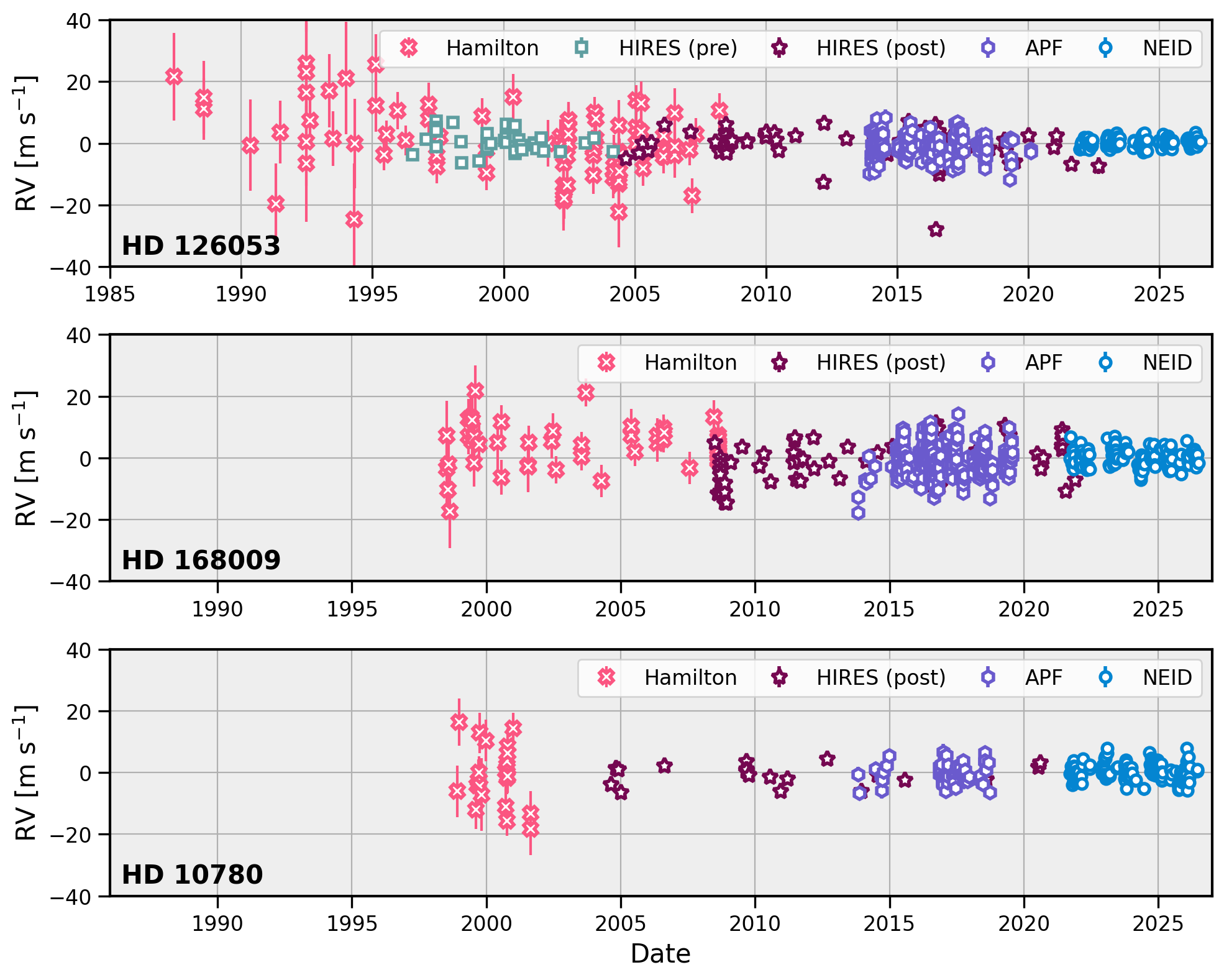}
    \caption{Radial velocity time series measurements for \targetx (upper), \targety (middle), and \targetz (lower). The NEID RVs shown here were calculated with \texttt{SERVAL} as described in Section \ref{sec:rv_serval}. The time series for each instrument is shifted to have a median of $0$ m~s$^{-1}$.}
    \label{fig:all_rvs}
\end{figure}

To extend the observing baselines and assess the long-term RV behavior for each of our targets, we supplement the NEID data with archival measurements from the HIgh Resolution Echelle Spectrograph \citep[HIRES;][]{Vogt1994} on the Keck I telescope at Maunakea, the Levy spectrograph on the Automated Planet Finder (APF) facility \citep{Vogt2014} at Lick Observatory, and the Hamilton spectrograph at Lick Observatory. All three of these spectrographs make use of a molecular iodine absorption cell to produce a rest-frame reference spectrum against which relative RVs are calculated, following the method first introduced by \citet{Butler1996}. We describe the data in Sections \ref{sec:HIRES}, \ref{sec:APF}, and \ref{sec:Hamilton} and show the RVs in \autoref{fig:all_rvs}.

\subsubsection{HIRES} \label{sec:HIRES}

As described in \citet{Rosenthal2021}, the HIRES observations for our targets were initially collected as part of the California \& Carnegie Planet Search \citep[pre-2004;][]{Cumming2008} and later as part of two coincident but separate surveys: the California Planet Search \citep{Howard2010} and Lick-Carnegie Exoplanet Survey \citep{Butler2017}.
HIRES RVs from these surveys were computed with two independent pipelines and published separately by \citet{Rosenthal2021} as part of the California Legacy Survey (CLS) as well as by \citet{Teklu2025}.
The public version of the CLS data set is complete only until 2020, while the \citet{Teklu2025} data set includes RVs through 2022.
We adopt the latter, as the extended baseline overlaps with the first year of NETS data for each of our targets.
The details of the HIRES RVs are summarized in Table \ref{tab:rv_stats}.

\subsubsection{APF} \label{sec:APF}

The CLS data set also includes RV measurements from APF. \targetx and \targety were both observed heavily as part of the APF-50 survey \citep{Fulton2017} from 2013 to 2020, while \targetz received relatively sparse coverage over approximately the same time frame. For each star, we identify and remove several instances in the CLS catalog in which RV measurements were duplicated (back-to-back entries with identical timestamps, velocities, and measurement uncertainties). Three points are removed for \targetx, nine points are removed for \targety, and one point is removed for \targetz. The number of remaining APF measurements and the observing baselines are summarized in Table \ref{tab:rv_stats}.

\subsubsection{Hamilton Spectrograph} \label{sec:Hamilton}

We also make use of RVs collected as part of the Lick Planet Search \citep{Fischer2014} using the Hamilton spectrograph at Lick Observatory. As with the APF data, we take the Hamilton RVs directly from CLS catalog. \targetx and \targety each have a baseline of just over a decade, while \targetz was only observed for three years. The Lick Planet Search RV coverage for each target is summarized in Table \ref{tab:rv_stats}.

\section{Properties of Target Stars}\label{sec:target_stars}

For this work, we adopt stellar parameters from \citet{Brewer2016}. These values are listed in Table \ref{tab:star_params}.
\targetx, \targety, and \targetz are bright, nearby stars with rich histories of spectroscopic characterization, and we note that derived parameters are expected to vary across different data sets and measurement techniques \citep[see, e.g.,][]{Blanco-Cuaresma2019,Soubiran2022,Freckelton2026}.
We confirm that the \citet{Brewer2016} results are consistent with the literature consensus, with the notable exception of stellar age, which we therefore exclude from Table \ref{tab:star_params}.
In the following subsections, we comment on this discrepancy as well as on other relevant properties of each star.

\begin{deluxetable*}{lccccc}
\tablecaption{Summary of Stellar Parameters for \targetx, \targety, and \targetz \label{tab:star_params}}
\tablehead{Parameter & \multicolumn{3}{c}{Value} & Description & Ref.} 
\startdata
\hline
\hspace{-0.2cm} Identifiers:  & \targetx & \targety & \targetz & & \\
& HIP\,70319 & HIP\,89474 & HIP\,8362 & & \\
& GJ\,547 & GJ\,9622 & GJ\,75 & & \\
\multicolumn{6}{l}{\hspace{-0.2cm} Coordinates and Parallax:} \\
~~~$\alpha_{\mathrm{J2016}}$ & 14:23:15.52 & 18:15:32.35 & 1:47:46.24 & Right Ascension (RA) & Gaia DR3\\
~~~$\delta_{\mathrm{J2016}}$ & +1:14:21.99 & +45:12:31.70 & +63:51:05.07 & Declination (Dec) & Gaia DR3\\
~~~$\varpi$  & $57.271 \pm 0.038$ &  $42.935 \pm 0.016$ &  $99.590 \pm 0.044$ & Parallax (mas) & Gaia DR3 \\
~~~$d$  & $17.45\pm0.01$ & $23.28\pm0.01$ & $10.038\pm0.004$ & Distance (pc) & BJ21 \\
\multicolumn{6}{l}{\hspace{-0.2cm} Stellar Parameters:}\\
~~~$T_{\rm eff}$ & $5714\pm25$ & $5808\pm25$ & $5344\pm25$ & Effective Temperature (K) &  B16\\
~~~$L_\star$ & $0.81\pm0.04$ & $1.32\pm0.06$ & $0.52^{+0.03}_{-0.02}$ & Luminosity (${\rm L}_\odot$) & B16\\
~~~$R_\star$ & $0.92\pm0.03$ & $1.14\pm0.03$ & $0.84\pm0.02$ & Stellar Radius (${\rm R}_\odot$)   &  B16\\
~~~$M_\star$ & $0.91_{-0.04}^{+0.03}$ & $1.05_{-0.03}^{+0.02}$ & $0.90\pm0.03$ & Stellar Mass (${\rm M}_\odot$)  &  B16\\
~~~$\log g$ & $4.47\pm0.04$ & $4.35\pm0.03$ & $4.54\pm0.03$ & Surface Gravity ($\log$ (cm~s$^{-2}$)) &  B16\\
~~~[Fe/H]& $-0.31\pm0.01$ & $0.03\pm0.01$ & $0.01\pm0.01$ & Metallicity (dex) &  B16\\
\hline
\enddata
\tablenotetext{}{References: Gaia DR3 \citep{GaiaCollaboration2022}, B16 \citep{Brewer2016}, BJ21 \citep{Bailer-Jones2021}}
\end{deluxetable*}

\subsection{HD 126053}\label{sec:stellar_HD126053}

\targetx is a solar-type (G1.5V, $T_{\rm eff}=5714$ K, $\log g$ = 4.47), solar-mass ($M_\star =0.91\ {\rm M}_\odot$) star at a distance of $d=17.45$ pc.  \citet{Brewer2016} report an age of $6.5_{-2.8}^{+2.6}$ Gyr, but both \citet{Casali2020} and \citet{Ramirez2012} find the star to be older, with age estimates of $9.3\pm2.05$ Gyr and $10.35^{+1.86}_{-2.76}$ Gyr, respectively. All three of these ages were derived from isochrone fits, albeit with different grids for the \citet{Casali2020} result \citep[Yale-Potsdam;][]{Spada2017} and the \citet{Brewer2016} and \citet{Ramirez2012} results \citep[Yonsei-Yale;][]{Demarque2004}. While the values are consistent with each other given the large uncertainties, it is evident that the age of the star is not well constrained.

We also took note of literature measurements of the stellar rotation period and activity cycle, both of which can influence our interpretation of detected RV signals. No robust rotation measurements exist, but \citet{Hempelmann2016} report a low-significance detection of a 26 d period based on an analysis of \CaII H \& K data from the TIGRE facility. \citet{Isaacson2024} derive a 17.53 yr activity cycle based on the HIRES S-index data, while \citet{Schmitt2026} analyze a combination of data from TIGRE and the Mount Wilson survey \citep{Wilson1978} and find no strong evidence for cyclic variations. \citet{Schmitt2026} instead classify \targetx as a Maunder minimum candidate given the lack of a coherent cycle and the low median activity level. \citet{Baum2022} also diagnose the S-index time series as ``flat'' based on a combination of Mount Wilson and HIRES data, albeit with a shorter HIRES baseline than that used by \citet{Isaacson2024}.

\subsection{HD 168009}\label{sec:stellar_HD168009}

\targety is a solar-type (G1V, $T_{\rm eff}=5808$ K, $\log g$ = 4.35), solar-mass ($M_\star =1.05\ {\rm M}_\odot$) star at a distance of $d=23.28$ pc. As with \targetx, the age given by \citet{Brewer2016}, $5.4_{-1.2}^{+1.6}$ Gyr, is younger than that derived by other studies. \citet{Ramirez2012} derive an age of $7.39^{+1.82}_{-0.60}$ Gyr (Yonsei-Yale isochrones) and \citet{Casagrande2011} report ages of $9.20_{-1.21}^{+1.11}$ Gyr \citep[BaSTI isochrones;][]{Pietrinferni2004,Pietrinferni2006,Pietrinferni2009} and  $8.10_{-1.92}^{+1.52}$ Gyr \citep[Padova isochrones;][]{Bertelli2008,Bertelli2009}. There is a greater discrepancy in these measurements than in those of \targetx, and it is also clear in this case that the age of the star is poorly constrained.

Literature measurements of rotational modulation for \targety differ as well.
A 6 d rotation period was derived by \citet{Hempelmann2016} using TIGRE \CaII H \& K data, while \citet{Hirsch2021} identify a significant 30 d signal in the HIRES S-index data.
We suspect the latter result is more likely to be the true rotation period given the long (17.52 yr) activity cycle period reported by \citet{Isaacson2024} based on their HIRES S-index measurements. Solar-type stars are found to follow well-established relations between rotation period and activity cycle period \citep[e.g.,][]{Bohm-Vitense2007,Metcalfe2017,Mittag2023}, where a 6 d rotation period would be an extreme outlier for a cycle exceeding 15 yr. This is confirmed by our analysis in Section \ref{sec:activity_fitting}.

\citet{Soubiran2004} and \citet{Mahdi2016} identify \targety as a solar analog based on its spectroscopic resemblance to the Sun and the close match in fundamental parameters and individual elemental abundances. However, \citet{Mahdi2016} rule this star out as a potential solar sibling due to its significantly different Galactic space velocities and older age. \targety is nevertheless a compelling planet search target, as the planetary systems of older solar analogs will inform how planet formation outcomes might have differed at earlier epochs.


\subsection{HD 10780}\label{sec:stellar_HD10780}

\targetz is a K0V star with effective temperature $T_{\rm eff}=5344$ K, surface gravity $\log g = 4.54$, and mass of $M_\star = 0.90\,{\rm M}_\odot$. At a distance of just $d=10.04$ pc, \targetz is among the highest priority targets for direct imaging with facilities such as the Habitable Worlds Observatory \citep[HWO;][]{Feinberg2024}. Following \citet{Mamajek2024}, we calculate the Earth-equivalent insolation distance (EEID) to be $r_{\rm EEID} = 0.721$ au ($\theta_{\rm EEID} = 71.8$ mas). Adopting an inner working angle of 70 mas and scaling the conservative habitable zone limits of $0.99$ au to $1.7$ au \citep[moist greenhouse to maximum greenhouse, as given by][]{Kopparapu2013} to the luminosity of \targetz, HWO will be able to image an Earth-sized planet anywhere in the habitable zone at contrasts of $>10^{-10}$.

The \citet{Brewer2016} age of $4.4_{-2.7}^{+3.4}$ Gyr is again the youngest of the isochrone-derived ages, though still consistent with all other measurements. Other isochrone-derived ages include a measurement from \citet{daSilva2015} of $6.2^{+4.1}_{-3.9}$ Gyr (Yonsei-Yale isochrones) and two measurements from \citet{Casagrande2011} of $6.21_{-4.53}^{+4.80}$ Gyr (Padova isochrones) and $6.50_{-4.64}^{+4.78}$ Gyr (BaSTI isochrones). Gyrochronological ages compiled by \citet{Ware2026}, however, indicate the star is significantly younger than the Sun. \citet{Barnes2007} present a rotation-age relation with which they calculate \targetz to be just $1.945^{+0.28}_{-0.28}$ Gyr old. Using the rotation-age relation derived by \citet{Mamajek2008}, \citet{Maldonado2010} calculate an age of 2.53 Gyr (no uncertainties given). Both of these results use a stellar rotation period of 23 d as determined by \citet{Baliunas1996} from Mount Wilson S-index measurements. \citet{Ramirez2012} also use the \citet{Barnes2007} rotation-age relation but report a different age of $1.90^{+0.40}_{-0.40}$ Gyr, as they adopt a rotation period of 21.7 d as derived by \citet{Gaidos2000} from photometric measurements with the 0.75 m automated photoelectric telescopes at Fairborn Observatory. 

In addition to the rotation periods from \citet{Baliunas1996} and \citet{Gaidos2000}, \citet{Olspert2018} carried out an independent analysis of the Mount Wilson data and derived a similar period of $22.14\pm0.55$ d. \citet{Olspert2018} also calculate an activity cycle period of $7.53\pm0.16$ yr. This is consistent with the findings of \citet{Baliunas1995}, who simply establish a cycle lower limit of 7 yr, but differs somewhat from the $9.69$ yr activity cycle derived by \citet{Schmitt2026} using the combined Mount Wilson + TIGRE data set. \citet{Schmitt2026} do, however, report a secondary periodogram peak at 7.55 yr. \citet{Baum2022} analyze the combined Mount Wilson + HIRES S-index measurements for this star and classify it as variable with no clear periodic cycle detection. These results point to a variable, or complex, activity cycle for \targetz, which can lead to the detection of multiple or different cycle periods over different baselines \citep{Olah2009}. Complex cycles are predominantly found for stars younger than 3 Gyr \citep{Olah2016}, which suggests the gyrochronology age results are more accurate.

\subsection{TESS Photometric Variability}\label{sec:tessphot}

To complement the literature record, we searched for rotational variability using autocorrelation functions computed from the quality-filtered, sector-normalized TESS-Gaia Light Curve  \citep[TGLC;][]{Han2023} aperture photometry and PSF photometry. To avoid assuming that active regions remain coherent across the multi-year TESS baseline, the autocorrelation functions were evaluated separately within contiguous observing campaigns and then combined. \targetx has only one observed sector and therefore does not provide sufficient baseline to test the 26 d rotation period  reported by \citet{Hempelmann2016}. For \targety, which was observed in 10 sectors, we do not find a consistent autocorrelation feature near either 6 d or 30 d. For \targetz, the PSF extraction exhibits a feature near 26 d, but this is not reproduced in the aperture extraction. We therefore do not identify any strong, extraction-independent rotational modulation in the TESS photometry. Rotational photometric variations for these stars are either not common to the aperture and PSF extractions, affected by the limited or discontinuous TESS sampling, or significantly discrepant from the expected rotation periods.

\section{NEID Radial Velocity Calculation}\label{sec:neidrv}

\subsection{CCF}\label{sec:rv_ccf}

As noted in Section \ref{sec:neid}, the NEID pipeline automatically computes and reported CCF-based RVs. While inspecting these velocities, we identified an error in the CCF calculation in version 1.5 of the DRP\footnote{This CCF error is present in all versions of the DRP from 1.0.0 through 1.5.3.}. The NEID CCD contains several columns of bad pixels, which are reported as \textsf{NaN}s in the output spectra. At the start of the CCF calculation, a linear interpolation is performed over each \textsf{NaN} region. This interpolation manifests as time-varying shape deformation for any spectral line that intersects a \textsf{NaN} region, as the interpolation will capture a different part of the line at each epoch due to the combination of barycentric motion and the changing stellar RV signal. Line shape changes are well known to pull the peak of the CCF and bias any resulting RV measurements. Because barycentric motion dominates the observed Doppler shifts for our targets, the net result is the introduction of an annual signal in the CCF RVs reported by the NEID DRP. The amplitude of this effect is $1-2$ m~s$^{-1}$ for the targets we are focusing on in this work, but the exact structure of the signal depends on the systemic velocity of the star (which dictates which lines intersect each bad pixel region) and on the amplitude of the barycentric throw. We revise the DRP CCF algorithm such that any line that will intersect a bad pixel region at any barycentric phase (vacuum wavelength + systemic velocity shift $\pm$ 30 km~s$^{-1}$) is removed from the CCF mask, and we then recompute the CCF RVs for each of our targets. These independently calculated values are used in place of the pipeline RVs for the remainder of this work.

\begin{figure}
    \centering
    \includegraphics[width=\linewidth]{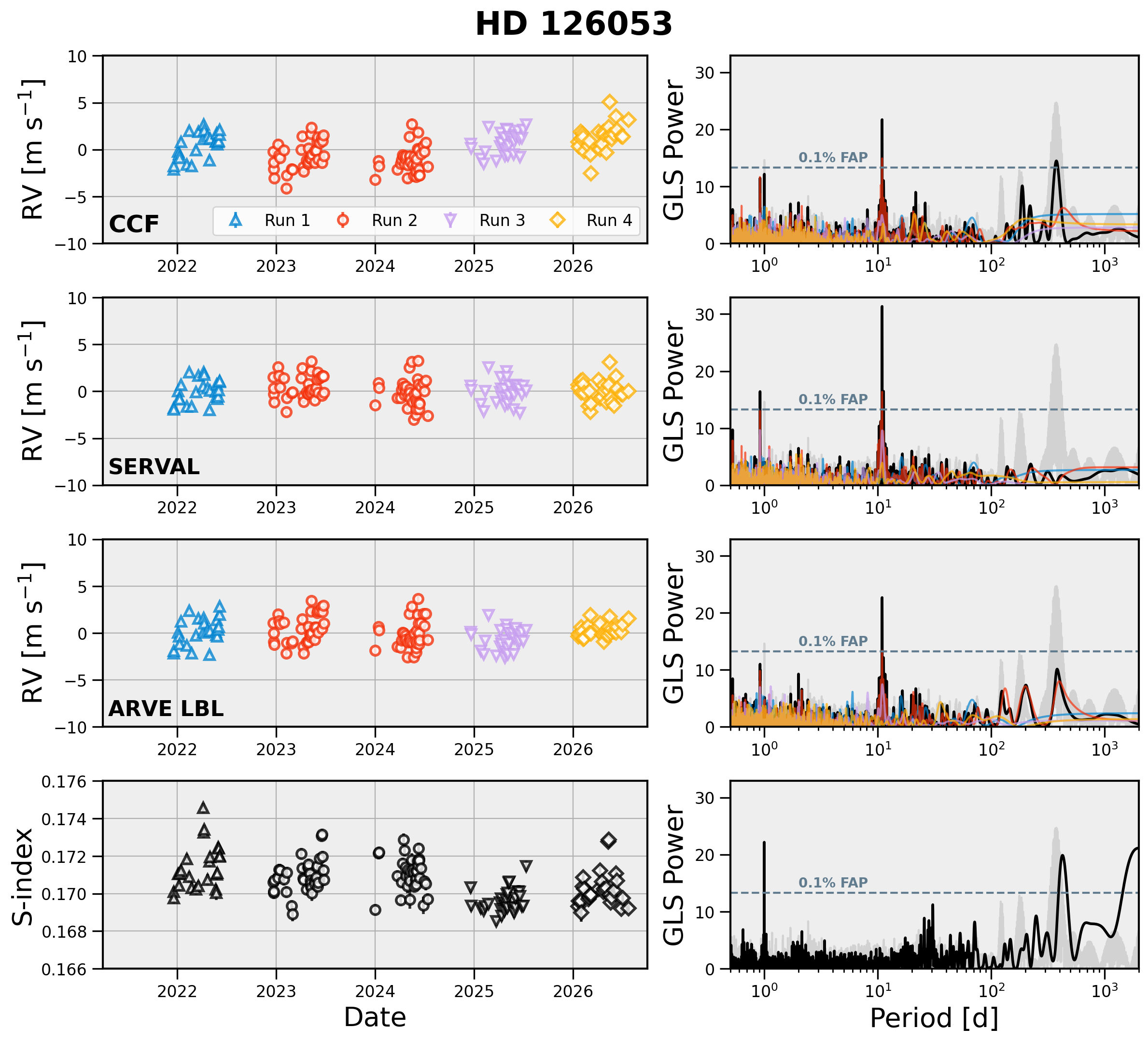}
    \caption{\targetx NEID RVs and S-index measurements (left) and GLS periodograms (right).  We separate the RVs into Run 1 (blue), Run 2 (red), Run 3 (purple), and Run 4 (yellow) based on known changes to the instrument zero point. The GLS periodogram of the combined time series is shown in black, and the window function is plotted in grey in the background. The CCF, SERVAL, and ARVE RV calculations all yield detections of a $P=10.9$ d signal. This signal is not present in the S-index data, but these data do contain a $\sim 1$ yr signal that is also evident in the CCF and ARVE RVs.}
    \label{fig:HD126053_rvs}
\end{figure}

\begin{figure}
    \centering
    \includegraphics[width=\linewidth]{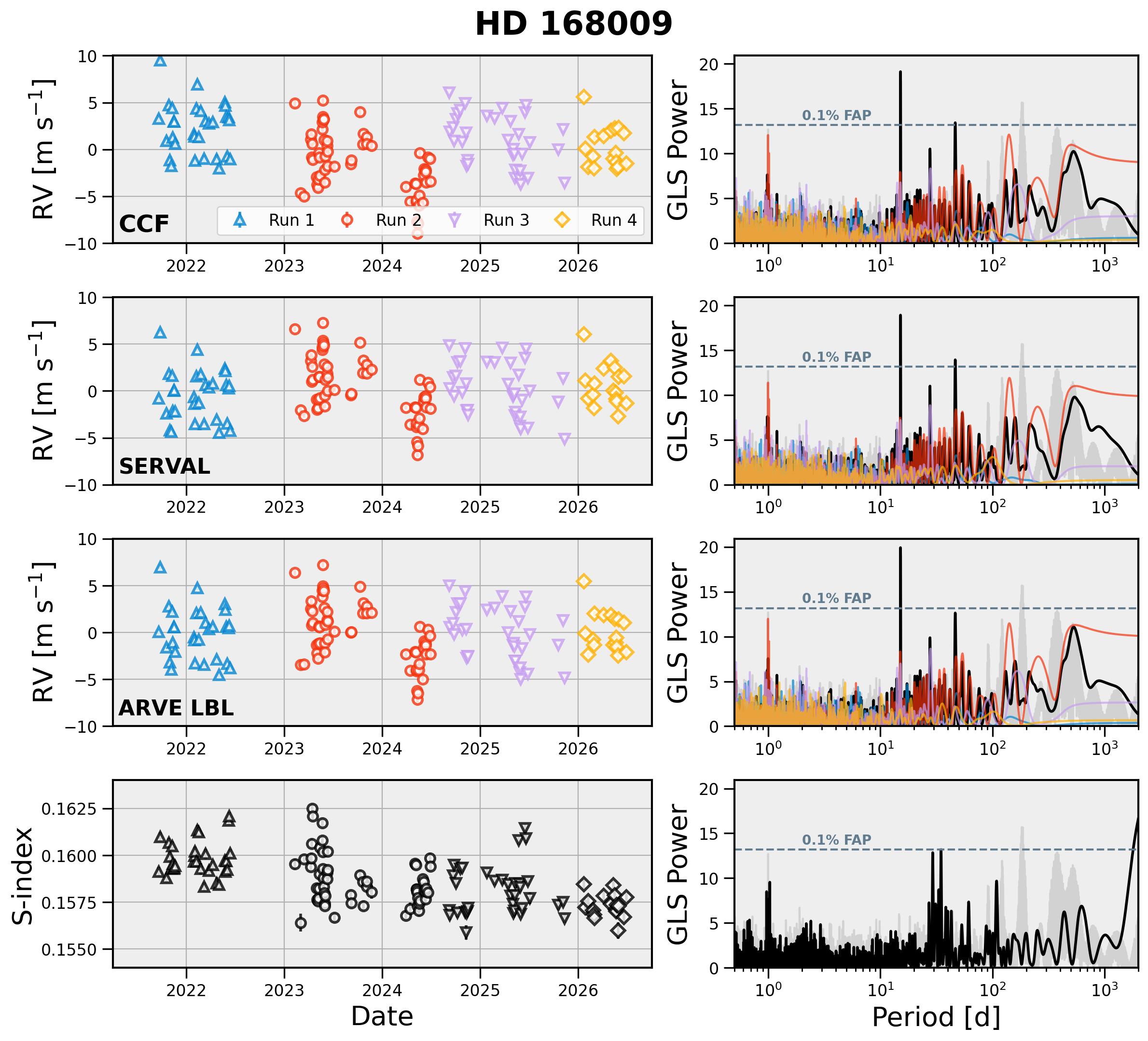}
    \caption{Same as \autoref{fig:HD126053_rvs} but for \targety. All three RV reductions yield signals at $P=15.1$ d, $P = 46.6$ d, and $P=27.7$ d, and there is evidence for a $P\sim30$ d signal in the S-index periodogram.}
    \label{fig:HD168009_rvs}
\end{figure}

\begin{figure}
    \centering
    \includegraphics[width=\linewidth]{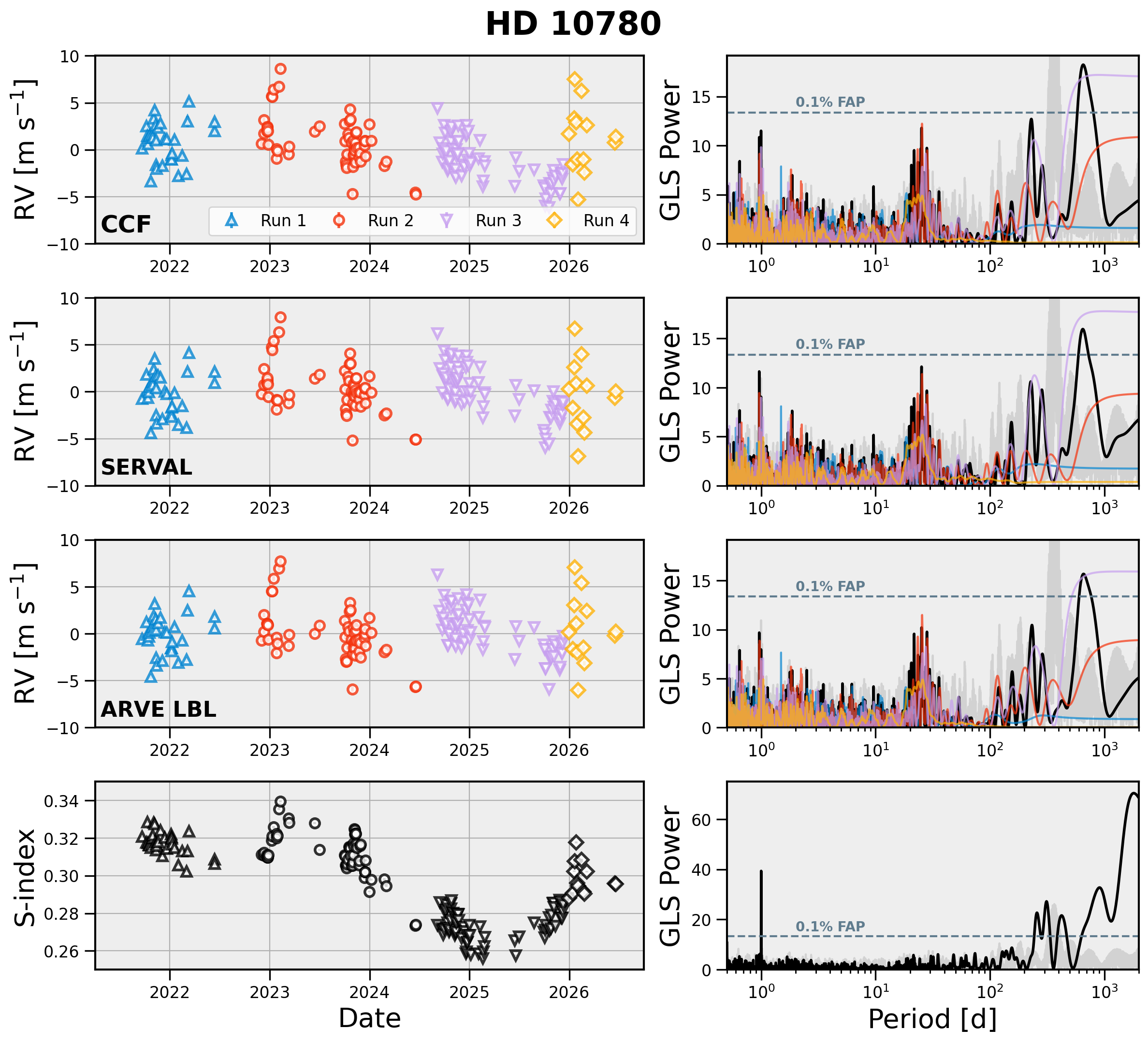}
    \caption{Same as \autoref{fig:HD126053_rvs} but for \targetz. All three RV reductions produce a broad GLS peak near $P\sim650$ d, as well as a power excess near $P=25$ d. The S-index time series is dominated by long period power, consistent with literature measurements of a long-term activity cycle..}
    \label{fig:HD10780_rvs}
\end{figure}

\subsection{SERVAL}\label{sec:rv_serval}

We also process the wavelength-calibrated spectra for our targets with SpEctrum Radial Velocity AnaLyser \citep[SERVAL;][]{Zechmeister2018}. SERVAL uses the template-matching method \citep{Anglada-Escude2012} to compute RVs by comparing each spectrum to a high-S/N reference spectrum. Our implementation of SERVAL includes some changes to tailor the algorithm to NEID data as described in \citet{Stefansson2022}. An underlying assumption for template-matching methods such as SERVAL is that the spectra being compared share the same fundamental shape. We therefore process each NEID Run separately, as there is evidence that the instrumental line spread function changed between runs \citep{Gupta2025}.

\subsection{ARVE}\label{sec:rv_arve}

We calculate a third set of RVs for each star using the Analyzing Radial Velocity Elements (\texttt{ARVE}) software package \citep{AlMoulla2025}, which includes both a CCF-based RV method and a line-by-line, template-based RV method. The latter, which we employ here, functions by measuring the velocities of individual segments of lines. A weighted average RV is calculated for the full spectrum after removing outliers.

We compute line-by-line and weighted average RVs using ARVE's default functionality with the following exceptions. First, instead of applying a single barycentric correction to the entire spectrum, we apply separate corrections to individual echelle orders to account for chromatic variations in the flux-weighted midpoint of each exposure. These barycentric corrections were computed by the NEID DRP using \texttt{barycorrpy} \citep{Kanodia2018} and were provided as a standard pipeline product. In addition, we set a more stringent threshold for telluric contamination of $0.1\%$ in flux (default $1\%$).
We also select only the lines contained in the corresponding NEID DRP line masks, which have been vetted for RV stability, and we restrict the analysis to lines that lie within the free spectral range of each echelle order.
As with SERVAL, we separate the NEID data into different runs before extracting RVs with ARVE.

\subsection{Comparison of NEID RV measurements}\label{sec:rvcomp}

The CCF, SERVAL, and ARVE RVs for \targetx, \targety, and \targetz are shown in Figures \ref{fig:HD126053_rvs}, \ref{fig:HD168009_rvs}, and \ref{fig:HD10780_rvs}, respectively. We compare the single measurement precisions returned by each of these methods and find that SERVAL achieves the best precision, followed by the ARVE LBL calculation and then our modified CCF calculation. The relative performance of the three methods is consistent from run to run and from star to star.

We also compute the generalized Lomb-Scargle (GLS) periodogram \citep{Lomb1976,Scargle1982,Zechmeister2009} for each RV time series, first for each individual run and then for the combined data set for each star. False alarm probabilities (FAP) are calculated following Equation 24 of \citet{Zechmeister2009}. The template-matching methods (SERVAL and ARVE) inherently remove run-to-run zero point offsets because the RVs are computed relative to an average spectrum rather than an absolute wavelength reference. To enable a balanced comparison, we also subtract the median of each run from the CCF RVs before calculating the combined GLS. The periodograms are shown alongside the corresponding RV time series data in Figures \ref{fig:HD126053_rvs}, \ref{fig:HD168009_rvs}, and \ref{fig:HD10780_rvs}. We also compute and show the GLS periodograms of the NEID S-index measurements for comparison. No offsets have been applied in this last calculation, as we neither expect nor observe changes in the S-index zero point between NEID runs.

\subsubsection{HD 126053}\label{sec:rvcomp_HD126053}

For \targetx, SERVAL returns a median single measurement precision of 25 cm~s$^{-1}$ across the full NEID data set, which outperforms our CCF calculation (median precision 44 cm~s$^{-1}$) and the ARVE LBL calculation  (median precision 34 cm~s$^{-1}$). In \autoref{fig:HD126053_rvs}, we show that all three methods yield a strong signal (FAP $\ll 0.1\%$) at a period of $P=10.91$ d. This matches the periodic signal first identified by \citet{Gupta2025} in their analysis of the NEID RVs through July 2024. A second peak at $P=0.92$ d exceeds the $0.1\%$ FAP threshold in the SERVAL time series and is present at a lower level in the CCF and LBL data, but this is best explained by the 1 d sampling alias of the former signal. The CCF and LBL results also show power at 1 yr and harmonics thereof. This might be partially attributed to sampling, as indicated by the coincident power in the window function, but its absence in the SERVAL reduction suggests an alternate cause. Another possibility is telluric contamination, which varies annually due to Earth's barycentric motion and is treated more conservatively by SERVAL than by the other reductions. SERVAL masks spectral regions that are susceptible to contamination, making tellurics less likely to print through to the RVs. A significant S-index peak at 424 d suggests the annual RV signal could instead be driven by stellar activity. This signal is investigated in Section \ref{sec:HD126053yr}. The NEID S-index periodogram also exhibits a weak power excess between 25--32 d, which is consistent with the marginal detection of a 26 d rotation period by \citet{Hempelmann2016}.

\begin{figure*}
    \centering
    \includegraphics[width=0.8\linewidth]{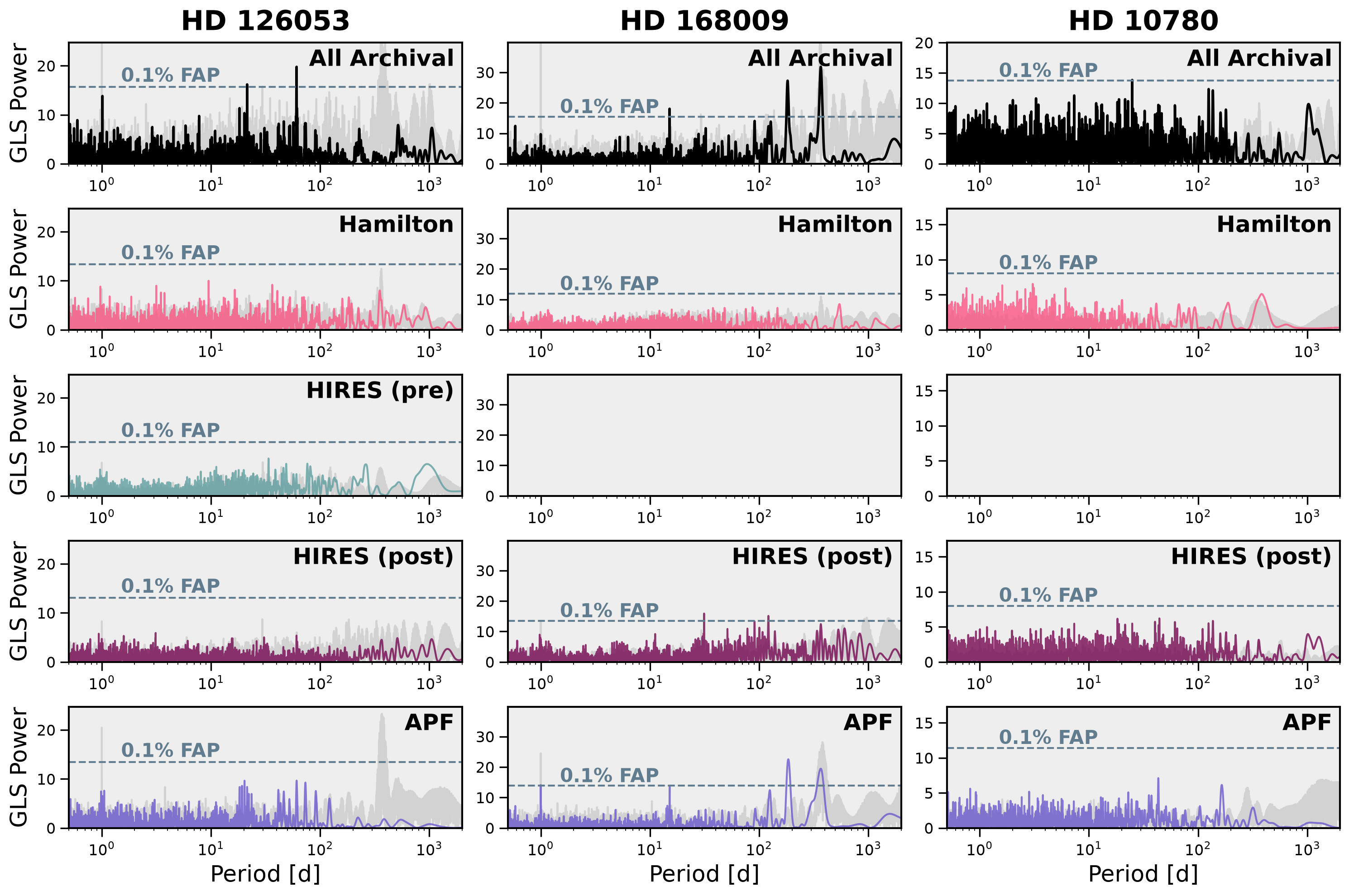}
    \caption{GLS periodograms of the Hamilton, HIRES, and APF RVs for \targetx (left), \targety (center), and \targetz (right). We show the periodograms of the combined time series in the upper panels. For \targetx, we detect signals at $21$ d and $61$ d in the combined periodogram. For \targety, signals are detected at $15$ d,  $182$ d, and $367$ d in the APF and combined periodograms. For \targetz, only a weak $25$ d signal is detected in the combined data set.}
    \label{fig:archival_gls}
\end{figure*}

The GLS periodograms of the Hamilton, HIRES, and APF RVs are shown in \autoref{fig:archival_gls}. There is no strong evidence for periodic signals in any of these data sets on their own, but two signals emerge at 21.4 d and 60.8 d in the combined time series. We run a simple leave-one-out test with each spectrograph and find that the power at 21.4 d is driven almost entirely by APF and the power at 60.8 d carries contributions from APF and the post-upgrade HIRES data. Neither of these signals show pronounced power in the NEID data. We explore this discrepancy further in Section \ref{sec:rvsearch}.

\subsubsection{HD 168009}\label{sec:rvcomp_HD168009}

For \targety, SERVAL consistently performs better than the other reductions, with a median measurement precision of 22 cm~s$^{-1}$ (compared to 39 cm~s$^{-1}$ for the CCF RVs and 29 cm~s$^{-1}$ for the ARVE LBL RVs). We detect a prominent signal (FAP $\ll 0.1\%$) at a period of $P=15.14$ d in each of the CCF, SERVAL, and ARVE RVs, as well as secondary and tertiary peaks at $46.6$ d and $27.7$ d. We attribute the $15.14$ d signal to the planet \planety \citep{Rosenthal2021}; this signal was also successfully recovered in the earlier NEID RV analysis carried out by \citet{Gupta2025}. Though it is not detected at a significant level, we note that the $27.7$ d signal is very close to the most significant periodogram peak in the residuals of the \citet{Rosenthal2021} fit to \planety in the CLS data. The $46.6$ d signal has not previously appeared in the literature. We discuss each of these in more detail in the following sections.
As with \targetx, we also see significant power at the 1 d sampling alias and near annual harmonics. The frequency structure is consistent across all three RV reductions.

The GLS periodograms of the archival RVs for \targety are shown in \autoref{fig:archival_gls}. The $15.14$ d signal from \planety is detected in the APF data and in the combined archival time series, as expected given that these are the RVs with which the planet was first discovered. This signal is not independently detected in the Hamilton or HIRES data, but HIRES detects signals at $P=31.3$ d and $P=121.3$ d. There are also two significant signals at $P=367$ d and $P=182$ d; \citet{Rosenthal2021} classify these as false positives resulting from APF systematics and we adopt the same interpretation.

We note that the 27.7 d RV signal detected in the NEID data and the 31.3 d signal detected in HIRES are close to the 30 d activity signal identified by \citet{Hirsch2021}. The NEID S-index periodogram (\autoref{fig:HD168009_rvs}) also exhibits power in this vicinity, with distinct peaks at $29.3$ d and $34.8$ d (FAP $=0.2\%$ in both cases). We consider a possible activity-driven origin in Sections \ref{sec:analysis} and \ref{sec:discussion}.

\subsubsection{HD 10780}\label{sec:rvcomp_HD10780}

The RVs and GLS periodograms for \targetz are shown in \autoref{fig:HD10780_rvs} and \autoref{fig:archival_gls}. We determine the median measurement precisions across the combined NEID Runs to be 18 cm~s$^{-1}$ for SERVAL, 33 cm~s$^{-1}$ for the CCF RVs, and 25 cm~s$^{-1}$ for the ARVE LBL RVs. The GLS periodogram is dominated by a broad peak centered on $P\approx650$ d with FAP$\ll 0.1\%$.
\citet{Gupta2025} reported the detection of a periodic RV signal at $P = 25.6$ d in the combined Run 1 and Run 2 NEID data. There is a power excess at this period in all three of our NEID RV reductions. However, this power does not exceed the $0.1\%$ FAP threshold, reaching peak significances of just FAP$=0.57\%$ (CCF), FAP$=0.65\%$ (SERVAL), and FAP$=6.29\%$ (ARVE). The lower strength is not surprising given that a separate signal dominates the long-term RV behavior. \citet{Gupta2025} found that the 25.6 d signal was stronger after accounting for a long-term linear RV trend, and we speculate that both the trend and the 650 d signal seen here are tied to the stellar activity cycle which can clearly be seen in the NEID S-index time series (\autoref{fig:HD10780_rvs}).

As we show in \autoref{fig:archival_gls}, a similar $24.9$ d signal is present in the combined archival data set for this star.
Using a leave-one-out test for each instrument, we find that this signal is not dominated by contributions from any single spectrograph.

\section{Analysis} \label{sec:analysis}

\subsection{Periodogram Grid Search}\label{sec:rvsearch}

We begin our analysis by replicating the \rvsearch \citep{Rosenthal2021} iterative periodogram search carried out by \citet{Gupta2025} and extending it to the full Run 1 -- Run 4 NEID data set, first excluding (Section \ref{sec:rvsearch_neid}) and then including (Section \ref{sec:rvsearch_all}) archival RVs.  We adopt a similar set up to \citet{Gupta2025}, which is in turn based on the original implementation as introduced by \citet{Rosenthal2021}, with some minor changes. We summarize our implementation here.

We run \rvsearch over a frequency grid with $f_{\rm min} = \frac{1}{2\tau}$~d$^{-1}$, $f_{\rm max}=\frac{1}{2}$~d$^{-1}$, and $\Delta f =\frac{1}{2\pi\tau}$~d$^{-1}$, where $\tau$ is the total observing baseline for each star. During the first iteration, \rvsearch steps through this grid and computes the difference between the Bayesian Information Criterion (BIC) of the best-fit sinusoid at each period and that of a zero-planet model. A $\Delta$BIC periodogram is constructed and a full Keplerian fit is performed for the most significant peak exceeding a $0.1\%$ FAP threshold. This FAP threshold is determined empirically as described by \citet{Rosenthal2021}. If no peaks exceed this threshold, the search concludes. But if a peak is found, the best-fit Keplerian is subtracted and the grid search is repeated, comparing the $\Delta$BIC between $n+1$ and $n$ planet models. Each fit includes a white-noise jitter term and RV zero point offset for each instrument and each NEID Run. We do not include any long-term trends or curvature terms, as these are expected to be highly degenerate with relative zero point offsets. We use the NEID RVs as computed with SERVAL for each search. The archival RVs are binned into nightly visits to reduce short-timescale scatter before we run \rvsearch.

\subsubsection{\rvsearch Analysis of NEID RVs}\label{sec:rvsearch_neid}

\rvsearch detects one signal for \targetx; the best-fit Keplerian orbit has $P=10.91$ d and $K=1.14$ m~s$^{-1}$ and is consistent with the signal identified by \citet{Gupta2025}. The most prominent peak in the residuals is at $P=28.1$ d, but the $\Delta$BIC falls well short of the $0.1\%$ false alarm threshold. 

For \targety, we detect two signals in the NEID RVs. The first has $P=15.14$ d and $K=2.30$ m~s$^{-1}$ and is again consistent with the signal identified by \citet{Gupta2025} as well as with the planet parameters given by \citet{Hirsch2021} and \citet{Rosenthal2021}. The second has $P=27.75$ d and $K=1.97$ m~s$^{-1}$. This coincides with a weak peak in the GLS periodogram of the NEID RVs but also with potential activity signals, as described in Section \ref{sec:rvcomp_HD168009}. The 46.6 d signal seen in the GLS of the NEID RVs is notably absent here. While there is isolated power near this period in each iteration of the $\Delta$BIC periodogram, it never surpasses the false alarm threshold. 

For \targetz, the first signal recovered has $P=14.6$ yr, $K=6.66$ m~s$^{-1}$, and a moderate eccentricity ($e=0.4$). This is followed by a $P=25.26$ d, $K=1.38$ m~s$^{-1}$ signal. Detections of eccentric signals with periods that exceed the observing baseline should typically be treated with caution, particularly in cases such as this where the baseline spans multiple unconstrained zero point offsets. The periapse phase of the 14.6 yr orbit is well-sampled and the semi-amplitude is large relative to the measurement error, lending some confidence that this is not simply a sampling artifact, but a lack of sampling at other phases indicates the period is poorly constrained. In addition, both of the detected signals have timescales comparable to previous measurements of the stellar activity cycle and stellar rotation period (Section \ref{sec:stellar_HD10780}), suggesting that these variations are not induced by an orbiting planet but rather intrinsic to the star.

\subsubsection{\rvsearch Analysis of All RVs}\label{sec:rvsearch_all}

With the addition of archival RVs for \targetx, the \rvsearch results are nearly unchanged. We still detect one prominent signal at $P=10.91$ d, albeit with a slightly lower amplitude of $K=1.10$ m~s$^{-1}$, and the residual periodogram has a peak at $P=28.1$ d that does not exceed the false alarm threshold. This result demonstrates that although the archival data do not exhibit significant power at 10.91 d on their own, they are not at odds with the signal seen in the NEID RVs. Further, there is no evidence for the 21 d and 61 d signals that were present in the archival data, suggesting that these are inconsistent with the higher precision NEID RVs and thus likely to be either transient or spurious.

The archival data substantially alter the \rvsearch results for \targety. While the 15.14 d signal from \planety is detected first and a 27.77 d signal is still present, the latter is preceded by a detection at 184.05 d and followed by a marginal non-detection (FAP$=0.11\%$) at 380 d. As we note in Section \ref{sec:rvcomp_HD168009}, the power at these annual harmonics appears to be driven by the APF data, which supports the conclusion from \citet{Rosenthal2021} that these stem from APF systematics. We confirm this by repeating the search with just NEID, HIRES, and Hamilton RVs. The results are nearly identical to the NEID-only case; the two detected signals have 
$P=15.14$ d and $K=2.26$ m~s$^{-1}$ and
$P=27.76$ d and $K=1.93$ m~s$^{-1}$. 
Given the inconsistency between the APF data and the data from the other spectrographs, we omit the APF RVs from our subsequent analysis of this system.

For \targetz, \rvsearch again recovers an eccentric long-period signal ($P=2.05$ yr, $K=3.0$ m~s$^{-1}$) and a weaker signal close to the stellar rotation period ($P=25.37$ d, $K=1.3$ m~s$^{-1}$). The difference between the long-period signal detected in the NEID-only RVs and that detected here further points to stellar activity, as the complex stellar cycle can not be consistently explained by a single Keplerian signal.

\subsection{Stellar Activity Fitting}\label{sec:activity_fitting}

We fit and compare three models of the NEID S-index time series for each target to determine which model best describes the corresponding stellar activity variations over the NEID baseline. These are: white noise ``jitter'' only (WN), white noise jitter + stellar rotation (WN+Rot), and white noise jitter + stellar rotation + stellar activity cycle (WN+Rot+Cyc). Rotational modulation and stellar activity cycles are modeled using quasiperiodic (QP) Gaussian process (GP) kernels. Model fitting is carried out with the nested sampler routine \texttt{NestedSamplers.jl} \citep{Lucas2023} using a random walk with 500 live points in each step. The cutoff threshold is set to $\Delta \log Z$ = 0.01, where $\log Z$ is the natural logarithm of the Bayesian evidence.

We adopt broad priors for the amplitude ($\alpha$), harmonic complexity ($\Gamma$), and decay timescale ($\tau$) hyperparameters for both the rotation and activity cycle kernels. A broad prior is also used for the activity cycle period ($P_{\rm cyc}$), but we impose narrower priors on the stellar rotation period ($P_{\rm rot}$) hyperparameters based on literature measurements as discussed in Section \ref{sec:target_stars}. The complete fitting basis and prior distributions are given in Table \ref{tab:activity_fit}.

\begin{deluxetable*}{lccccc}
\tablecaption{Stellar Activity Fitting Results for \targetx, \targety, and \targetz \label{tab:activity_fit}}
\tablehead{Parameter & Prior & \multicolumn{3}{c}{Posterior} & Unit} 
\startdata
\hline
& & \targetx & \targety & \targetz & \\
\hline
$\log \alpha_{\rm rot}$ & $\mathcal{U}(-10,5)$ & $-7.027^{+0.080}_{-0.077}$ & $-6.80^{+0.10}_{-0.08}$& $-4.90^{+0.11}_{-0.11}$& \\
$\log P_{\rm rot}$ & $\mathcal{N}(\log P_{\rm lit}, 0.2)$ & $3.361^{+0.070}_{-0.058}$ & $3.402^{+0.035}_{-0.030}$& $3.053^{+0.027}_{-0.025}$ & log d \\
$\log \tau_{\rm rot}$ & $\mathcal{N}(4,1)$ & $2.98^{+0.12}_{-0.14}$& $3.58^{+0.43}_{-0.21}$& $3.252^{+0.089}_{-0.074}$ & log d \\
$\log \Gamma_{\rm rot}$ & $\mathcal{N}(0,1)$ & $0.178^{+0.26}_{-0.31}$& $0.33^{+0.29}_{-0.31}$ & $-0.55^{+0.25}_{-0.26}$ & \\
$\alpha_{\rm rot}$ & derived & $0.00089^{+0.00007}_{-0.00007}$& $0.00111^{+0.00012}_{-0.00009}$& $0.00742^{+0.00087}_{-0.00076}$ & \\
$P_{\rm rot}$ & derived & $28.8^{+2.1}_{-1.6}$& $30.1^{+1.1}_{-0.9}$& $21.18^{+0.59}_{-0.53}$ & d\\
$\tau_{\rm rot}$ & derived & $19.7^{+2.5}_{-2.6}$ & $35.7^{+12.6}_{-6.8}$ & $25.9^{+2.4}_{-1.8}$ & d\\
$\Gamma_{\rm rot}$ & derived & $1.19^{+0.35}_{-0.32}$& $1.39^{+0.46}_{-0.37}$& $0.58^{+0.17}_{-0.13}$ & \\
$\log \alpha_{\rm cyc}$ & $\mathcal{U}(-15,10)$ & \nodata & \nodata & $-3.80^{+0.39}_{-0.35}$ & \\
$\log P_{\rm cyc}$ & $\mathcal{U}(0,3.912)$ & \nodata & \nodata & $1.76^{+0.44}_{-0.27}$& log yr\\
$\log \tau_{\rm cyc}$ & $\mathcal{U}(0,3.912)$ & \nodata & \nodata &  $0.96^{+1.37}_{-0.65}$& log yr\\
$\log \Gamma_{\rm cyc}$ & $\mathcal{N}(0,1)$ & \nodata & \nodata & $0.19^{+0.54}_{-0.62}$& \\
$\alpha_{\rm cyc}$ & derived & \nodata & \nodata & $0.022^{+0.011}_{-0.007}$ & \\
$P_{\rm cyc}$ & derived & \nodata & \nodata & $5.8^{+3.3}_{-1.4}$ & yr\\
$\tau_{\rm cyc}$ & derived & \nodata & \nodata & $2.6^{+7.6}_{-1.2}$& yr\\
$\Gamma_{\rm cyc}$ & derived & \nodata & \nodata & $1.21^{+0.87}_{-0.57}$ & \\
$\log \sigma_{\rm WN}$ & $\mathcal{U}(-10,10)$ & $-8.69^{+0.20}_{-0.25}$ & $-8.27^{+0.26}_{-0.38}$& $-6.502^{+0.094}_{-0.086}$ & \\
$\sigma_{\rm WN}$ & derived & $0.00017^{+0.00004}_{-0.00004}$  & $0.00026^{+0.00008}_{-0.00005}$ & $0.00150^{+0.00015}_{-0.00012}$ & \\
\hline
\multicolumn{2}{l}{$\Delta \log Z_{({\rm WN+Rot})-({\rm WN)}}$} & \textbf{71.4} & \textbf{67.5} & 251.2 & \\
\multicolumn{2}{l}{$\Delta \log Z_{({\rm WN+Rot+Cyc})-({\rm WN)}}$} & 71.2 & 66.9 & \textbf{265.4} & \\
\hline
\enddata
\tablenotetext{}{The $\log P_{\rm rot}$ priors are centered on 26 d for \targetx, 30 d for \targety, and 22 d for \targetz.}
\end{deluxetable*}

Using the Bayesian evidence to measure fit quality, the best model for both \targetx and \targety is the WN+Rot model. For \targetx, this model has $\Delta \log Z = 71.4$ relative to the WN model and $\Delta \log Z = 0.2$ relative to the WN+Rot+Cyc model. For \targety, the WN+Rot model has $\Delta \log Z = 67.5$ relative to the WN model and $\Delta \log Z = 0.6$ relative to the WN+Rot+Cyc model. We note that the $\Delta \log Z$ metric only marginally favors the WN+Rot model over the WN+Rot+Cyc model in both cases, but because WN+Rot is the lower-complexity model, we find the margin sufficient to select this as the preferred activity model for each of these stars. In addition, we find that the choice of model does not significantly affect our treatment of short-timescale activity variations. The posteriors on the rotation kernel hyperparameters from the WN+Rot+Cyc and WN+Rot fits are consistent to within one standard deviation. For \targetz, the WN+Rot+Cyc model has $\Delta \log Z = 265.4$ relative to the WN model and $\Delta \log Z = 14.2$ relative to the WN+Rot model. We therefore adopt the WN+Rot+Cyc model, which is strongly preferred.

The posteriors on the adopted model parameters for each star are given in Table \ref{tab:activity_fit}. The best-fit activity cycle for \targetz, 5.8$^{+3.3}_{-1.4}$ yr, is consistent with the value derived by \citet{Olspert2018} but slightly lower than the dominant period identified by \citet{Schmitt2026}. This discrepancy is likely due to the relatively short NEID baseline, which limits the fidelity of our result. The rotation periods for each star are also consistent with literature values. This is expected given that we imposed Gaussian priors centered on the literature values, but we highlight that the posteriors are significantly narrower than the priors (by a factor of three for \targetx, a factor of six for \targety, and a factor of eight for \targetz), so the results are not highly sensitive to the priors, i.e., the activity signal is indeed modulated at these periods in the NEID data.

\subsection{Radial Velocity Fitting and Model Comparison}\label{sec:rvfit}

We model the RVs for each star using the \textsf{exoplanet} package \citep{exoplanet:zenodo} to fit Keplerian orbits and the Hamiltonian Monte Carlo sampler PyMC \citep{pymc2023} for GP regression and parameter estimation. The fitting basis for each planet consists of the semi-amplitude ($K$), orbital period ($P$), time of periastron ($T_p$), and eccentricity and argument of periastron resampled on the unit disk as $\sqrt{e}\cos\omega$ and $\sqrt{e}\sin\omega$. Each GP component is modeled using a QP kernel as in Section \ref{sec:activity_fitting}. We also include separate systemic velocity offsets ($\gamma$) and jitter terms ($\sigma$) for each instrument, and each run of NEID data is treated as an independent instrument. Uniform, non-informative priors are used for the orbital parameters, offset and jitter terms, and GP amplitudes and harmonic complexities. We impose Gaussian priors on the timescale hyperparameters of the rotation GP, $P_{\rm rot}$ and $\tau_{\rm rot}$, with means and widths informed by the posteriors from the S-index fit. For the activity cycle, which is only used for \targetz, we adopt the same uniform priors on  $\log P_{\rm cyc}$ and $\log \tau_{\rm cyc}$ as in the S-index fit, as these timescales are not well constrained by the NEID activity measurements. The complete set of fit parameters and priors is listed in Tables \ref{tab:HD126053_fit}, \ref{tab:HD168009_fit}, and \ref{tab:HD10780_fit}.

We consider constant velocity, single-planet (1P), rotation GP (GP$_{\rm rot}$), and single-planet + rotation GP models for each star. For \targety, we also consider a second planet (2P), and for \targetz, we test models both with and without an activity cycle GP component (GP$_{\rm cyc}$). After computing the posterior samples, we use the maximum-likelihood solution to calculate the BIC and perform model comparison. Further details are provided in the Appendix. Models with smaller BIC values are preferred. Each fit is first run using only the NEID data and then repeated with the full RV time series, and the archival data are again binned into nightly visits.
Our analysis focuses on the fits to the full RV time series for each star, with the NEID-only fits serving as a consistency check.

\subsubsection{HD 126053 Model Comparison}\label{sec:modelcomp_HD126053}

\begin{figure}
    \centering
    \includegraphics[width=\linewidth]{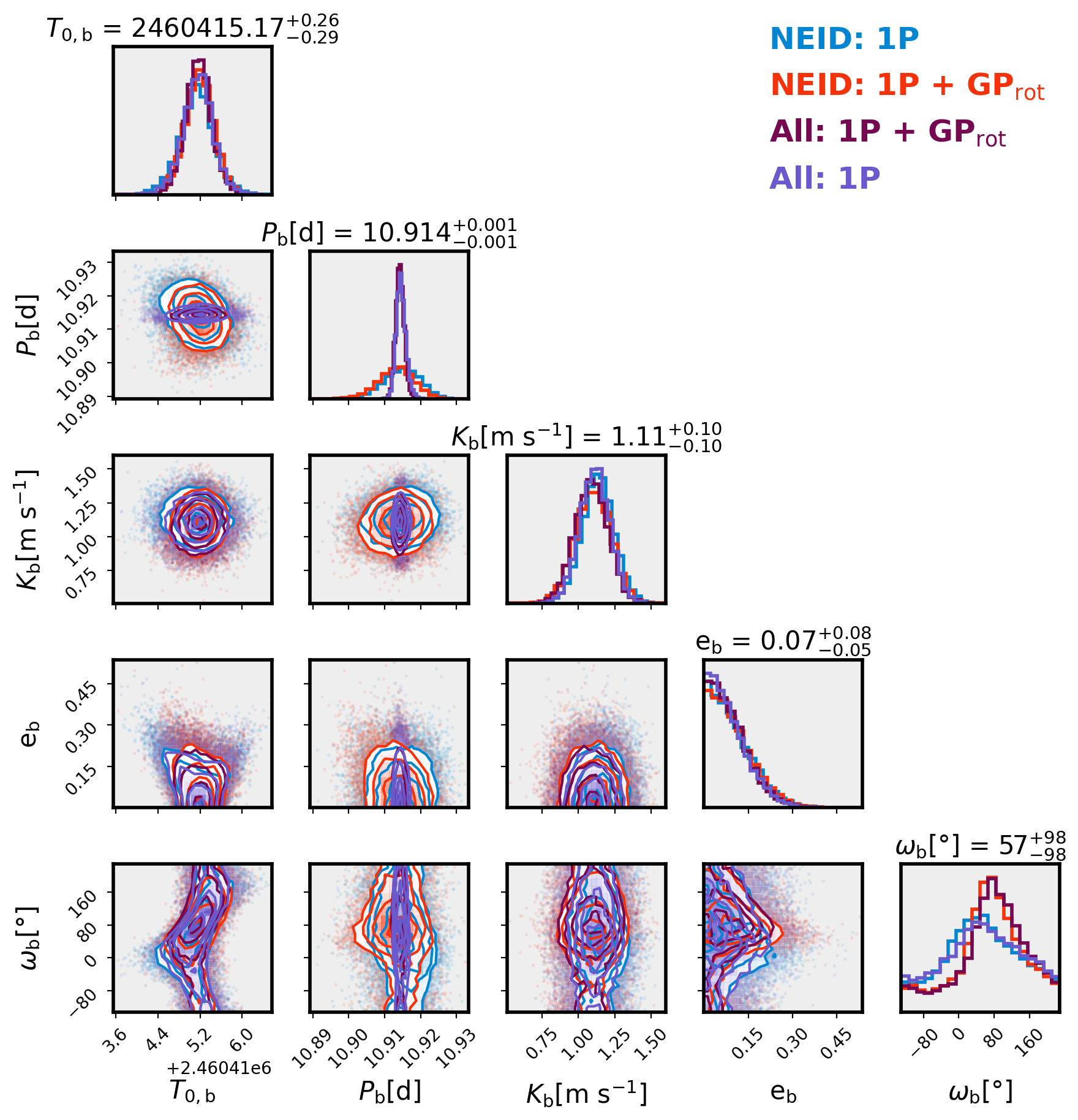}
    \caption{Posterior distributions for the orbital elements of \planetx for the single-planet (1P; blue), single-planet + GP (1P+GP$_{\rm rot}$; red) fits to the NEID RVs and the 1P (purple) and 1P+GP$_{\rm rot}$ (maroon) fits to the full RV time series. The posteriors are consistent across all fits.}
    \label{fig:HD126053_corner}
\end{figure}

The 1P and 1P+GP$_{\rm rot}$ models are strongly preferred over the constant velocity and GP$_{\rm rot}$ models for \targetx. While the 1P+GP$_{\rm rot}$ model has a slightly higher likelihood, with $\Delta \log L_{(1{\rm P+GP}_{\rm rot}) -(1{\rm P})} = 9.4$, the $\Delta$BIC swings in favor of the 1P model due to the added complexity penalty from the GP ($\Delta$BIC$_{(1{\rm P+GP}_{\rm rot}) -(1{\rm P})} = 6.0$). We show the posteriors on the Keplerian orbital parameters for the 1P and 1P+GP$_{\rm rot}$ fits in \autoref{fig:HD126053_corner}. The posterior distributions are fully consistent with each other, demonstrating that the best-fit orbit is insensitive to the inclusion of a GP component. We interpret these results as a confident detection of \planetx with $P_b=10.91$ d and $K_b = 1.11$ m~s$^{-1}$, and we adopt the 1P model as our preferred RV solution.

\begin{deluxetable}{lcrrrc}
\tablecaption{Best-fit RV Model Parameters for \targetx \label{tab:HD126053_fit}}
\tablehead{Parameter & Prior & & \multicolumn{1}{c}{Posterior} && Unit} 
\startdata
\hline
$T_{0,b}$ & $\mathcal{U}(2460408,2460422)$ & &$2460415.17^{+0.26}_{-0.29}$ & &BJD \\
$P_b$ & $\mathcal{U}(10,12)$ & &$10.9145^{+0.0014}_{-0.0012}$ & & d \\
$\sqrt{e}\cos\omega_b$ & $\mathcal{U}(-1,1)$ & &$0.04^{+0.22}_{-0.23}$ & &\nodata \\
$\sqrt{e}\sin\omega_b $& $\mathcal{U}(-1,1)$ & &$0.07^{+0.19}_{-0.22}$ & &\nodata \\
$e_b $& derived &&$ 0.07^{+0.08}_{-0.05}$ & &\nodata \\
$\omega_b $ & derived && $56^{+98}_{-97}$ & & deg \\
$ K_b$ & $\mathcal{U}(0,10) $&& $1.11^{+0.10}_{-0.10}$ &  &m~s$^{-1}$ \\
$ m_b \sin i $& derived && $3.60^{+0.34}_{-0.34}$ & &${\rm M}_\oplus $ \\
$ a_b$ & derived && $0.0933^{+0.0010}_{-0.0010}$ & & au \\
$ \gamma_{\rm NEID, Run 1} $& $\mathcal{U}(-10,10)$ && $-0.41^{+0.22}_{-0.23}$ & & m~s$^{-1}$ \\
$ \gamma_{\rm NEID, Run 2} $& $\mathcal{U}(-10,10)$ && $0.09^{+0.12}_{-0.12}$ & & m~s$^{-1}$ \\
$ \gamma_{\rm NEID, Run 3} $& $\mathcal{U}(-10,10)$ && $-0.01^{+0.14}_{-0.14}$ & & m~s$^{-1}$ \\
$ \gamma_{\rm NEID, Run 4} $& $\mathcal{U}(-10,10)$ && $0.29^{+0.19}_{-0.18}$ & & m~s$^{-1}$ \\
$ \gamma_{\rm Hamilton} $&$ \mathcal{U}(-10,10)$ && $0.21^{+1.18}_{-1.20}$ &  &m~s$^{-1}$ \\
$ \gamma_{\rm HIRES-pre} $& $\mathcal{U}(-10,10)$ &&$ 0.87^{+0.49}_{-0.50}$ & & m~s$^{-1}$ \\
$ \gamma_{\rm HIRES-post} $& $\mathcal{U}(-10,10)$& & $-0.97^{+0.58}_{-0.56}$ & & m~s$^{-1}$ \\
$ \gamma_{\rm APF} $& $\mathcal{U}(-10,10)$ && $-1.16^{+0.29}_{-0.30}$ & & m~s$^{-1}$ \\
$ \sigma_{\rm NEID, Run 1} $& $\mathcal{U}(0,10)$ && $1.09^{+0.19}_{-0.16}$ & & m~s$^{-1}$ \\
$ \sigma_{\rm NEID, Run 2} $& $\mathcal{U}(0,10)$ && $1.01^{+0.10}_{-0.08}$ & &  m~s$^{-1}$ \\
$ \sigma_{\rm NEID, Run 3} $&$ \mathcal{U}(0,10)$ && $0.71^{+0.12}_{-0.10}$ & & m~s$^{-1}$ \\
$ \sigma_{\rm NEID, Run 4} $& $\mathcal{U}(0,10)$ && $0.94^{+0.16}_{-0.13}$ & & m~s$^{-1}$ \\
$ \sigma_{\rm Hamilton} $& $\mathcal{U}(0,10)$ && $6.27^{+1.28}_{-1.18}$ & & m~s$^{-1}$ \\
$ \sigma_{\rm HIRES-pre}$ & $\mathcal{U}(0,10)$ && $2.82^{+0.44}_{-0.37}$ & & m~s$^{-1}$ \\
$ \sigma_{\rm HIRES-post} $& $\mathcal{U}(0,10)$ && $4.78^{+0.46}_{-0.41}$ & & m~s$^{-1}$ \\
$ \sigma_{\rm APF} $& $\mathcal{U}(0,10)$ && $3.36^{+0.23}_{-0.21}$ & & m~s$^{-1}$ \\
\hline
\enddata
\tablenotetext{}{The posteriors given here correspond to the single-planet (1P) fit to the full RV time series.}
\end{deluxetable}

\subsubsection{HD 168009 Model Comparison}\label{sec:modelcomp_HD168009}

\begin{figure*}
    \centering
    \includegraphics[width=0.495\linewidth]{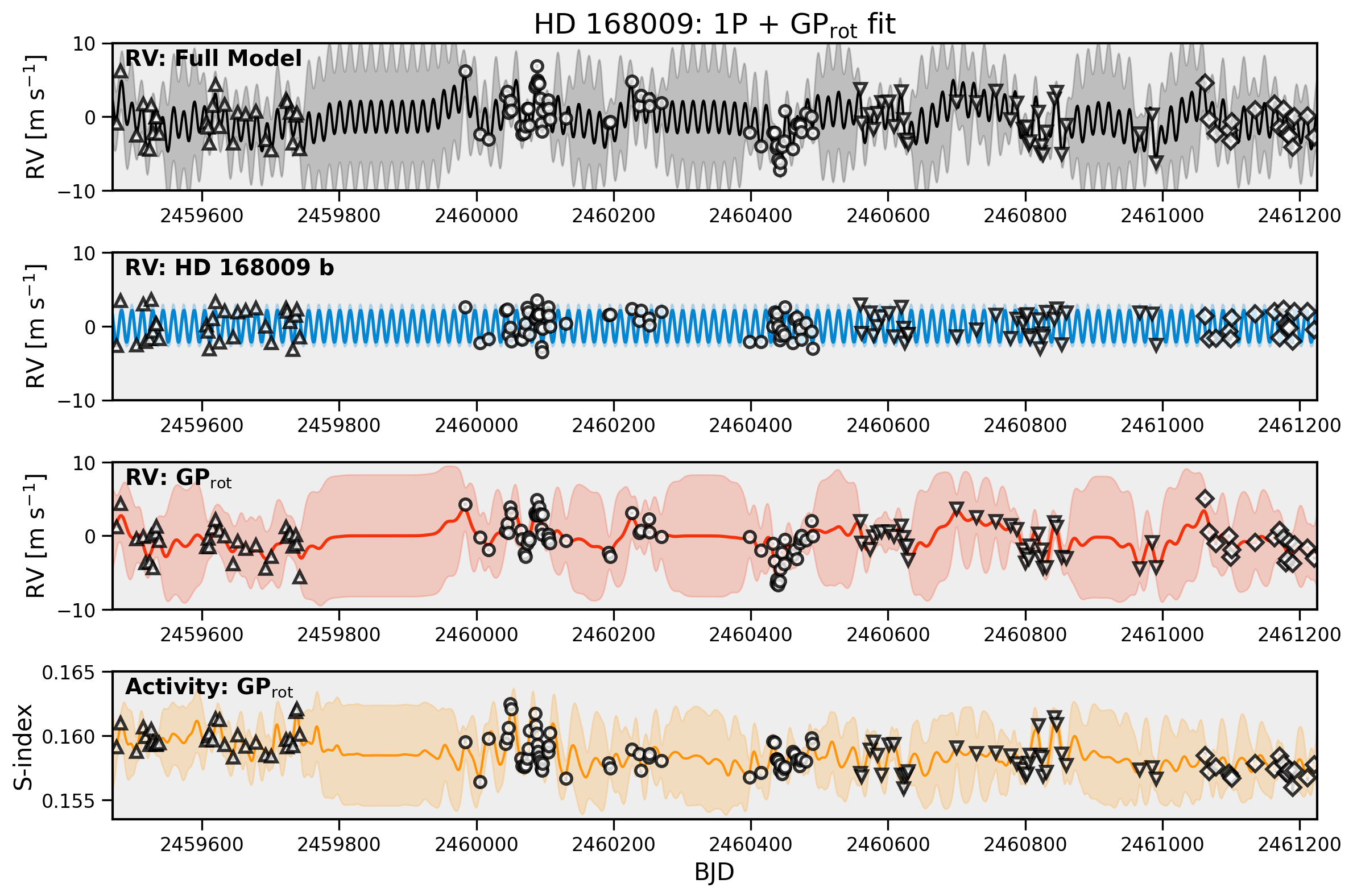}
    \includegraphics[width=0.495\linewidth]{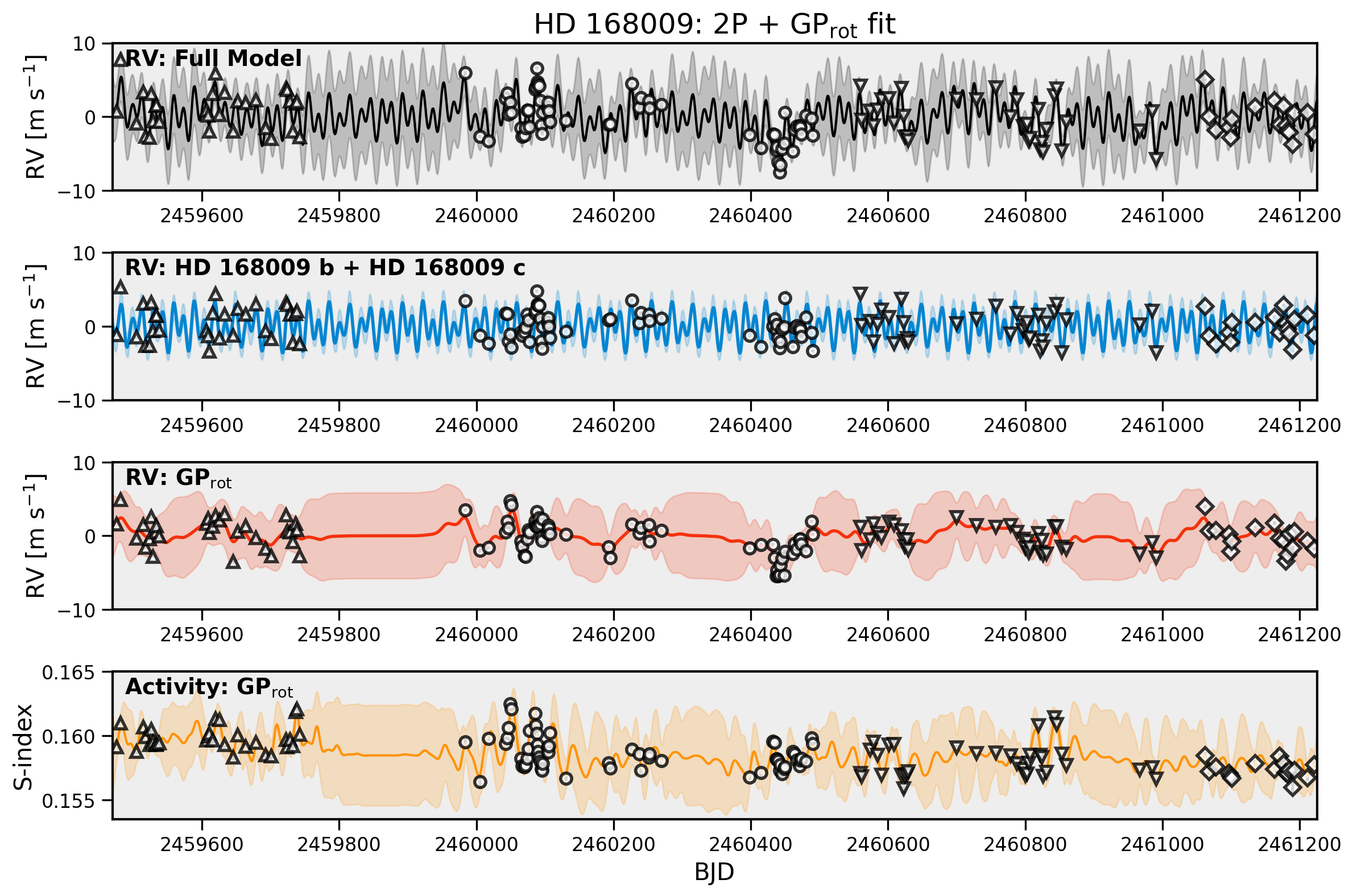}
    \caption{Best-fit 1P+GP$_{\rm rot}$ (left) and 2P+GP$_{\rm rot}$ (right) RV models for \targety. The full models are shown in the top row, followed by the isolated Keplerian components (second row), the GP components (third row), and the GP$_{\rm rot}$ fit to the S-index data (bottom row).}
    \label{fig:HD168009_rv_models}
\end{figure*}

\begin{figure*}
    \centering
    \includegraphics[width=\linewidth]{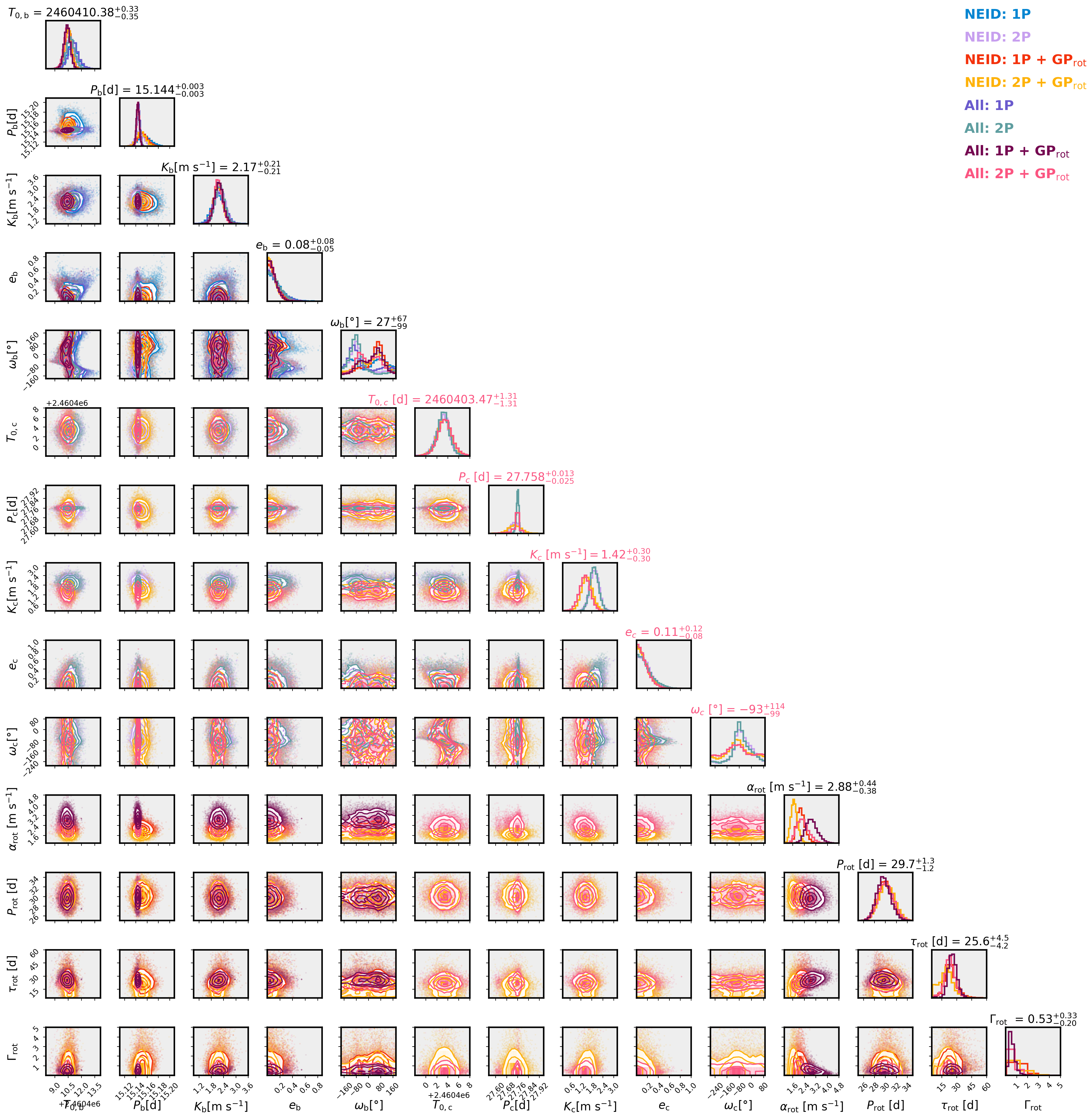}
    \caption{Posterior distributions for the orbital elements of \planety, \planetz, and the best-fit GP$_{\rm rot}$ component for \targety. We show the posteriors for the single-planet (1P; blue), two-planet (2P; lavender), single-planet + GP (1P+GP$_{\rm rot}$; red), and two-planet + GP (1P+GP$_{\rm rot}$; yellow) fits to the NEID RVs and the 1P (purple), 2P (green), 1P+GP$_{\rm rot}$ (maroon), and 2P+GP$_{\rm rot}$ (pink) fits to the full RV time series. The posteriors on most parameters are consistent across all fits. Discrepant results for $\omega_b$, $K_c$, and $\alpha_{\rm rot}$ are discussed in Section \ref{sec:modelcomp_HD168009}.}
    \label{fig:HD168009_corner}
\end{figure*}

For \targety, the 1P+GP$_{\rm rot}$ model and 2P+GP$_{\rm rot}$ model outperform the constant velocity, GP$_{\rm rot}$, 1P, and 2P models. As with \targetx, the higher complexity model (in this case 2P+GP$_{\rm rot}$) has a higher likelihood, but the less complex 1P+GP$_{\rm rot}$ model is favored by a margin of $\Delta$BIC$_{(2{\rm P+GP}_{\rm rot}) -(1{\rm P+GP}_{\rm rot})} = 1.0$. We show the RV time series for the best-fit 1P+GP$_{\rm rot}$ and 2P+GP$_{\rm rot}$ models over the NEID baseline in \autoref{fig:HD168009_rv_models}, along with the isolated Keplerian and GP components and the fit to the NEID S-index measurements. The temporal structures of the GP components of these two fits are similar, albeit with different amplitudes, and both of them resemble the S-index time series. We compare the posteriors on all of the 1P and 2P model fits in \autoref{fig:HD168009_corner}. The best-fit orbital parameters and GP hyperparameters are mostly insensitive to our choice of model, with only $\alpha_{\rm rot}$, $K_c$, and $\omega_b$ varying perceptibly from model to model. The posteriors on these parameters are nevertheless consistent to better than 1-$\sigma$. We attribute the differences in $\omega_b$ to the low eccentricity, which naturally impedes any strong constraints on the orientation of the orbit. The amplitude parameters $\alpha_{\rm rot}$ and $K_c$ vary in response to the addition of a second planet or a GP component, respectively, due to the proximity of the stellar rotation period to the orbital period.

We report a confident detection of \planety with $P_b=15.14$ d and $K=2.17$ m~s$^{-1}$ (Table \ref{tab:HD168009_fit}).  The orbital parameters are well constrained and insensitive to our choice of model. Our two-planet fits also exhibit good convergence resulting in a $4.7\sigma$ semi-amplitude measurement for a candidate companion, \planetz, at $P_c=27.8$ d. Because the $\Delta$BIC between the 1P+GP$_{\rm rot}$ model and the 2P+GP$_{\rm rot}$ model is so small, we also calculate the evidence ratio using the learned harmonic mean estimator as implemented in \texttt{harmonic} \citep{McEwen2021}. We find $\Delta \log Z = 1.3$ in favor of the 1P+GP$_{\rm rot}$ model, so the two model comparison metrics agree that there is insufficient evidence to confirm a second planet in this system based on the RVs alone. We investigate this further in Section \ref{sec:HD168009c}.

\begin{deluxetable}{lcrrrrrc}
\tablewidth{\linewidth}
\tablecaption{Best-fit RV Model Parameters for \targety \label{tab:HD168009_fit}}
\tablehead{Parameter & Prior & & & \multicolumn{3}{c}{Posterior} & Unit} 
\startdata
\hline
& & & & 1P + GP$_{\rm rot}$ & &  2P + GP$_{\rm rot}$ & \\
\hline
$T_{0,b}$ & $\mathcal{U}(2460405,2460415)$ & & & $2460410.38^{+0.33}_{-0.35}$ & & $2460410.37^{+0.35}_{-0.38}$  & BJD \\
$P_b$ & $\mathcal{U}(14,16)$ & & & $15.1437^{+0.0027}_{-0.0026}$ & & $15.1439^{+0.0026}_{-0.0025}$& d \\
$\sqrt{e}\cos\omega_b$ & $\mathcal{U}(-1,1)$ & & & $0.11^{+0.15}_{-0.19}$ & & $0.11^{+0.16}_{-0.19}$ & \nodata \\
$\sqrt{e}\sin\omega_b $& $\mathcal{U}(-1,1)$ & & & $0.07^{+0.22}_{-0.27}$ & & $-0.00^{+0.24}_{-0.26}$ & \nodata \\
$e_b $& derived & & & $0.08^{+0.08}_{-0.05}$ & &  $0.08^{+0.08}_{-0.05}$ & \nodata \\
$\omega_b $ & derived & & & $27^{+66}_{-99}$ & & $0^{+87}_{-80}$ & deg \\
$ K_b$ & $\mathcal{U}(0,10) $ & & & $2.17^{+0.21}_{-0.21}$ & & $2.14^{+0.20}_{-0.19}$ &  m~s$^{-1}$ \\
$ m_b \sin i $& derived & & & $8.61^{+0.85}_{-0.86}$ & & $8.48^{+0.81}_{-0.76}$ & ${\rm M}_\oplus $ \\
$ a_b$ & derived & & & $0.1218^{+0.0011}_{-0.0012}$ & & $0.1217^{+0.0011}_{-0.0011}$& au \\
$T_{0,c}$ & $\mathcal{U}(2460398,2460408)$ & & & \nodata & & $2460403.47^{+1.31}_{-1.31}$  & BJD \\
$P_c$ & $\mathcal{U}(22,32)$ & & & \nodata & & $27.7579^{+0.0130}_{-0.0249}$& d \\
$\sqrt{e}\cos\omega_c$ & $\mathcal{U}(-1,1)$ & & & \nodata & & $-0.01^{+0.24}_{-0.25}$& \nodata \\
$\sqrt{e}\sin\omega_c$ & $\mathcal{U}(-1,1)$ & & & \nodata & & $-0.08^{+0.29}_{-0.28}$& \nodata \\
$e_c $& derived & & & \nodata & &  $0.11^{+0.12}_{-0.08}$ & \nodata \\
$\omega_c $ & derived & & & \nodata & & $-93^{+114}_{-99}$ & deg \\
$ K_c$ & $\mathcal{U}(0,10) $ & & & \nodata & & $1.42^{+0.30}_{-0.30}$ &  m~s$^{-1}$ \\
$ m_c \sin i $& derived & & & \nodata & & $6.83^{+1.46}_{-1.50}$ & ${\rm M}_\oplus $ \\
$ a_c$ & derived & & & \nodata & & $0.1823^{+0.0017}_{-0.0017}$& au \\
$\alpha_{\rm rot}$ & $\mathcal{U}(0,10)$ & & & $2.88^{+0.44}_{-0.38}$ & & $2.26^{+0.45}_{-0.37}$ & m~s$^{-1}$ \\
$P_{\rm rot}$ & $\mathcal{N}(30.1,1.6)$ & & & $29.7^{+1.3}_{-1.2}$ & & $30.0^{+1.5}_{-1.3}$ & d \\
$\tau_{\rm rot}$ & $\mathcal{N}(35.7,18.9)$ & & & $25.6^{+4.5}_{-4.2}$ & & $22.9^{+5.1}_{-4.8}$ & d \\
$\log \Gamma_{\rm rot}$ & $\mathcal{U}(-4,4)$ & & & $-0.63^{+0.48}_{-0.48}$ & & $-0.69^{+0.62}_{-0.66}$ & \nodata \\
$\Gamma_{\rm rot}$ & derived & & & $0.53^{+0.33}_{-0.20}$ & & $0.50^{+0.43}_{-0.24}$ & \nodata \\
$ \gamma_{\rm NEID, Run 1} $& $\mathcal{U}(-10,10)$ & & & $0.64^{+1.26}_{-1.14}$&  & $0.19^{+1.01}_{-0.90}$ & m~s$^{-1}$ \\
$ \gamma_{\rm NEID, Run 2} $& $\mathcal{U}(-10,10)$ & & & $0.35^{+0.99}_{-0.98}$&  & $0.40^{+0.76}_{-0.75}$ & m~s$^{-1}$ \\
$ \gamma_{\rm NEID, Run 3} $& $\mathcal{U}(-10,10)$ & & & $0.47^{+1.03}_{-1.06}$&  & $0.40^{+0.81}_{-0.81}$ &  m~s$^{-1}$ \\
$ \gamma_{\rm NEID, Run 4} $& $\mathcal{U}(-10,10)$ & & & $0.71^{+1.52}_{-1.49}$&  & $0.91^{+1.16}_{-1.18}$ &   m~s$^{-1}$ \\
$ \gamma_{\rm Hamilton} $& $\mathcal{U}(-10,10)$ & & & $4.46^{+1.13}_{-1.13}$ & & $4.42^{+1.04}_{-1.06}$ &   m~s$^{-1}$ \\
$ \gamma_{\rm HIRES-post} $& $\mathcal{U}(-10,10)$ & & & $-0.43^{+0.67}_{-0.69}$ & & $-0.38^{+0.63}_{-0.64}$  & m~s$^{-1}$ \\
$ \sigma_{\rm NEID, Run 1} $& $\mathcal{U}(0,10)$ & & & $1.62^{+0.35}_{-0.29}$&  & $1.60^{+0.33}_{-0.27}$  & m~s$^{-1}$ \\
$ \sigma_{\rm NEID, Run 2} $& $\mathcal{U}(0,10)$ & & & $1.03^{+0.17}_{-0.13}$&  & $1.07^{+0.18}_{-0.14}$  & m~s$^{-1}$ \\
$ \sigma_{\rm NEID, Run 3} $& $\mathcal{U}(0,10)$ & & & $0.86^{+0.24}_{-0.19}$&  & $0.89^{+0.22}_{-0.18}$  & m~s$^{-1}$ \\
$ \sigma_{\rm NEID, Run 4} $& $\mathcal{U}(0,10)$ & & & $1.36^{+0.45}_{-0.34}$&  & $1.27^{+0.42}_{-0.32}$ & m~s$^{-1}$ \\
$ \sigma_{\rm Hamilton} $& $\mathcal{U}(0,10)$ & & & $0.35^{+1.13}_{-0.25}$ & & $0.37^{+1.36}_{-0.28}$  & m~s$^{-1}$ \\
$ \sigma_{\rm HIRES-post} $& $\mathcal{U}(0,10)$ & & & $3.50^{+0.46}_{-0.42}$ & & $3.83^{+0.49}_{-0.46}$  & m~s$^{-1}$ \\
\hline
\enddata
\tablenotetext{}{The 1P + GP$_{\rm rot}$ model is preferred based on the $\Delta$BIC metrics given in Table \ref{tab:modelcomp}.}
\end{deluxetable}

\subsubsection{HD 10780 Model Comparison}\label{sec:modelcomp_HD10780}

\begin{figure}
    \centering
    \includegraphics[width=\linewidth]{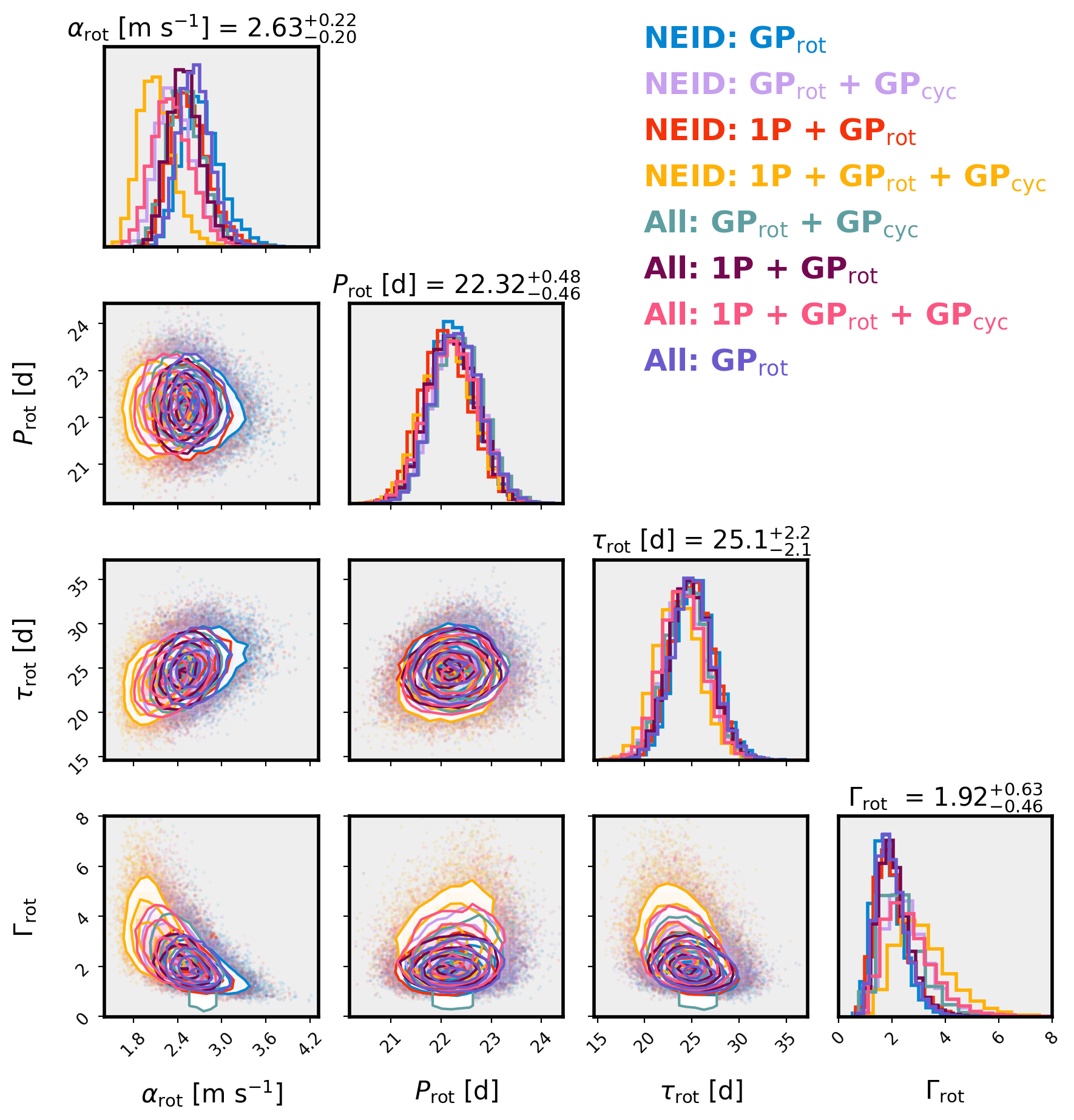}
    \caption{Posterior distributions for parameters of the stellar rotation GP component for \targetz. We show the posteriors for the rotation-only (GP$_{\rm rot}$; blue), cycle + rotation (GP$_{\rm rot}$+GP$_{\rm cyc}$; lavender), single-planet + rotation (1P+GP$_{\rm rot}$; red), and single-planet + cycle + rotation (1P+GP$_{\rm rot}$+GP$_{\rm cyc}$; yellow) fits to the NEID RVs and the GP$_{\rm rot}$ (purple), GP$_{\rm rot}$+GP$_{\rm cyc}$ (green), 1P+GP$_{\rm rot}$ (maroon), and 1P+GP$_{\rm rot}$+GP$_{\rm cyc}$ (pink) fits to the full RV time series. The posteriors are consistent across all fits.}
    \label{fig:HD10780_corner}
\end{figure}

We find that the RV variations for \targetz are best characterized with our single-component GP$_{\rm rot}$ model, which has $\Delta$BIC $=-117.8$ relative to the constant velocity model and $\Delta$BIC = $-9.5$ relative to the next best model (1P+GP$_{\rm rot}$). The addition of a Keplerian component (1P) and/or an activity cycle GP term (GP$_{\rm cyc}$) improves the likelihood of the fit, but the significance of the $\Delta$BIC is reduced due to the complexity penalty from the added parameters.  We therefore do not find sufficient evidence for a 25 d Keplerian signal in the \targetz RVs. It is also notable that the activity cycle term does not appreciably improve the fit given the strong detection of a cycle in the S-index data and the detection of long-period RV signals in our \rvsearch analysis. We attribute this to the degeneracy between long-term variations and instrumental zero point offsets, as the unconstrained $\gamma$ terms will absorb velocity shifts that would otherwise be captured by an activity cycle.
The posteriors on the GP$_{\rm rot}$ model hyperparameters are shown in \autoref{fig:HD10780_corner}; the results are consistent across all tested models, and the best-fit rotation period and decay timescale are in line with the result of our fit to the S-index time series.

\begin{deluxetable}{lcrc}
\tablewidth{\linewidth}
\tablecaption{Best-fit RV Model Parameters for \targetz \label{tab:HD10780_fit}}
\tablehead{Parameter & Prior & \multicolumn{1}{c}{Posterior} & Unit} 
\startdata
\hline
$\alpha_{\rm rot}$ & $\mathcal{U}(0,10)$ & $2.63^{+0.22}_{-0.20}$ &  m~s$^{-1}$ \\
$P_{\rm rot}$ & $\mathcal{N}(21.18,0.89)$ & $22.32^{+0.48}_{-0.46}$ & d \\
$\tau_{\rm rot}$ & $\mathcal{N}(25.9,3.6)$ & $25.1^{+2.2}_{-2.1}$ & d \\
$\log \Gamma_{\rm rot}$ & $\mathcal{U}(-4,4)$ & $0.65^{+0.29}_{-0.28}$ & \nodata \\
$\Gamma_{\rm rot}$ & derived &  $1.92^{+0.63}_{-0.46}$ & \nodata \\
$ \gamma_{\rm NEID, Run 1} $& $\mathcal{U}(-10,10)$ & $0.02^{+0.98}_{-0.92}$ & m~s$^{-1}$ \\
$ \gamma_{\rm NEID, Run 2} $& $\mathcal{U}(-10,10)$ & $-0.18^{+0.73}_{-0.76}$ &  m~s$^{-1}$ \\
$ \gamma_{\rm NEID, Run 3} $& $\mathcal{U}(-10,10)$ & $-0.05^{+0.71}_{-0.74}$ &  m~s$^{-1}$ \\
$ \gamma_{\rm NEID, Run 4} $& $\mathcal{U}(-10,10)$ & $0.53^{+1.31}_{-1.26}$ &  m~s$^{-1}$ \\
$ \gamma_{\rm Hamilton} $& $\mathcal{U}(-10,10)$ &$ 1.32^{+2.01}_{-2.00}$ &  m~s$^{-1}$ \\
$ \gamma_{\rm HIRES-post} $& $\mathcal{U}(-10,10)$ & $-1.16^{+0.73}_{-0.72}$ &  m~s$^{-1}$ \\
$ \gamma_{\rm APF} $& $\mathcal{U}(-10,10)$ & $-0.34^{+0.63}_{-0.63}$ &  m~s$^{-1}$ \\
$ \sigma_{\rm NEID, Run 1} $& $\mathcal{U}(0,10)$ & $0.61^{+0.26}_{-0.20}$ & m~s$^{-1}$ \\
$ \sigma_{\rm NEID, Run 2} $& $\mathcal{U}(0,10)$ & $0.57^{+0.13}_{-0.12}$&  m~s$^{-1}$ \\
$ \sigma_{\rm NEID, Run 3} $&$ \mathcal{U}(0,10)$ & $0.69^{+0.14}_{-0.12}$ &  m~s$^{-1}$ \\
$ \sigma_{\rm NEID, Run 4} $& $\mathcal{U}(0,10)$ & $1.30^{+0.73}_{-0.62}$&  m~s$^{-1}$ \\
$ \sigma_{\rm Hamilton}$ & $\mathcal{U}(0,10)$ & $3.35^{+3.06}_{-3.02}$ &  m~s$^{-1}$ \\
$ \sigma_{\rm HIRES-post} $& $\mathcal{U}(0,10)$ & $0.74^{+0.95}_{-0.60}$ & m~s$^{-1}$ \\
$ \sigma_{\rm APF} $& $\mathcal{U}(0,10)$ & $0.42^{+0.85}_{-0.32}$ &  m~s$^{-1}$ \\
\hline
\enddata
\tablenotetext{}{The posteriors given here correspond to the GP-only (GP$_{\rm rot}$) fit to the full RV time series.}
\end{deluxetable}

\section{Discussion}\label{sec:discussion}

\subsection{HD 126053 b and HD 168009 b}\label{sec:HD126053b}

We detect and confirm \planetx, a low-mass ($m \sin i = 3.60^{+0.34}_{-0.34}\,{\rm M}_\oplus$) planet with an orbital period of $P =10.9146^{+0.0014}_{-0.0012}$ d and an RV semi-amplitude of $K=1.11^{+0.10}_{-0.10}$ m~s$^{-1}$. 
We also recover the signal from \planety, with $P=15.1437^{+0.0027}_{-0.0026}$ d, $m \sin i =8.61^{+0.85}_{-0.86}\,{\rm M}_\oplus$, and $K = 2.17^{+0.21}_{-0.21}$ m~s$^{-1}$. 
Our best-fit parameters for \planety are consistent with those reported in the initial discovery by \citet{Rosenthal2021}.

\begin{figure*}
    \centering
    \includegraphics[width=0.49\linewidth]{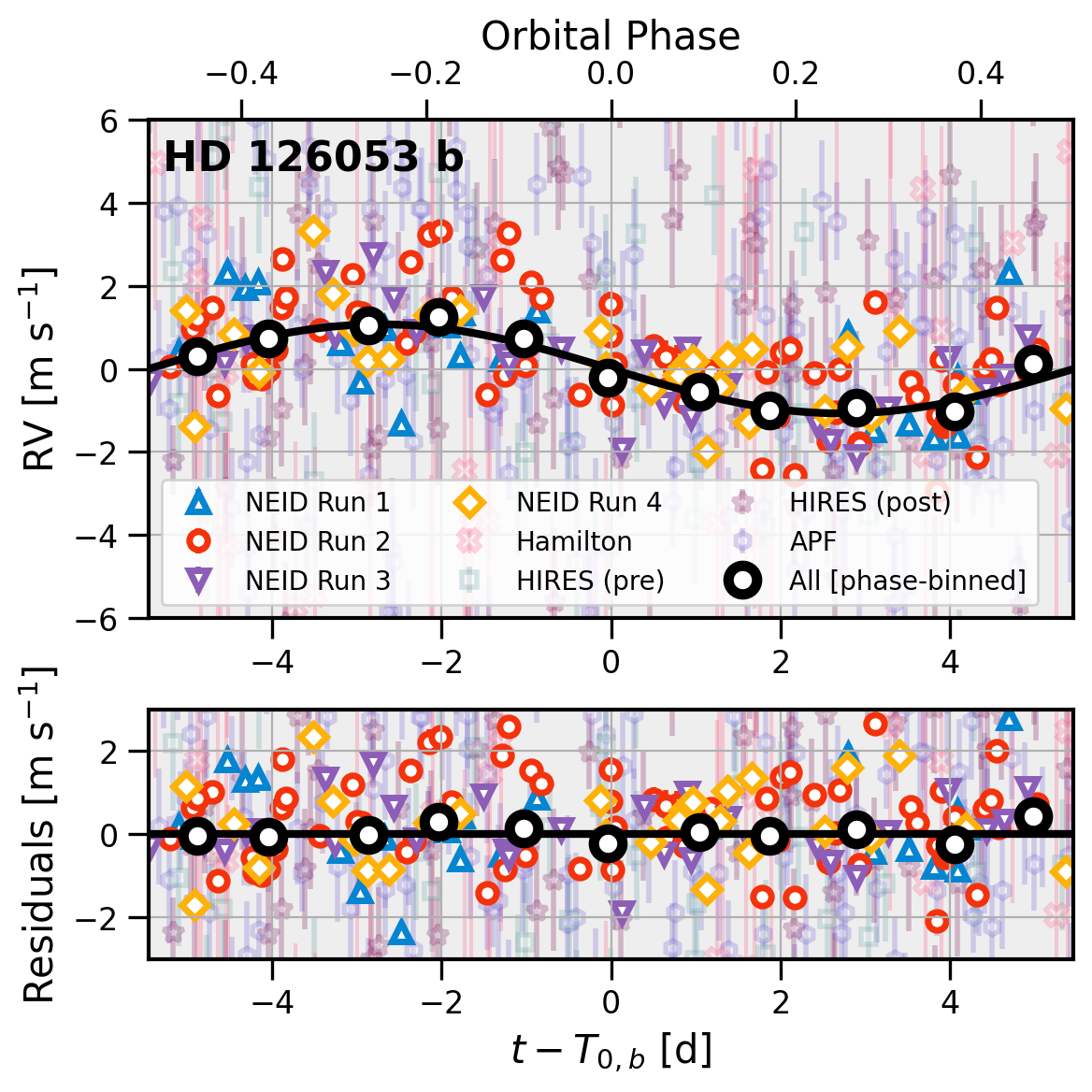}
    \includegraphics[width=0.49\linewidth]{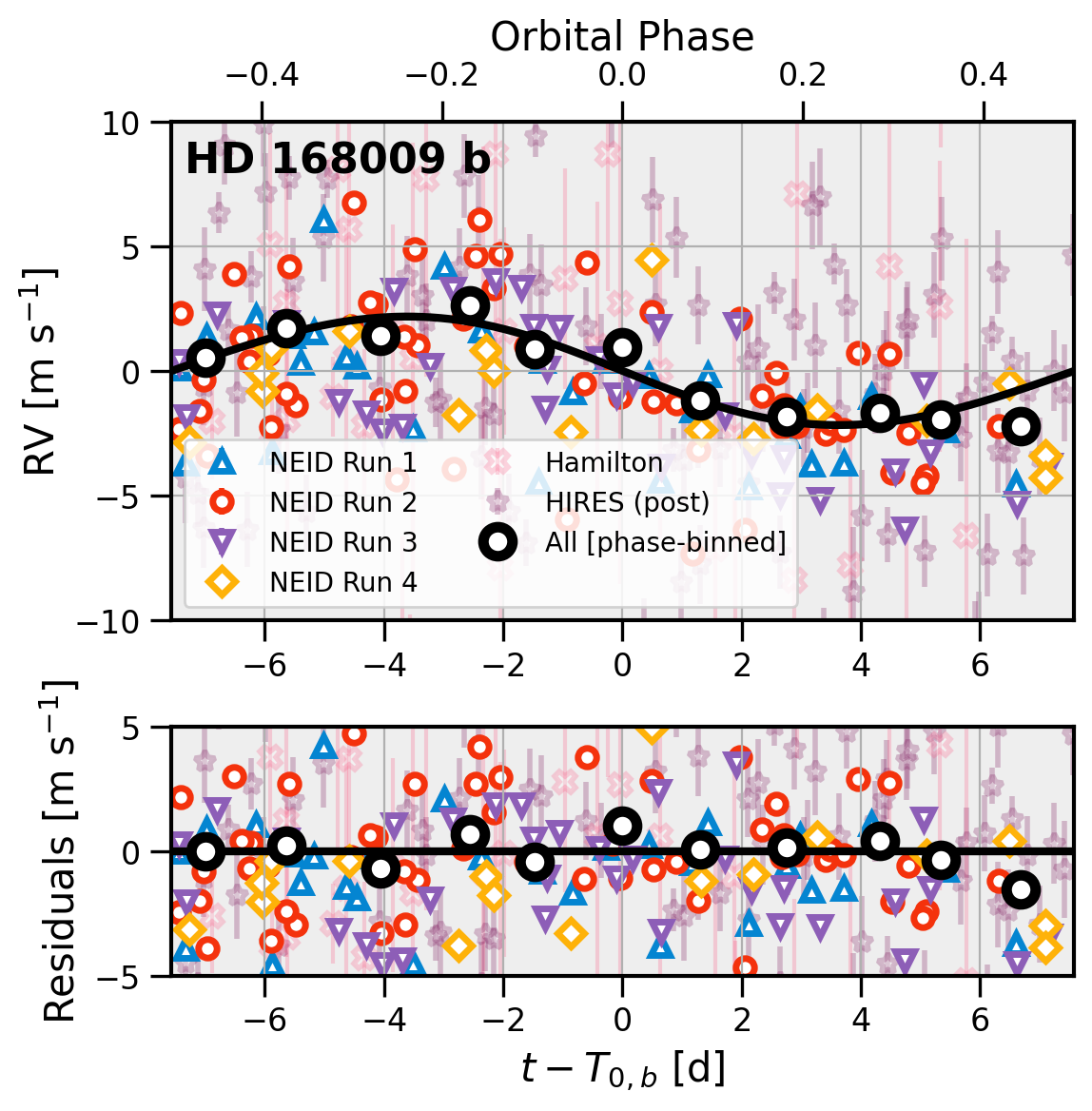}
    \caption{Phase-folded RVs and best-fit orbit model for \planetx (left) and \planety (right). The RVs and orbit model are shown in the upper panels and the residuals are shown in the lower panels. The data markers for each instrument are the same as in Figures \ref{fig:all_rvs} and \ref{fig:HD126053_rvs}. We also plot the phase-binned RVs for each system (black circles). For \planetx, the binned data closely follow the orbit model and the residual scatter is small. For \planety, we see deviations at some phases; this is likely due to activity-induced RV variations, as the stellar rotation period is approximately twice the orbital period.}
    \label{fig:single_planet_phase}
\end{figure*}

\begin{figure}
    \centering
    \includegraphics[width=\linewidth]{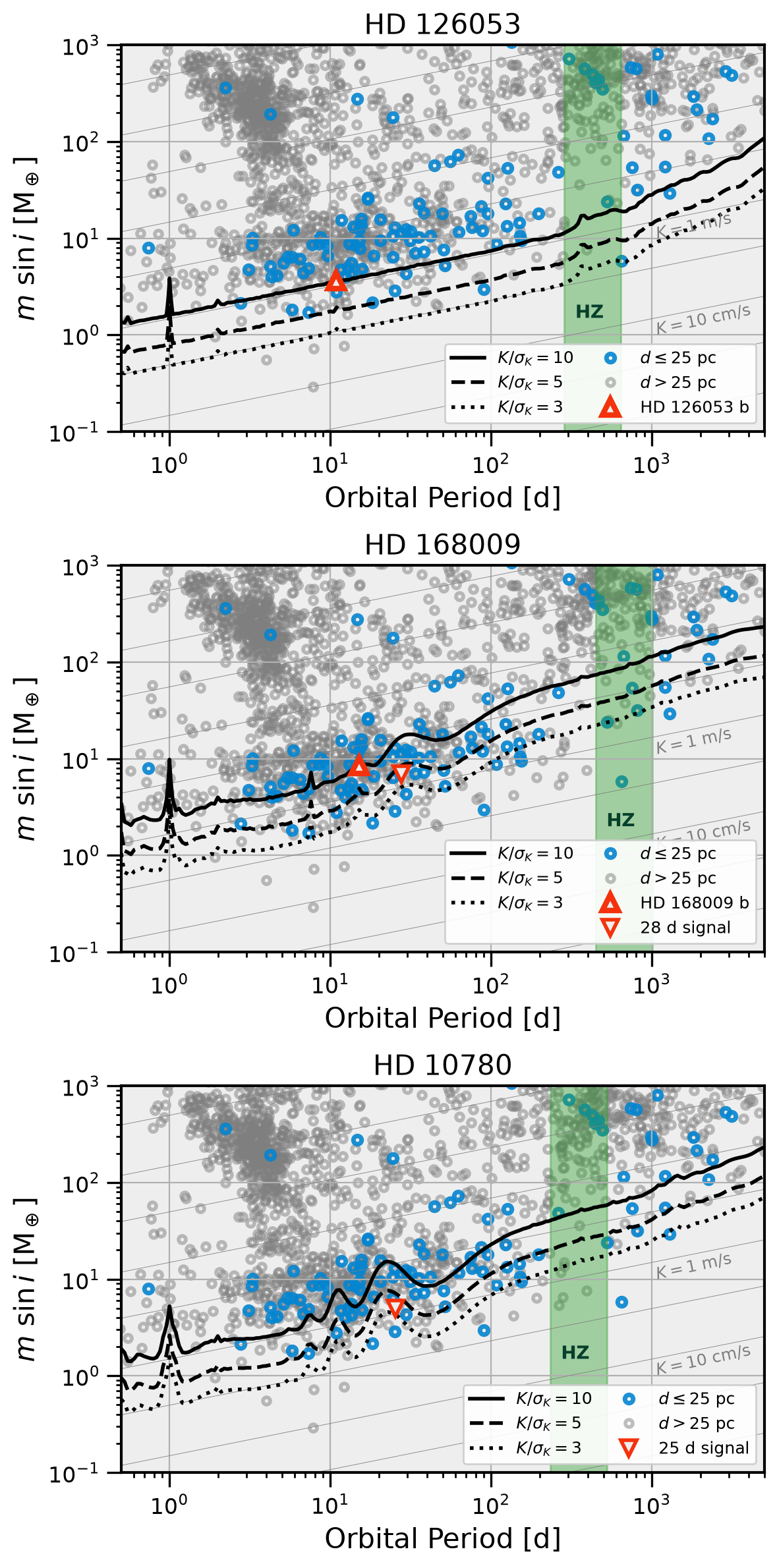}
    \caption{Comparison of detected planets and candidate signals (red) to the known exoplanet population within 25 pc (blue) and beyond 25 pc (grey) and to the detection sensitivity limits (black) as calculated via the Fisher information content of the RV data. We also show the location of the conservative habitable zone in green. The detection sensitivity for \targetx follows lines of constant $K$ at short periods before falling off at longer periods, while the sensitivity for \targety and \targetz is characterized by added structure near the stellar rotation period and harmonics thereof.}
    \label{fig:sensitivity}
\end{figure}

We show the phase-folded RV orbits of \planetx and \planety in \autoref{fig:single_planet_phase} and we place both planets alongside the known population of planetary companions to Sun-like stars ($4000$ K $< T_{\rm eff}< 6500$ K) in \autoref{fig:sensitivity}. The known exoplanet properties are taken from the NASA Exoplanet Archive \citep{Christiansen2025} and filtered to only include planets with a (minimum) mass precision of better than 50\%. \planetx lies close to the lower envelope of known exoplanets in mass-period space, and it is among the least massive planets within 25 pc.  The physical properties of \planetx and \planety can not be further characterized without first breaking the mass-inclination degeneracy, but they are both likely to be members of the most common class of known exoplanets, i.e., short-period planets between the sizes of Earth and Neptune \citep{Christiansen2025}.

\subsection{HD 126053 1-yr signal}\label{sec:HD126053yr}

\begin{figure}
    \centering
    \includegraphics[width=\linewidth]{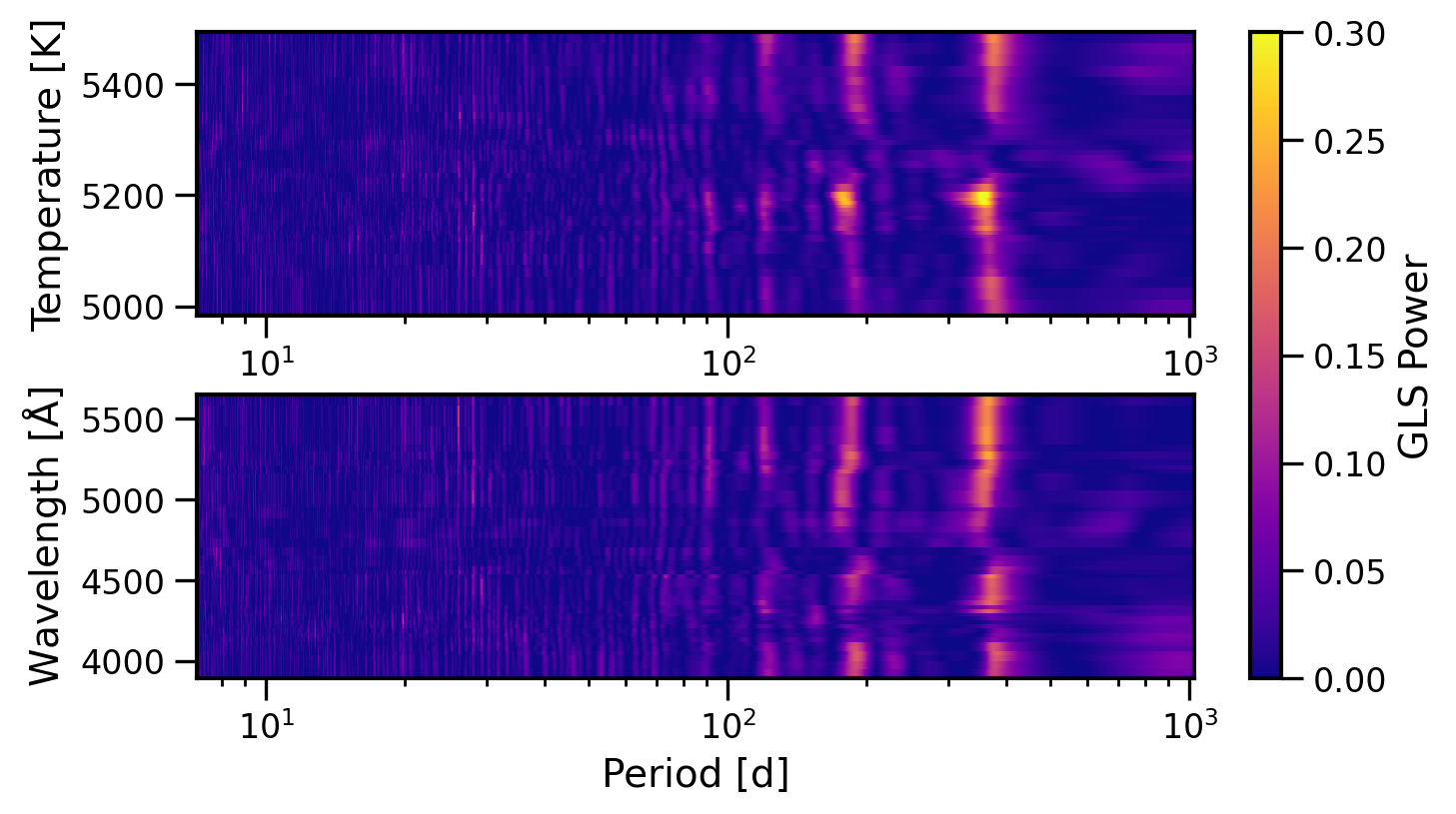}
    \caption{GLS power of the residual line-by-line RVs for \targetx as a function of formation temperature (upper) and wavelength (lower). There is clear chromatic variation in the signal at $P\sim1$ yr.}
    \label{fig:HD126053_lbl}
\end{figure}

In Section \ref{sec:neidrv}, we found evidence for a $\sim 1$ yr signal in the CCF and ARVE RVs for \targetx (\autoref{fig:HD126053_rvs}). This signal was absent from the SERVAL RVs and therefore was not considered in our analysis in Section \ref{sec:analysis}. To assess whether we were correct to dismiss this signal, we investigate the chromaticity using the line-by-line velocity measurements from ARVE. We sort the lines by the associated formation temperature given in the ARVE mask and then compute the weighted mean velocity in temperature bins from 4700 K to 5550 K. The bin widths and number of lines per bin are adjusted such that the cumulative photon noise precision of each binned velocity measurement is $\lesssim 1$ m~s$^{-1}$.
We repeat this process for wavelength bins from 3800 \AA\ to 6200 \AA. We then subtract the \planetx signal from the wavelength-dependent and temperature-dependent RV measurements and compute the GLS of the residuals in each bin (\autoref{fig:HD126053_lbl}). 

The amplitude of the $P\sim1$ yr signal varies with both wavelength and formation temperature, and the signal shifts to shorter periods at longer wavelengths. Chromatic changes such as these are usually evidence that RV variations are produced by tellurics or stellar activity, which affect lines differentially, rather than by an orbiting body, which will induce a uniform shift for the entire spectrum \citep[e.g.,][]{Dumusque2018,AlMoulla2024}. The presence of a similar periodic signal in the NEID S-index time series points to stellar activity, while the 1-yr periodicity could be a sign of telluric contamination. Regardless of the specific source of the variations, we conclude that the annual signal is not planet-induced and need not be considered in our RV analysis in this work.

\subsection{HD 168009 c candidate signal}\label{sec:HD168009c}

\begin{figure*}
    \centering
    \includegraphics[width=0.8\linewidth]{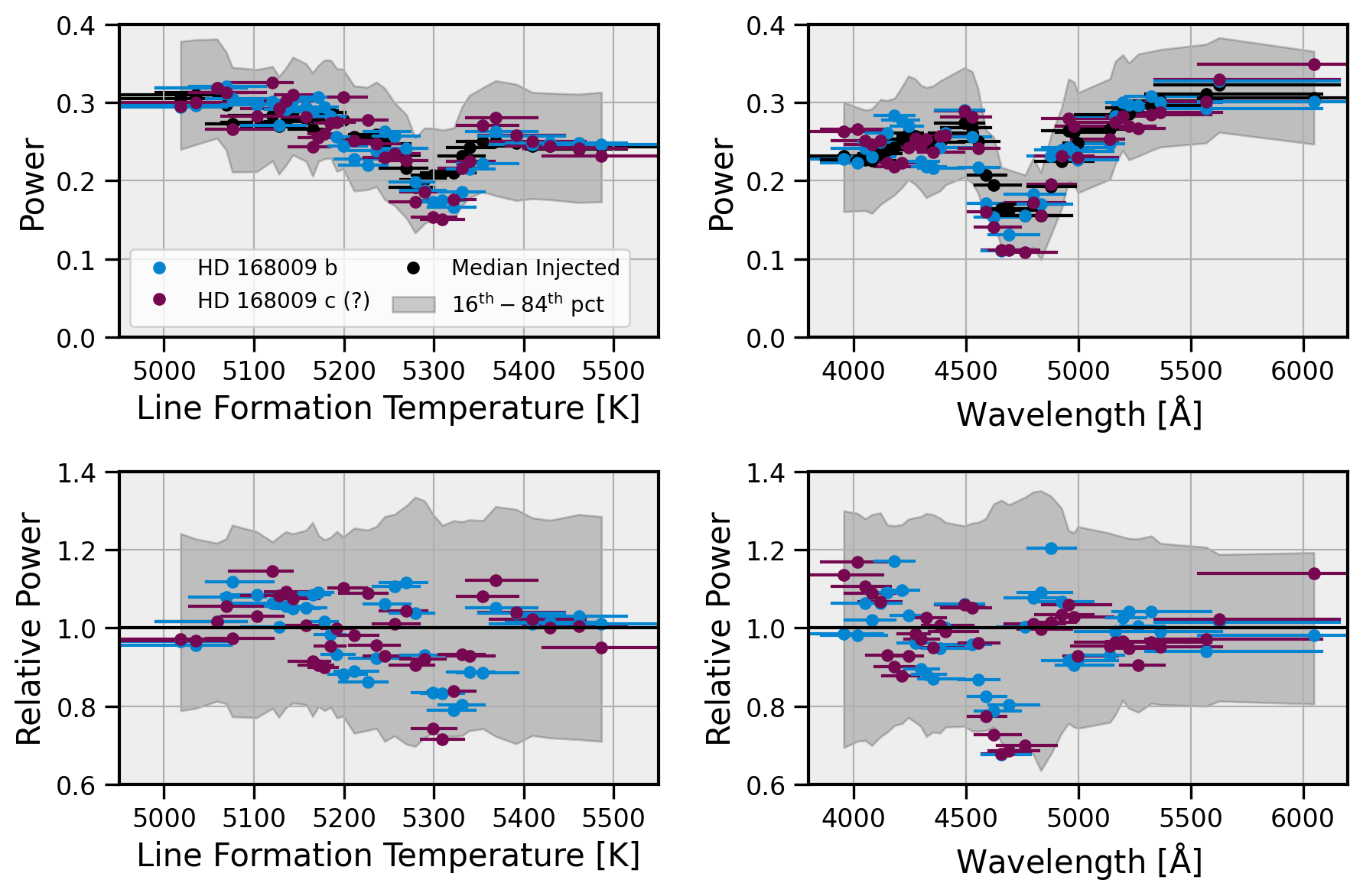}
    \caption{GLS power of the line-by-line RVs for \targety as a function of formation temperature (left) and wavelength (right). The power at the period of \planety is shown in blue and the residual power at the period of \planetz is shown in purple. The median power of injected achromatic signals is shown in black and the 16\textsuperscript{th}--84\textsuperscript{th} percentile confidence intervals are plotted as a grey shaded region. We show the observed power in the upper panels and the power relative to the injected signals in the lower panels. The horizontal error bars represent the width of each temperature (or wavelength) bin, and the markers represent the median temperature (or wavelength) in each bin. The chromatic dependence of the \planety and \planetz signals closely tracks that of the injected achromatic signals, which we interpret as evidence that the observed power is achromatic as well.}
    \label{fig:HD168009_lbl}
\end{figure*}

Based on our analysis of the \targety RVs in Section \ref{sec:modelcomp_HD168009}, we were unable to confirm the candidate Keplerian signal at $P\sim28$ d, as both the evidence ratio and the BIC prefer models that attributed this signal to a GP component.
The model comparison, however, does not constitute conclusive proof that stellar activity is responsible for this signal.
We repeat the analysis described in Section \ref{sec:HD126053yr} for \targety, and we examine the chromatic dependence of both the \planety signal at 15 d (in the observed RVs) and the candidate \planetz signal at 28 d (in the residuals to the \planety orbit).
Chromatic changes in the power levels of both signals are apparent (\autoref{fig:HD168009_lbl}). However, these variations have far smaller amplitudes than those described in Section \ref{sec:HD126053yr} and the periods remain stable.
In addition, the 15 d and 28 d signals appear to have similar wavelength dependence and temperature dependence. This suggests the chromatic variations may be due to differences in the noise level rather than in the amplitude of the RV signals. To investigate this possibility and calibrate the significance of any variations, we subtract both planetary signals from the line-by-line RV time series and inject achromatic periodic signals into the residuals.
We inject sinusoids with $K = 2$ m~s$^{-1}$ and periods drawn from a uniform distribution, $\mathcal{U}(10,100)$ d, and we calculate the power at the injected period.
In \autoref{fig:HD168009_lbl}, we show the median power and 16\textsuperscript{th}--84\textsuperscript{th} percentile confidence intervals as a function of temperature and wavelength for a set of 500 injection trials. 
The temperature dependence and wavelength dependence of the observed signals are reflected in the power of the injected orbits, with power deficits near 5300 K and 4700 \AA, a weak inverse correlation between signal strength and temperature, and a weak positive correlation with wavelength. The consistent presence of the power deficits suggests a larger amount of RV jitter in these temperature and wavelength bins, which will suppress the strength of an achromatic signal.
The ratio of the observed power of the \planety and \planetz signals to the power of the injected signals falls within the 16\textsuperscript{th}--84\textsuperscript{th} percentile range in nearly every temperature and wavelength bin, but the 5300 K and 4700 \AA\ deficits are slightly more pronounced for the candidate \planetz. Still, we find the signal strength ratios to be sufficiently achromatic.

As a second test of the nature of the 28 d signal, we calculate the Spearman correlation coefficient, $\rho$, for the NEID RVs and S-index measurements for \targety. We first calculate the correlation within each NEID Run, and then for the combined RVs and S-index measurements. For this test, the S-index values are shifted such that each run has the same median value before they are combined. These shifts are necessary to remove long-term changes in the activity level and isolate variations on the stellar rotation timescale. While we did not recover the activity cycle for this star in Section \ref{sec:activity_fitting}, the activity level does appear to be slowly decreasing over the NEID baseline (\autoref{fig:HD168009_rvs}).

\begin{figure*}
    \centering
    \includegraphics[width=0.8\linewidth]{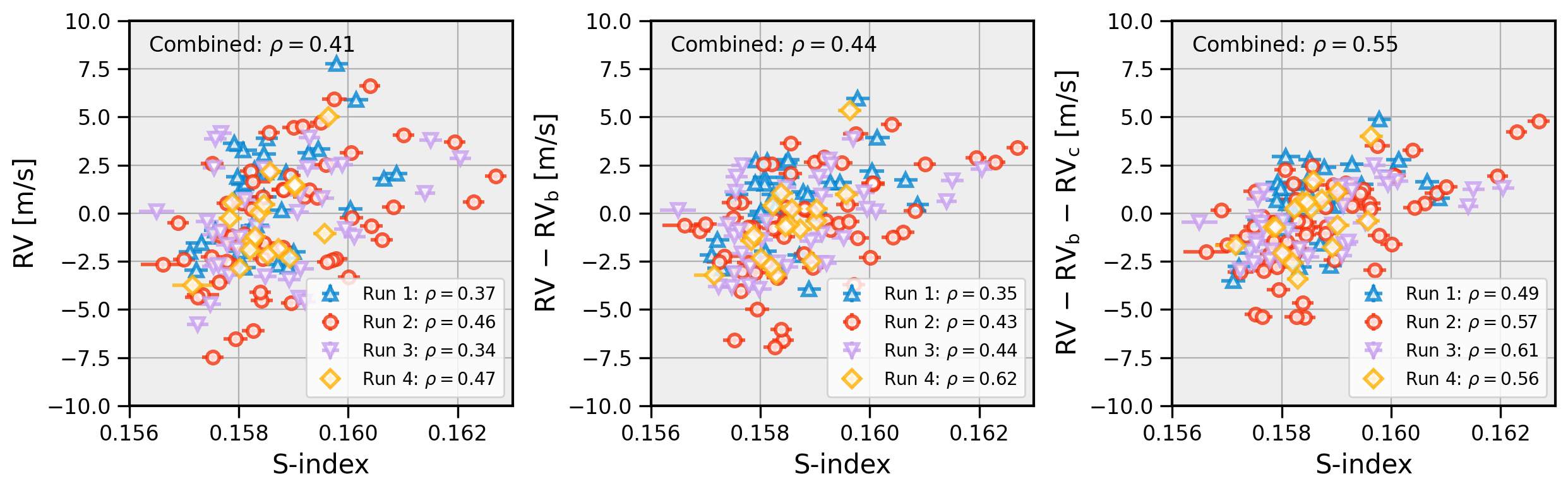}
    \caption{Activity-RV correlation for \targety. We compare the correlation of activity with the RVs (left), the residuals after subtracting \planety (center), and the residuals after subtracting both \planety and \planetz (right). The correlation strength increases slightly after removing \planety and increases significantly after removing \planetz. This result disfavors a stellar activity-driven origin for the 28 d signal.}
    \label{fig:HD168009_corr}
\end{figure*}

The correlation coefficients for the observed RVs, the residuals to the best-fit orbit for \planety, and the residuals to both \planety and \planetz are shown in \autoref{fig:HD168009_corr}. A decrease in $\rho$ after subtracting the candidate \planetz signal would be compelling evidence that this signal is driven by stellar activity. But there is a large \textit{increase} in the correlation strength after we subtract \planetz; the 28 d signal is evidently uncorrelated with stellar activity. While this can be interpreted as proof that \planetz is real and not simply a product of stellar rotation, it could also simply be evidence that the activity indicators and activity-driven RVs are out of phase, as is commonly observed for Sun-like stars \citep[e.g.,][]{CollierCameron2019,Burrows2024}.

\subsection{HD 10780 stellar activity}

\begin{figure}
    \centering
    \includegraphics[width=\linewidth]{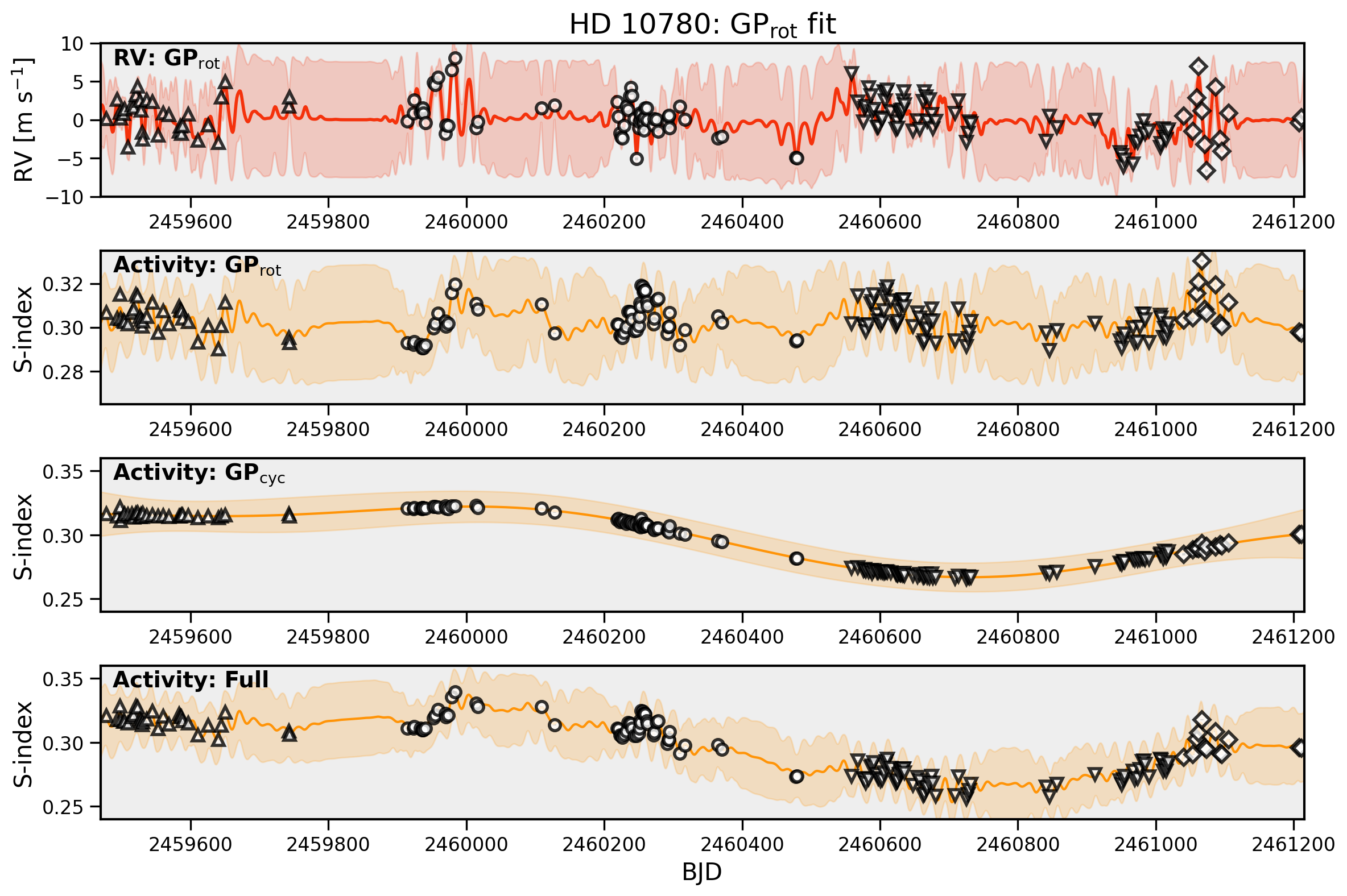}
    \caption{Best-fit RV and activity models for \targetz. The RV model consists of a single GP component. The best-fit activity model includes separate components for rotation (second panel) and the activity cycle (third panel), and we show the full activity model in the lower panel.}
    \label{fig:HD10780_ts}
\end{figure}

We show the GP$_{\rm rot}$ fit to the \targetz RV time series in \autoref{fig:HD10780_ts}, along with the rotation period and activity cycle components of our fit to the NEID S-index time series for this target. The rotationally-modulated RV variability is well constrained. We find a best-fit rotation period of $P_{\rm rot} = 22.32^{+0.48}_{-0.46}$ d, which agrees with the 23 d, 21.7 d, and 22.14 d periods measured by \citet{Baliunas1995}, \citet{Gaidos2000}, and \citet{Olspert2018}, respectively.
However, due to the frequent RV zero point offsets between runs and instruments, we are unable to constrain the response of the RVs to the stellar activity cycle. This might be resolved if the cycle period could be precisely measured, but this is challenging given the relatively short NEID baseline and the dearth of other activity data contemporaneous with the full RV baseline. We considered including S-index measurements from HIRES as published by \citet{Isaacson2024}, but the HIRES coverage is too sparse for this to yield an appreciable improvement. \targetz was also observed with the HKP-2 spectrograph from 1977 to 1995 as part of the long-term Mount Wilson stellar activity monitoring survey \citep{Wilson1978}, with S-index measurements compiled and published by \citet{Baum2022}.  These data have no overlap with the RV measurements considered in this work, however, and given the likely complex stellar cycle, we refrain from combining these with the NEID data which commence more than 20 years later. A more promising line of future work would be jointly fitting the NEID RVs and S-index measurements with multi-dimensional GPs \citep[e.g.,][]{Rajpaul2015,Tran2023}, such that the activity cycle serves to constrain the RV zero point offsets. In spite of the unconstrained cycle, the rotation-only fit nevertheless provides a good model for the activity-induced RV variability on short timescales, which can be useful in assessing exoplanet sensitivity limits as we show in the following section.

\subsection{Companion sensitivity limits}\label{sec:sensitivity}

The goal of NETS is to achieve sensitivity to Earth-sized planets in the habitable zones of nearby Sun-like stars. Neither \planetx nor \planety meet these criteria, but the results of our RV analysis in this work can be used to place informed limits on exoplanet sensitivity for \targetx, \targety, and \targetz. We evaluate the progress that has been made in this direction by calculating the Fisher information content \citep{Fisher1922} of the full RV time series for each star following the methods outlined in \citet{Gupta2024}. We refer the reader to \citet{NETSI} for a detailed explanation of how this method is applied. In brief, the Fisher information matrix is given as
\begin{equation}
    B_{i,j} = \left(\frac{\partial\mu}{\partial \theta_i}\right)^T C^{-1}\left(\frac{\partial\mu}{\partial \theta_j}\right),
\end{equation}
where $\mu$ is the radial velocity model, $\theta$ is the parameter vector, and $C$ is the covariance matrix. Here, $\mu$ is the sum of Keplerian signals from all $n$ known planets in each system and a circular orbital model for a hypothetical planet $n+1$. The parameter vector consists of the orbital elements and instrument-dependent systemic velocity offsets. We construct the covariance matrix using our best-fit ``noise'' model for each star. For \targety and \targetz, this is the sum of the photon noise, the derived instrument-specific white noise jitter, and the covariance term of the GP$_{\rm rot}$ kernel. Because the 1P model was preferred over the 1P+GP$_{\rm rot}$ model for \targetx, the covariance matrix for this system includes only photon noise and white noise jitter. 

We use the Fisher information matrix to calculate the uncertainty on the orbital parameters as
\begin{equation}
    \sigma_{\theta_i}^2 = B_{i,i}^{-1}.
\end{equation}
for a mass-period grid of $5 $d $\leq P<5000$ d and $0.1\, {\rm M}_\oplus \leq m \sin i \leq 1000\, {\rm M}_\oplus$ for planet $n+1$. We marginalize over the orbital phase and the remaining elements of $\theta$ are fixed to their maximum likelihood values from our model fits in \ref{sec:rvfit}. The 3-, 5-, and 10-$\sigma$ contours on the expected detection sensitivity $K_{n+1}/\sigma_{K_{n+1}}$ as a function of mass and period are shown in \autoref{fig:sensitivity}. Based on the injection-recovery comparison performed by \citet{Gupta2024}, we adopt a $K/\sigma_K\geq5$ as our expected detection threshold, i.e., we can rule out the presence of planets with minimum masses above the 5-$\sigma$ sensitivity level.

Among known exoplanets discovered with the RV method and orbiting FGK stars, 27 have a minimum mass of $<5\,{\rm M}_\oplus$ and 29  have $K<1.5$ m~s$^{-1}$, but \planetx is the only planet of this mass with no known companions. Further, our Fisher information results show that this planet is likely the most massive planet in its system with $P<100$ d, and we are already sensitive to planets with $m \sin i =6\,{\rm M}_\oplus$ at the inner edge of the conservative habitable zone. Any additional planets in this system must have lower masses or long periods, and the host star is sufficiently RV-quiet that we should have a good chance to detect such planets if \targetx continues to be monitored with NEID or similar EPRV spectrographs.

The sensitivity results for \targety shed some light on the difficulty of assessing the nature of \planetz. We show in \autoref{fig:sensitivity} that the sensitivity decreases for orbital periods near the stellar rotation period of $P_{\rm rot} = 
1$ d due to the correlated noise contribution from stellar activity. A planet with $m \sin i$ and $P$ equal to the nominal values for \planetz from  our 2P+GP$_{\rm rot}$ model is expected to have $K/\sigma_K = 4.2$. This is comparable to the measured precision on $K_c$ and it demonstrates the existing data are insufficient to reach the $5$-$\sigma$ threshold for a confident detection. Continued monitoring will be necessary to more conclusively detect or reject the planet candidate. However, this system is likely a poor target if we aim to reach Earth-mass sensitivity in the habitable zone, as we are currently only sensitive to habitable-zone planets with $K>3$ m~s$^{-1}$  at the 5-$\sigma$ level. 

The sensitivity limits for \targetz are similarly affected by stellar activity. Even if the 25 d candidate planet considered in this work were real, we would be insensitive to it due to its proximity to the stellar rotation period at $P_{\rm rot} = 22.3$ d.
Although this is a high priority target for HWO, our prospects for discovering low-mass planets in the habitable zone of \targetz with RVs will require drastic improvements in jitter mitigation.
The existing data limit sensitivity in this period range to 20--30 M$_\oplus$ (\autoref{fig:sensitivity}).
This is a marked improvement relative to previous habitable zone mass limits for this star \citep[35--50 M$_\oplus$;][]{Harada2025} and brings us closer to ruling out more massive planets that could threaten the dynamical stability of potential Earth twins \citep{Kane2024,Kane2024B}.
But for both \targety and \targetz, detection and mass measurement of small planets in the habitable zone will undoubtedly require methods to clean and remove stellar activity signals at the spectral level. One would also need to continue to monitor \targetz and place stronger constraints on the activity cycle to accurately assess planet sensitivity.

\section{Summary}\label{sec:summary}

In this work, we have characterized the stellar activity and RV signals for the nearby, Sun-like stars \targetx, \targety, and \targetz, which are being monitored with NEID as part of the NETS program \citep{Gupta2021}. We detect and confirm \planetx, which is among the least massive planets orbiting a Sun-like star within 25 pc, and we confirm the planetary nature of \planety. We also detect the RV signature of stellar rotation for \targety and \targetz, as well as a candidate exoplanet signal near the rotation period of \targety. Our analysis of the \targety RVs in the time domain and at the spectral level is insufficient to confirm whether this candidate is real, but we also find no compelling evidence to attribute the signal to stellar activity.

We calculate the exoplanet sensitivity limits for each star using the Fisher information content of the RV measurements and show that \targetx is a very promising target if we aim to achieve sensitivity to low-mass planets in the habitable zone, while \targety and \targetz currently prove challenging due to their prominent activity-induced variability. Continued RV monitoring of each of these stars will nevertheless be valuable. For \targety, additional RV data will amplify our ability to disentangle planet-induced and activity-induced signals, allowing us to either confirm \planetz as a planet  or conclusively attribute this signal to stellar activity. \targetz is a high-priority target for direct imaging exoplanet searches with facilities such as HWO. Extending the NEID observing baseline should enable stricter constraints on the stellar activity cycle, which will then allow us to better model and remove this signal and to accurately predict exoplanet sensitivity limits for this star.

\begin{acknowledgments}
Support for this work was provided in part by NASA through the NASA Hubble Fellowship grant HST-HF2-51605.001-A awarded by the Space Telescope Science Institute, which is operated by the Association of Universities for Research in Astronomy, Inc., for NASA, under contract NAS5-26555.

This paper contains data taken with the NEID instrument, which was funded by the NASA-NSF Exoplanet Observational Research (NN-EXPLORE) partnership and built by Pennsylvania State University. 
NEID is installed on the WIYN telescope, which is operated by the NSF National Optical-Infrared Astronomy Research Laboratory, and the
NEID archive is operated by the NASA Exoplanet Science Institute at the California Institute of Technology. NEID is funded by NASA through JPL contract 1547612 and the NEID Data Reduction Pipeline is funded through JPL contract 1644767.
NN-EXPLORE is managed by the Jet Propulsion Laboratory, California Institute of Technology under contract with the National Aeronautics and Space Administration. We thank the NEID Queue Observers and WIYN Observing Associates for their skillful execution of our observations.
Part of this work was performed at the Jet Propulsion Laboratory, California Institute of Technology, sponsored by the United States Government under the Prime Contract 80NM0018D0004 between Caltech and NASA. S.M. \& E.F. acknowledge support from NASA grant 80NSSC26K0554. 
The Center for Exoplanets and Habitable Worlds is supported by Penn State and its Eberly College of Science.

Based in part on observations at Kitt Peak National Observatory, NSF’s NOIRLab, managed by the Association of Universities for Research in Astronomy (AURA) under a cooperative agreement with the National Science Foundation. Astronomers are honored to be permitted to conduct scientific research on the sacred mountain located in the homelands of the Schuk Toak District within the Tohono O'odham Nation. We honor their past, present, and future generations, who have lived here for time immemorial and will forever call this place home.

\end{acknowledgments}

\facilities{WIYN(NEID), Keck (HIRES), Lick (Hamilton, APF), TESS}

\software{
        \texttt{ARVE} \citep{AlMoulla2025},
        \texttt{astropy} \citep{AstropyCollaboration2013,AstropyCollaboration2018,AstropyCollaboration2022},
        \texttt{barycorrpy} \citep{Kanodia2018},
        \texttt{exoplanet} \citep{exoplanet:zenodo}
        \texttt{harmonic} \citep{McEwen2021},
        \texttt{matplotlib} \citep{Hunter2007},
        \texttt{NestedSamplers.jl} \citep{Lucas2023},
        \texttt{numpy} \citep{Harris2020},
        \texttt{PyMC} \citep{pymc2023},
        \texttt{RadVel} \citep{Fulton2018},
        \texttt{RVSearch} \citep{Rosenthal2021}, \texttt{scipy} \citep{Oliphant2007},
        \texttt{SERVAL} \citep{Zechmeister2018}
          }

\appendix \label{appendix}

\section{Model Comparison Metrics}

Model likelihoods, $L$, are computed as part of the PyMC orbit fitting routine described in Section \ref{sec:rvfit}. In addition to the full fitting basis, these likelihoods account for measurement uncertainties and, where appropriate, a GP covariance penalty. We use $\log L$ to compute the BIC of each model as
\begin{equation}
    {\rm BIC} = k_{\rm params} \log n_{\rm obs} - 2\log L,
\end{equation}
where $k_{\rm params}$ is the number of parameters in the fitting basis and $n_{\rm obs}$ is the number of RV measurements. In Table \ref{tab:modelcomp}, we list the log-likelihoods and BIC values relative to the constant velocity model for each star.

\begin{deluxetable}{llrrrr}
\tablewidth{\linewidth}
\tablecaption{Model Comparison Metrics for \targetx, \targety, and \targetz \label{tab:modelcomp}}
\tablehead{Star & Model & \multicolumn{2}{c}{NEID RVs} &  \multicolumn{2}{c}{All RVs}} 
\startdata
\hline
&  & $\Delta\log L$ & $\Delta$BIC & $\Delta\log L$ & $\Delta$BIC \\
\hline
\targetx &GP$_{\rm rot}$&$21.7$&$-23.0$&$24.7$&$-24.6$\\
&\textbf{1P}&$42.5$&$-59.7$&$48.8$&$-66.7$\\
&1P\,+\,GP$_{\rm rot}$&$49.9$&$-54.2$&$58.2$&$-60.7$\\
\hline
\targety &GP$_{\rm rot}$&$33.2$&$-46.6$&$41.1$&$-59.7$\\
&1P&$24.9$&$-24.9$&$30.9$&$-33.6$\\
&\textbf{1P\,+\,GP}$_{\rm rot}$&$59.8$&$-75.0$&$75.6$&$-100.6$\\
&2P&$53.2$&$-57.0$&$62.0$&$-67.8$\\
&2P\,+\,GP$_{\rm rot}$&$74.7$&$-80.0$&$89.2$&$-99.6$\\
\hline
\targetz&\textbf{GP}$_{\rm rot}$&$71.5$&$-122.4$&$70.0$&$-117.8$\\
&GP$_{\rm rot}$\,+\,GP$_{\rm cyc}$&$72.0$&$-102.7$&$70.2$&$-95.9$\\
&1P&$13.9$&$-2.1$&$14.8$&$$-1.8$$\\
&1P\,+\,GP$_{\rm rot}$&$78.3$&$-110.3$&$79.2$&$-108.3$\\
&1P\,+\,GP$_{\rm cyc}$&$38.9$&$-31.4$&$37.4$&$-24.8$\\
&1P\,+\,GP$_{\rm rot}$\,+\,GP$_{\rm cyc}$&$81.5$&$-96.1$&$77.5$&$-82.6$\\
\hline
\enddata
\tablenotetext{}{The $\Delta\log L$ and $\Delta$BIC values are reported relative to the constant velocity model. The preferred model for each star is identified with bold text.}
\end{deluxetable}

\bibliography{references}{}

\begin{thebibliography}{}
\expandafter\ifx\csname natexlab\endcsname\relax\def\natexlab#1{#1}\fi
\providecommand{\url}[1]{\href{#1}{#1}}
\providecommand{\dodoi}[1]{doi:~\href{http://doi.org/#1}{\nolinkurl{#1}}}
\providecommand{\doeprint}[1]{\href{http://ascl.net/#1}{\nolinkurl{http://ascl.net/#1}}}
\providecommand{\doarXiv}[1]{\href{https://arxiv.org/abs/#1}{\nolinkurl{https://arxiv.org/abs/#1}}}

\bibitem[{O. Abril-Pla {et~al.}(2023)Abril-Pla, Andreani, Carroll, Dong, Fonnesbeck, Kochurov, Kumar, Lao, Luhmann, Martin, Osthege, Vieira, Wiecki, \& Zinkov}]{pymc2023}
Abril-Pla, O., Andreani, V., Carroll, C., {et~al.} 2023, \bibinfo{title}{{PyMC}: A Modern and Comprehensive Probabilistic Programming Framework in {P}ython,} {PeerJ} Computer Science, 9, \dodoi{10.7717/peerj-cs.1516}

\bibitem[{K. {Al Moulla}(2025){Al Moulla}}]{AlMoulla2025}
{Al Moulla}, K. 2025, \bibinfo{title}{{ARVE: Analyzing Radial Velocity Elements: I. The Code},} \aap, 701, A266, \dodoi{10.1051/0004-6361/202554897}

\bibitem[{K. {Al Moulla} {et~al.}(2024){Al Moulla}, {Dumusque}, \& {Cretignier}}]{AlMoulla2024}
{Al Moulla}, K., {Dumusque}, X., \& {Cretignier}, M. 2024, \bibinfo{title}{{Measuring precise radial velocities on individual spectral lines. IV. Stellar activity correlation with line formation temperature},} \aap, 683, A106, \dodoi{10.1051/0004-6361/202348150}

\bibitem[{G. {Anglada-Escud{\'e}} \& R.~P. {Butler}(2012){Anglada-Escud{\'e}} \& {Butler}}]{Anglada-Escude2012}
{Anglada-Escud{\'e}}, G., \& {Butler}, R.~P. 2012, \bibinfo{title}{{The HARPS-TERRA Project. I. Description of the Algorithms, Performance, and New Measurements on a Few Remarkable Stars Observed by HARPS},} \apjs, 200, 15, \dodoi{10.1088/0067-0049/200/2/15}

\bibitem[{G. {Anglada-Escud{\'e}} {et~al.}(2016){Anglada-Escud{\'e}}, {Amado}, {Barnes}, {Berdi{\~n}as}, {Butler}, {Coleman}, {de La Cueva}, {Dreizler}, {Endl}, {Giesers}, {Jeffers}, {Jenkins}, {Jones}, {Kiraga}, {K{\"u}rster}, {L{\'o}pez-Gonz{\'a}lez}, {Marvin}, {Morales}, {Morin}, {Nelson}, {Ortiz}, {Ofir}, {Paardekooper}, {Reiners}, {Rodr{\'\i}guez}, {Rodr{\'\i}guez-L{\'o}pez}, {Sarmiento}, {Strachan}, {Tsapras}, {Tuomi}, \& {Zechmeister}}]{Anglada-Escude2016}
{Anglada-Escud{\'e}}, G., {Amado}, P.~J., {Barnes}, J., {et~al.} 2016, \bibinfo{title}{{A terrestrial planet candidate in a temperate orbit around Proxima Centauri},} \nat, 536, 437, \dodoi{10.1038/nature19106}

\bibitem[{ {Astropy Collaboration} {et~al.}(2013){Astropy Collaboration}, {Robitaille}, {Tollerud}, {Greenfield}, {Droettboom}, {Bray}, {Aldcroft}, {Davis}, {Ginsburg}, {Price-Whelan}, {Kerzendorf}, {Conley}, {Crighton}, {Barbary}, {Muna}, {Ferguson}, {Grollier}, {Parikh}, {Nair}, {Unther}, {Deil}, {Woillez}, {Conseil}, {Kramer}, {Turner}, {Singer}, {Fox}, {Weaver}, {Zabalza}, {Edwards}, {Azalee Bostroem}, {Burke}, {Casey}, {Crawford}, {Dencheva}, {Ely}, {Jenness}, {Labrie}, {Lim}, {Pierfederici}, {Pontzen}, {Ptak}, {Refsdal}, {Servillat}, \& {Streicher}}]{AstropyCollaboration2013}
{Astropy Collaboration}, {Robitaille}, T.~P., {Tollerud}, E.~J., {et~al.} 2013, \bibinfo{title}{{Astropy: A community Python package for astronomy},} \aap, 558, A33, \dodoi{10.1051/0004-6361/201322068}

\bibitem[{ {Astropy Collaboration} {et~al.}(2018){Astropy Collaboration}, {Price-Whelan}, {Sip{\H{o}}cz}, {G{\"u}nther}, {Lim}, {Crawford}, {Conseil}, {Shupe}, {Craig}, {Dencheva}, {Ginsburg}, {VanderPlas}, {Bradley}, {P{\'e}rez-Su{\'a}rez}, {de Val-Borro}, {Aldcroft}, {Cruz}, {Robitaille}, {Tollerud}, {Ardelean}, {Babej}, {Bach}, {Bachetti}, {Bakanov}, {Bamford}, {Barentsen}, {Barmby}, {Baumbach}, {Berry}, {Biscani}, {Boquien}, {Bostroem}, {Bouma}, {Brammer}, {Bray}, {Breytenbach}, {Buddelmeijer}, {Burke}, {Calderone}, {Cano Rodr{\'\i}guez}, {Cara}, {Cardoso}, {Cheedella}, {Copin}, {Corrales}, {Crichton}, {D'Avella}, {Deil}, {Depagne}, {Dietrich}, {Donath}, {Droettboom}, {Earl}, {Erben}, {Fabbro}, {Ferreira}, {Finethy}, {Fox}, {Garrison}, {Gibbons}, {Goldstein}, {Gommers}, {Greco}, {Greenfield}, {Groener}, {Grollier}, {Hagen}, {Hirst}, {Homeier}, {Horton}, {Hosseinzadeh}, {Hu}, {Hunkeler}, {Ivezi{\'c}}, {Jain}, {Jenness}, {Kanarek}, {Kendrew}, {Kern}, {Kerzendorf}, {Khvalko}, {King}, {Kirkby}, {Kulkarni},
  {Kumar}, {Lee}, {Lenz}, {Littlefair}, {Ma}, {Macleod}, {Mastropietro}, {McCully}, {Montagnac}, {Morris}, {Mueller}, {Mumford}, {Muna}, {Murphy}, {Nelson}, {Nguyen}, {Ninan}, {N{\"o}the}, {Ogaz}, {Oh}, {Parejko}, {Parley}, {Pascual}, {Patil}, {Patil}, {Plunkett}, {Prochaska}, {Rastogi}, {Reddy Janga}, {Sabater}, {Sakurikar}, {Seifert}, {Sherbert}, {Sherwood-Taylor}, {Shih}, {Sick}, {Silbiger}, {Singanamalla}, {Singer}, {Sladen}, {Sooley}, {Sornarajah}, {Streicher}, {Teuben}, {Thomas}, {Tremblay}, {Turner}, {Terr{\'o}n}, {van Kerkwijk}, {de la Vega}, {Watkins}, {Weaver}, {Whitmore}, {Woillez}, {Zabalza}, \& {Astropy Contributors}}]{AstropyCollaboration2018}
{Astropy Collaboration}, {Price-Whelan}, A.~M., {Sip{\H{o}}cz}, B.~M., {et~al.} 2018, \bibinfo{title}{{The Astropy Project: Building an Open-science Project and Status of the v2.0 Core Package},} \aj, 156, 123, \dodoi{10.3847/1538-3881/aabc4f}

\bibitem[{ {Astropy Collaboration} {et~al.}(2022){Astropy Collaboration}, {Price-Whelan}, {Lim}, {Earl}, {Starkman}, {Bradley}, {Shupe}, {Patil}, {Corrales}, {Brasseur}, {N{\"o}the}, {Donath}, {Tollerud}, {Morris}, {Ginsburg}, {Vaher}, {Weaver}, {Tocknell}, {Jamieson}, {van Kerkwijk}, {Robitaille}, {Merry}, {Bachetti}, {G{\"u}nther}, {Aldcroft}, {Alvarado-Montes}, {Archibald}, {B{\'o}di}, {Bapat}, {Barentsen}, {Baz{\'a}n}, {Biswas}, {Boquien}, {Burke}, {Cara}, {Cara}, {Conroy}, {Conseil}, {Craig}, {Cross}, {Cruz}, {D'Eugenio}, {Dencheva}, {Devillepoix}, {Dietrich}, {Eigenbrot}, {Erben}, {Ferreira}, {Foreman-Mackey}, {Fox}, {Freij}, {Garg}, {Geda}, {Glattly}, {Gondhalekar}, {Gordon}, {Grant}, {Greenfield}, {Groener}, {Guest}, {Gurovich}, {Handberg}, {Hart}, {Hatfield-Dodds}, {Homeier}, {Hosseinzadeh}, {Jenness}, {Jones}, {Joseph}, {Kalmbach}, {Karamehmetoglu}, {Ka{\l}uszy{\'n}ski}, {Kelley}, {Kern}, {Kerzendorf}, {Koch}, {Kulumani}, {Lee}, {Ly}, {Ma}, {MacBride}, {Maljaars}, {Muna}, {Murphy}, {Norman},
  {O'Steen}, {Oman}, {Pacifici}, {Pascual}, {Pascual-Granado}, {Patil}, {Perren}, {Pickering}, {Rastogi}, {Roulston}, {Ryan}, {Rykoff}, {Sabater}, {Sakurikar}, {Salgado}, {Sanghi}, {Saunders}, {Savchenko}, {Schwardt}, {Seifert-Eckert}, {Shih}, {Jain}, {Shukla}, {Sick}, {Simpson}, {Singanamalla}, {Singer}, {Singhal}, {Sinha}, {Sip{\H{o}}cz}, {Spitler}, {Stansby}, {Streicher}, {{\v{S}}umak}, {Swinbank}, {Taranu}, {Tewary}, {Tremblay}, {de Val-Borro}, {Van Kooten}, {Vasovi{\'c}}, {Verma}, {de Miranda Cardoso}, {Williams}, {Wilson}, {Winkel}, {Wood-Vasey}, {Xue}, {Yoachim}, {Zhang}, {Zonca}, \& {Astropy Project Contributors}}]{AstropyCollaboration2022}
{Astropy Collaboration}, {Price-Whelan}, A.~M., {Lim}, P.~L., {et~al.} 2022, \bibinfo{title}{{The Astropy Project: Sustaining and Growing a Community-oriented Open-source Project and the Latest Major Release (v5.0) of the Core Package},} \apj, 935, 167, \dodoi{10.3847/1538-4357/ac7c74}

\bibitem[{C.~A.~L. {Bailer-Jones} {et~al.}(2021){Bailer-Jones}, {Rybizki}, {Fouesneau}, {Demleitner}, \& {Andrae}}]{Bailer-Jones2021}
{Bailer-Jones}, C.~A.~L., {Rybizki}, J., {Fouesneau}, M., {Demleitner}, M., \& {Andrae}, R. 2021, \bibinfo{title}{{Estimating Distances from Parallaxes. V. Geometric and Photogeometric Distances to 1.47 Billion Stars in Gaia Early Data Release 3},} \aj, 161, 147, \dodoi{10.3847/1538-3881/abd806}

\bibitem[{S. {Baliunas} {et~al.}(1996){Baliunas}, {Sokoloff}, \& {Soon}}]{Baliunas1996}
{Baliunas}, S., {Sokoloff}, D., \& {Soon}, W. 1996, \bibinfo{title}{{Magnetic Field and Rotation in Lower Main-Sequence Stars: an Empirical Time-dependent Magnetic Bode's Relation?},} \apjl, 457, L99, \dodoi{10.1086/309891}

\bibitem[{S.~L. {Baliunas} {et~al.}(1995){Baliunas}, {Donahue}, {Soon}, {Horne}, {Frazer}, {Woodard-Eklund}, {Bradford}, {Rao}, {Wilson}, {Zhang}, {Bennett}, {Briggs}, {Carroll}, {Duncan}, {Figueroa}, {Lanning}, {Misch}, {Mueller}, {Noyes}, {Poppe}, {Porter}, {Robinson}, {Russell}, {Shelton}, {Soyumer}, {Vaughan}, \& {Whitney}}]{Baliunas1995}
{Baliunas}, S.~L., {Donahue}, R.~A., {Soon}, W.~H., {et~al.} 1995, \bibinfo{title}{{Chromospheric Variations in Main-Sequence Stars. II.},} \apj, 438, 269, \dodoi{10.1086/175072}

\bibitem[{A. {Baranne} {et~al.}(1996){Baranne}, {Queloz}, {Mayor}, {Adrianzyk}, {Knispel}, {Kohler}, {Lacroix}, {Meunier}, {Rimbaud}, \& {Vin}}]{Baranne1996}
{Baranne}, A., {Queloz}, D., {Mayor}, M., {et~al.} 1996, \bibinfo{title}{{ELODIE: A spectrograph for accurate radial velocity measurements.},} A\&As, 119, 373

\bibitem[{S.~A. {Barnes}(2007){Barnes}}]{Barnes2007}
{Barnes}, S.~A. 2007, \bibinfo{title}{{Ages for Illustrative Field Stars Using Gyrochronology: Viability, Limitations, and Errors},} \apj, 669, 1167, \dodoi{10.1086/519295}

\bibitem[{R. {Basant} {et~al.}(2025){Basant}, {Luque}, {Bean}, {Seifahrt}, {Brady}, {Zhao}, {Brown}, {Das}, {St{\"u}rmer}, {Kasper}, {Gupta}, \& {Stef{\'a}nsson}}]{Basant2025}
{Basant}, R., {Luque}, R., {Bean}, J.~L., {et~al.} 2025, \bibinfo{title}{{Four Sub-Earth Planets Orbiting Barnard's Star from MAROON-X and ESPRESSO},} \apjl, 982, L1, \dodoi{10.3847/2041-8213/adb8d5}

\bibitem[{A.~C. {Baum} {et~al.}(2022){Baum}, {Wright}, {Luhn}, \& {Isaacson}}]{Baum2022}
{Baum}, A.~C., {Wright}, J.~T., {Luhn}, J.~K., \& {Isaacson}, H. 2022, \bibinfo{title}{{Five Decades of Chromospheric Activity in 59 Sun-like Stars and New Maunder Minimum Candidate HD 166620},} \aj, 163, 183, \dodoi{10.3847/1538-3881/ac5683}

\bibitem[{G. {Bertelli} {et~al.}(2008){Bertelli}, {Girardi}, {Marigo}, \& {Nasi}}]{Bertelli2008}
{Bertelli}, G., {Girardi}, L., {Marigo}, P., \& {Nasi}, E. 2008, \bibinfo{title}{{Scaled solar tracks and isochrones in a large region of the Z-Y plane. I. From the ZAMS to the TP-AGB end for 0.15-2.5 \{M\}$_{☉}$ stars},} \aap, 484, 815, \dodoi{10.1051/0004-6361:20079165}

\bibitem[{G. {Bertelli} {et~al.}(2009){Bertelli}, {Nasi}, {Girardi}, \& {Marigo}}]{Bertelli2009}
{Bertelli}, G., {Nasi}, E., {Girardi}, L., \& {Marigo}, P. 2009, \bibinfo{title}{{Scaled solar tracks and isochrones in a large region of the Z-Y plane. II. From 2.5 to 20 M☉ stars},} \aap, 508, 355, \dodoi{10.1051/0004-6361/200912093}

\bibitem[{S. {Blanco-Cuaresma}(2019){Blanco-Cuaresma}}]{Blanco-Cuaresma2019}
{Blanco-Cuaresma}, S. 2019, \bibinfo{title}{{Modern stellar spectroscopy caveats},} \mnras, 486, 2075, \dodoi{10.1093/mnras/stz549}

\bibitem[{E. {B{\"o}hm-Vitense}(2007){B{\"o}hm-Vitense}}]{Bohm-Vitense2007}
{B{\"o}hm-Vitense}, E. 2007, \bibinfo{title}{{Chromospheric Activity in G and K Main-Sequence Stars, and What It Tells Us about Stellar Dynamos},} \apj, 657, 486, \dodoi{10.1086/510482}

\bibitem[{J.~M. {Brewer} {et~al.}(2016){Brewer}, {Fischer}, {Valenti}, \& {Piskunov}}]{Brewer2016}
{Brewer}, J.~M., {Fischer}, D.~A., {Valenti}, J.~A., \& {Piskunov}, N. 2016, \bibinfo{title}{{Spectral Properties of Cool Stars: Extended Abundance Analysis of 1,617 Planet-search Stars},} \apjs, 225, 32, \dodoi{10.3847/0067-0049/225/2/32}

\bibitem[{A. {Burrows} {et~al.}(2024){Burrows}, {Halverson}, {Siegel}, {Gilbertson}, {Luhn}, {Burt}, {Bender}, {Roy}, {Terrien}, {Vangstein}, {Mahadevan}, {Wright}, {Robertson}, {Ford}, {Stef{\'a}nsson}, {Ninan}, {Blake}, {McElwain}, {Schwab}, \& {Zhao}}]{Burrows2024}
{Burrows}, A., {Halverson}, S., {Siegel}, J.~C., {et~al.} 2024, \bibinfo{title}{{The Death of Vulcan: NEID Reveals That the Planet Candidate Orbiting HD 26965 Is Stellar Activity},} \aj, 167, 243, \dodoi{10.3847/1538-3881/ad34d5}

\bibitem[{J.~A. {Burt} {et~al.}(2025){Burt}, {Dumusque}, \& {Halverson}}]{Burt2025}
{Burt}, J.~A., {Dumusque}, X., \& {Halverson}, S. 2025, \bibinfo{title}{{Precise Radial Velocities},} arXiv e-prints, arXiv:2511.01954, \dodoi{10.48550/arXiv.2511.01954}

\bibitem[{R.~P. {Butler} {et~al.}(1996){Butler}, {Marcy}, {Williams}, {McCarthy}, {Dosanjh}, \& {Vogt}}]{Butler1996}
{Butler}, R.~P., {Marcy}, G.~W., {Williams}, E., {et~al.} 1996, \bibinfo{title}{{Attaining Doppler Precision of 3 M s-1},} \pasp, 108, 500, \dodoi{10.1086/133755}

\bibitem[{R.~P. {Butler} {et~al.}(2017){Butler}, {Vogt}, {Laughlin}, {Burt}, {Rivera}, {Tuomi}, {Teske}, {Arriagada}, {Diaz}, {Holden}, \& {Keiser}}]{Butler2017}
{Butler}, R.~P., {Vogt}, S.~S., {Laughlin}, G., {et~al.} 2017, \bibinfo{title}{{The LCES HIRES/Keck Precision Radial Velocity Exoplanet Survey},} \aj, 153, 208, \dodoi{10.3847/1538-3881/aa66ca}

\bibitem[{L. {Casagrande} {et~al.}(2011){Casagrande}, {Sch{\"o}nrich}, {Asplund}, {Cassisi}, {Ram{\'\i}rez}, {Mel{\'e}ndez}, {Bensby}, \& {Feltzing}}]{Casagrande2011}
{Casagrande}, L., {Sch{\"o}nrich}, R., {Asplund}, M., {et~al.} 2011, \bibinfo{title}{{New constraints on the chemical evolution of the solar neighbourhood and Galactic disc(s). Improved astrophysical parameters for the Geneva-Copenhagen Survey},} \aap, 530, A138, \dodoi{10.1051/0004-6361/201016276}

\bibitem[{G. {Casali} {et~al.}(2020){Casali}, {Spina}, {Magrini}, {Karakas}, {Kobayashi}, {Casey}, {Feltzing}, {Van der Swaelmen}, {Tsantaki}, {Jofr{\'e}}, {Bragaglia}, {Feuillet}, {Bensby}, {Biazzo}, {Gonneau}, {Tautvai{\v{s}}ien{\.{e}}}, {Baratella}, {Roccatagliata}, {Pancino}, {Sousa}, {Adibekyan}, {Martell}, {Bayo}, {Jackson}, {Jeffries}, {Gilmore}, {Randich}, {Alfaro}, {Koposov}, {Korn}, {Recio-Blanco}, {Smiljanic}, {Franciosini}, {Hourihane}, {Monaco}, {Morbidelli}, {Sacco}, {Worley}, \& {Zaggia}}]{Casali2020}
{Casali}, G., {Spina}, L., {Magrini}, L., {et~al.} 2020, \bibinfo{title}{{The Gaia-ESO survey: the non-universality of the age-chemical-clocks-metallicity relations in the Galactic disc},} \aap, 639, A127, \dodoi{10.1051/0004-6361/202038055}

\bibitem[{J.~L. {Christiansen} {et~al.}(2025){Christiansen}, {McElroy}, {Harbut}, {Ciardi}, {Crane}, {Good}, {Hardegree-Ullman}, {Kesseli}, {Lund}, {Lynn}, {Muthiar}, {Nilsson}, {Oluyide}, {Papin}, {Rivera}, {Swain}, {Susemiehl}, {Tam}, {van Eyken}, \& {Beichman}}]{Christiansen2025}
{Christiansen}, J.~L., {McElroy}, D.~L., {Harbut}, M., {et~al.} 2025, \bibinfo{title}{{The NASA Exoplanet Archive and Exoplanet Follow-up Observing Program: Data, Tools, and Usage},} \psj, 6, 186, \dodoi{10.3847/PSJ/ade3c2}

\bibitem[{A. {Collier Cameron} {et~al.}(2019){Collier Cameron}, {Mortier}, {Phillips}, {Dumusque}, {Haywood}, {Langellier}, {Watson}, {Cegla}, {Costes}, {Charbonneau}, {Coffinet}, {Latham}, {Lopez-Morales}, {Malavolta}, {Maldonado}, {Micela}, {Milbourne}, {Molinari}, {Saar}, {Thompson}, {Buchschacher}, {Cecconi}, {Cosentino}, {Ghedina}, {Glenday}, {Gonzalez}, {Li}, {Lodi}, {Lovis}, {Pepe}, {Poretti}, {Rice}, {Sasselov}, {Sozzetti}, {Szentgyorgyi}, {Udry}, \& {Walsworth}}]{CollierCameron2019}
{Collier Cameron}, A., {Mortier}, A., {Phillips}, D., {et~al.} 2019, \bibinfo{title}{{Three years of Sun-as-a-star radial-velocity observations on the approach to solar minimum},} \mnras, 487, 1082, \dodoi{10.1093/mnras/stz1215}

\bibitem[{A. {Cumming} {et~al.}(2008){Cumming}, {Butler}, {Marcy}, {Vogt}, {Wright}, \& {Fischer}}]{Cumming2008}
{Cumming}, A., {Butler}, R.~P., {Marcy}, G.~W., {et~al.} 2008, \bibinfo{title}{{The Keck Planet Search: Detectability and the Minimum Mass and Orbital Period Distribution of Extrasolar Planets},} \pasp, 120, 531, \dodoi{10.1086/588487}

\bibitem[{R. {da Silva} {et~al.}(2015){da Silva}, {Milone}, \& {Rocha-Pinto}}]{daSilva2015}
{da Silva}, R., {Milone}, A. d.~C., \& {Rocha-Pinto}, H.~J. 2015, \bibinfo{title}{{Homogeneous abundance analysis of FGK dwarf, subgiant, and giant stars with and without giant planets},} \aap, 580, A24, \dodoi{10.1051/0004-6361/201525770}

\bibitem[{P. {Demarque} {et~al.}(2004){Demarque}, {Woo}, {Kim}, \& {Yi}}]{Demarque2004}
{Demarque}, P., {Woo}, J.-H., {Kim}, Y.-C., \& {Yi}, S.~K. 2004, \bibinfo{title}{{Y$^{2}$ Isochrones with an Improved Core Overshoot Treatment},} \apjs, 155, 667, \dodoi{10.1086/424966}

\bibitem[{X. {Dumusque}(2018){Dumusque}}]{Dumusque2018}
{Dumusque}, X. 2018, \bibinfo{title}{{Measuring precise radial velocities on individual spectral lines. I. Validation of the method and application to mitigate stellar activity},} \aap, 620, A47, \dodoi{10.1051/0004-6361/201833795}

\bibitem[{L. {Feinberg} {et~al.}(2024){Feinberg}, {Ziemer}, {Ansdell}, {Crooke}, {Dressing}, {Mennesson}, {O'Meara}, {Pepper}, \& {Roberge}}]{Feinberg2024}
{Feinberg}, L., {Ziemer}, J., {Ansdell}, M., {et~al.} 2024, \bibinfo{title}{{The Habitable Worlds Observatory engineering view: status, plans, and opportunities},} in Society of Photo-Optical Instrumentation Engineers (SPIE) Conference Series, Vol. 13092, Space Telescopes and Instrumentation 2024: Optical, Infrared, and Millimeter Wave, ed. L.~E. {Coyle}, S.~{Matsuura}, \& M.~D. {Perrin}, 130921N, \dodoi{10.1117/12.3018328}

\bibitem[{D.~A. {Fischer} {et~al.}(2014){Fischer}, {Marcy}, \& {Spronck}}]{Fischer2014}
{Fischer}, D.~A., {Marcy}, G.~W., \& {Spronck}, J. F.~P. 2014, \bibinfo{title}{{The Twenty-five Year Lick Planet Search},} \apjs, 210, 5, \dodoi{10.1088/0067-0049/210/1/5}

\bibitem[{R.~A. {Fisher}(1922){Fisher}}]{Fisher1922}
{Fisher}, R.~A. 1922, \bibinfo{title}{{On the Mathematical Foundations of Theoretical Statistics},} Philosophical Transactions of the Royal Society of London Series A, 222, 309, \dodoi{10.1098/rsta.1922.0009}

\bibitem[{D. Foreman-Mackey {et~al.}(2021)Foreman-Mackey, Savel, Luger, Agol, Czekala, Price-Whelan, Hedges, Gilbert, Bouma, Brandt, \& Barclay}]{exoplanet:zenodo}
Foreman-Mackey, D., Savel, A., Luger, R., {et~al.} 2021, exoplanet-dev/exoplanet v0.5.1, \dodoi{10.5281/zenodo.1998447}

\bibitem[{A.~V. {Freckelton} {et~al.}(2026){Freckelton}, {Mortier}, {Bedell}, {Cretignier}, {Kolecki}, {Korn}, {Sousa}, {Tsantaki}, {Brewer}, {Buchhave}, {Davies}, {Gonz{\'a}lez Hern{\'a}ndez}, {Morrell}, {Nielsen}, {Passegger}, {Quirrenbach}, {Roy}, {Santos}, {Su{\'a}rez Mascare{\~n}o}, {Watson}, \& {Zhao}}]{Freckelton2026}
{Freckelton}, A.~V., {Mortier}, A., {Bedell}, M., {et~al.} 2026, \bibinfo{title}{{gr8stars II : judgement day for spectroscopic parameter model systematics},} arXiv e-prints, arXiv:2606.06234, \dodoi{10.48550/arXiv.2606.06234}

\bibitem[{B.~J. {Fulton}(2017){Fulton}}]{Fulton2017}
{Fulton}, B.~J. 2017, \bibinfo{title}{{APF-50: A robotic search for Earth's nearest neighbors},} PhD thesis, University of Hawai{\textquoteleft}i at M{\={a}}noa, United States

\bibitem[{B.~J. {Fulton} {et~al.}(2018){Fulton}, {Petigura}, {Blunt}, \& {Sinukoff}}]{Fulton2018}
{Fulton}, B.~J., {Petigura}, E.~A., {Blunt}, S., \& {Sinukoff}, E. 2018, \bibinfo{title}{{RadVel: The Radial Velocity Modeling Toolkit},} \pasp, 130, 044504, \dodoi{10.1088/1538-3873/aaaaa8}

\bibitem[{ {Gaia Collaboration}(2022){Gaia Collaboration}}]{GaiaCollaboration2022}
{Gaia Collaboration}. 2022, {VizieR Online Data Catalog: Gaia DR3 Part 1. Main source (Gaia Collaboration, 2022)},, VizieR On-line Data Catalog: I/355. Originally published in: 2023A\&A...674A...1G; doi:10.1051/0004-63 \dodoi{10.26093/cds/vizier.1355}

\bibitem[{E.~J. {Gaidos} {et~al.}(2000){Gaidos}, {Henry}, \& {Henry}}]{Gaidos2000}
{Gaidos}, E.~J., {Henry}, G.~W., \& {Henry}, S.~M. 2000, \bibinfo{title}{{Spectroscopy and Photometry of Nearby Young Solar Analogs},} \aj, 120, 1006, \dodoi{10.1086/301488}

\bibitem[{J.~I. {Gonz{\'a}lez Hern{\'a}ndez} {et~al.}(2024){Gonz{\'a}lez Hern{\'a}ndez}, {Su{\'a}rez Mascare{\~n}o}, {Silva}, {Stefanov}, {Faria}, {Tabernero}, {Sozzetti}, {Rebolo}, {Pepe}, {Santos}, {Cristiani}, {Lovis}, {Dumusque}, {Figueira}, {Lillo-Box}, {Nari}, {Benatti}, {Hobson}, {Castro-Gonz{\'a}lez}, {Allart}, {Passegger}, {Zapatero Osorio}, {Adibekyan}, {Alibert}, {Allende Prieto}, {Bouchy}, {Damasso}, {D'Odorico}, {Di Marcantonio}, {Ehrenreich}, {Lo Curto}, {Santos}, {Martins}, {Mehner}, {Micela}, {Molaro}, {Nunes}, {Palle}, {Sousa}, \& {Udry}}]{GonzalezHernandez2024}
{Gonz{\'a}lez Hern{\'a}ndez}, J.~I., {Su{\'a}rez Mascare{\~n}o}, A., {Silva}, A.~M., {et~al.} 2024, \bibinfo{title}{{A sub-Earth-mass planet orbiting Barnard's star},} \aap, 690, A79, \dodoi{10.1051/0004-6361/202451311}

\bibitem[{A.~F. {Gupta} \& M. {Bedell}(2024){Gupta} \& {Bedell}}]{Gupta2024}
{Gupta}, A.~F., \& {Bedell}, M. 2024, \bibinfo{title}{{Fishing for Planets: A Comparative Analysis of EPRV Survey Performance in the Presence of Correlated Noise},} \aj, 168, 29, \dodoi{10.3847/1538-3881/ad4ce6}

\bibitem[{A.~F. {Gupta} {et~al.}(2021){Gupta}, {Wright}, {Robertson}, {Halverson}, {Luhn}, {Roy}, {Mahadevan}, {Ford}, {Bender}, {Blake}, {Hearty}, {Kanodia}, {Logsdon}, {McElwain}, {Monson}, {Ninan}, {Schwab}, {Stef{\'a}nsson}, \& {Terrien}}]{Gupta2021}
{Gupta}, A.~F., {Wright}, J.~T., {Robertson}, P., {et~al.} 2021, \bibinfo{title}{{Target Prioritization and Observing Strategies for the NEID Earth Twin Survey},} \aj, 161, 130, \dodoi{10.3847/1538-3881/abd79e}

\bibitem[{A.~F. {Gupta} {et~al.}(2025{\natexlab{a}}){Gupta}, {Luhn}, {Wright}, {Mahadevan}, {Robertson}, {Krolikowski}, {Ford}, {Ca{\~n}as}, {Halverson}, {Lin}, {Kanodia}, {Fitzmaurice}, {Gilbertson}, {Bender}, {Blake}, {Dong}, {Giovinazzi}, {Logsdon}, {Monson}, {Ninan}, {Rajagopal}, {Roy}, {Schwab}, \& {Stef{\'a}nsson}}]{NETSI}
{Gupta}, A.~F., {Luhn}, J.~K., {Wright}, J.~T., {et~al.} 2025{\natexlab{a}}, \bibinfo{title}{{The NEID Earth Twin Survey. I. Confirmation of a 31 Day Planet Orbiting HD 86728},} \aj, 169, 1, \dodoi{10.3847/1538-3881/ad89bf}

\bibitem[{A.~F. {Gupta} {et~al.}(2025{\natexlab{b}}){Gupta}, {Fitzmaurice}, {Mahadevan}, {Robertson}, {Luhn}, {Wright}, {Logsdon}, {Krolikowski}, {Paredes}, {Bender}, {Giovinazzi}, {Lin}, {Blake}, {Ca{\~n}as}, {Ford}, {Halverson}, {Kanodia}, {McElwain}, {Monson}, {Ninan}, {Rajagopal}, {Roy}, {Schwab}, {Stef{\'a}nsson}, \& {Terrien}}]{Gupta2025}
{Gupta}, A.~F., {Fitzmaurice}, E., {Mahadevan}, S., {et~al.} 2025{\natexlab{b}}, \bibinfo{title}{{The NEID Earth Twin Survey. III. Survey Performance after Three Years on Sky},} \aj, 170, 264, \dodoi{10.3847/1538-3881/ae0339}

\bibitem[{S. {Halverson} {et~al.}(2016){Halverson}, {Terrien}, {Mahadevan}, {Roy}, {Bender}, {Stef{\'a}nsson}, {Monson}, {Levi}, {Hearty}, {Blake}, {McElwain}, {Schwab}, {Ramsey}, {Wright}, {Wang}, {Gong}, \& {Roberston}}]{Halverson2016}
{Halverson}, S., {Terrien}, R., {Mahadevan}, S., {et~al.} 2016, \bibinfo{title}{{A comprehensive radial velocity error budget for next generation Doppler spectrometers},} in Society of Photo-Optical Instrumentation Engineers (SPIE) Conference Series, Vol. 9908, Ground-based and Airborne Instrumentation for Astronomy VI, ed. C.~J. {Evans}, L.~{Simard}, \& H.~{Takami}, 99086P, \dodoi{10.1117/12.2232761}

\bibitem[{T. {Han} \& T.~D. {Brandt}(2023){Han} \& {Brandt}}]{Han2023}
{Han}, T., \& {Brandt}, T.~D. 2023, \bibinfo{title}{{TESS-Gaia Light Curve: A PSF-based TESS FFI Light-curve Product},} \aj, 165, 71, \dodoi{10.3847/1538-3881/acaaa7}

\bibitem[{C.~K. {Harada} {et~al.}(2025){Harada}, {Dressing}, {Turtelboom}, {Kane}, {Blunt}, {Dietrich}, {Hinkel}, {Li}, {Mamajek}, {Rice}, {Tuchow}, {Wittenmyer}, {Chin}, {Fernandez}, {Kulkarni}, {Lin}, {Liu}, {Liu}, {Nathan}, \& {Zbriger}}]{Harada2025}
{Harada}, C.~K., {Dressing}, C.~D., {Turtelboom}, E.~V., {et~al.} 2025, \bibinfo{title}{{SPORES-HWO. II. Companion Mass Limits and Updated Planet Properties for 120 Future Exoplanet Imaging Targets from 35 yr of Precise Doppler Monitoring},} \aj, 170, 343, \dodoi{10.3847/1538-3881/ae0b62}

\bibitem[{C.~R. {Harris} {et~al.}(2020){Harris}, {Millman}, {van der Walt}, {Gommers}, {Virtanen}, {Cournapeau}, {Wieser}, {Taylor}, {Berg}, {Smith}, {Kern}, {Picus}, {Hoyer}, {van Kerkwijk}, {Brett}, {Haldane}, {del R{\'\i}o}, {Wiebe}, {Peterson}, {G{\'e}rard-Marchant}, {Sheppard}, {Reddy}, {Weckesser}, {Abbasi}, {Gohlke}, \& {Oliphant}}]{Harris2020}
{Harris}, C.~R., {Millman}, K.~J., {van der Walt}, S.~J., {et~al.} 2020, \bibinfo{title}{{Array programming with NumPy},} \nat, 585, 357, \dodoi{10.1038/s41586-020-2649-2}

\bibitem[{A. {Hempelmann} {et~al.}(2016){Hempelmann}, {Mittag}, {Gonzalez-Perez}, {Schmitt}, {Schr{\"o}der}, \& {Rauw}}]{Hempelmann2016}
{Hempelmann}, A., {Mittag}, M., {Gonzalez-Perez}, J.~N., {et~al.} 2016, \bibinfo{title}{{Measuring rotation periods of solar-like stars using TIGRE. A study of periodic CaII H+K S-index variability},} \aap, 586, A14, \dodoi{10.1051/0004-6361/201526972}

\bibitem[{L.~A. {Hirsch} {et~al.}(2021){Hirsch}, {Rosenthal}, {Fulton}, {Howard}, {Ciardi}, {Marcy}, {Nielsen}, {Petigura}, {de Rosa}, {Isaacson}, {Weiss}, {Sinukoff}, \& {Macintosh}}]{Hirsch2021}
{Hirsch}, L.~A., {Rosenthal}, L., {Fulton}, B.~J., {et~al.} 2021, \bibinfo{title}{{Understanding the Impacts of Stellar Companions on Planet Formation and Evolution: A Survey of Stellar and Planetary Companions within 25 pc},} \aj, 161, 134, \dodoi{10.3847/1538-3881/abd639}

\bibitem[{A.~W. {Howard} {et~al.}(2010){Howard}, {Johnson}, {Marcy}, {Fischer}, {Wright}, {Bernat}, {Henry}, {Peek}, {Isaacson}, {Apps}, {Endl}, {Cochran}, {Valenti}, {Anderson}, \& {Piskunov}}]{Howard2010}
{Howard}, A.~W., {Johnson}, J.~A., {Marcy}, G.~W., {et~al.} 2010, \bibinfo{title}{{The California Planet Survey. I. Four New Giant Exoplanets},} \apj, 721, 1467, \dodoi{10.1088/0004-637X/721/2/1467}

\bibitem[{J.~D. {Hunter}(2007){Hunter}}]{Hunter2007}
{Hunter}, J.~D. 2007, \bibinfo{title}{{Matplotlib: A 2D Graphics Environment},} Computing in Science and Engineering, 9, 90, \dodoi{10.1109/MCSE.2007.55}

\bibitem[{H. {Isaacson} {et~al.}(2024){Isaacson}, {Howard}, {Fulton}, {Petigura}, {Weiss}, {Kane}, {Carter}, {Beard}, {Giacalone}, {Van Zandt}, {Murphy}, {Dai}, {Chontos}, {Polanski}, {Rice}, {Lubin}, {Brinkman}, {Rubenzahl}, {Blunt}, {Yee}, {MacDougall}, {Dalba}, {Tyler}, {Behmard}, {Angelo}, {Pidhorodetska}, {Mayo}, {Holcomb}, {Turtelboom}, {Hill}, {Bouma}, {Zhang}, {Crossfield}, \& {Saunders}}]{Isaacson2024}
{Isaacson}, H., {Howard}, A.~W., {Fulton}, B., {et~al.} 2024, \bibinfo{title}{{The California Legacy Survey. V. Chromospheric Activity Cycles in Main-sequence Stars},} \apjs, 274, 35, \dodoi{10.3847/1538-4365/ad676c}

\bibitem[{S.~R. {Kane} \& J.~A. {Burt}(2024){Kane} \& {Burt}}]{Kane2024B}
{Kane}, S.~R., \& {Burt}, J.~A. 2024, \bibinfo{title}{{Hic Sunt Dracones: Uncovering Dynamical Perturbers within the Habitable Zone},} \aj, 168, 279, \dodoi{10.3847/1538-3881/ad8a68}

\bibitem[{S.~R. {Kane} {et~al.}(2024){Kane}, {Li}, {Turnbull}, {Dressing}, \& {Harada}}]{Kane2024}
{Kane}, S.~R., {Li}, Z., {Turnbull}, M.~C., {Dressing}, C.~D., \& {Harada}, C.~K. 2024, \bibinfo{title}{{Dynamical Viability Assessment for Habitable Worlds Observatory Targets},} \aj, 168, 195, \dodoi{10.3847/1538-3881/ad6a50}

\bibitem[{S. {Kanodia} \& J. {Wright}(2018){Kanodia} \& {Wright}}]{Kanodia2018}
{Kanodia}, S., \& {Wright}, J. 2018, \bibinfo{title}{{Python Leap Second Management and Implementation of Precise Barycentric Correction (barycorrpy)},} Research Notes of the American Astronomical Society, 2, 4, \dodoi{10.3847/2515-5172/aaa4b7}

\bibitem[{R.~K. {Kopparapu} {et~al.}(2013){Kopparapu}, {Ramirez}, {Kasting}, {Eymet}, {Robinson}, {Mahadevan}, {Terrien}, {Domagal-Goldman}, {Meadows}, \& {Deshpande}}]{Kopparapu2013}
{Kopparapu}, R.~K., {Ramirez}, R., {Kasting}, J.~F., {et~al.} 2013, \bibinfo{title}{{Habitable Zones around Main-sequence Stars: New Estimates},} \apj, 765, 131, \dodoi{10.1088/0004-637X/765/2/131}

\bibitem[{N.~R. {Lomb}(1976){Lomb}}]{Lomb1976}
{Lomb}, N.~R. 1976, \bibinfo{title}{{Least-Squares Frequency Analysis of Unequally Spaced Data},} \apss, 39, 447, \dodoi{10.1007/BF00648343}

\bibitem[{M. {Lucas} {et~al.}(2023){Lucas}, {Kaur}, {Erlend Fjelde}, {Plavin}, {Ge}, \& {Pfiffer}}]{Lucas2023}
{Lucas}, M., {Kaur}, S., {Erlend Fjelde}, T., {et~al.} 2023, {TuringLang/NestedSamplers.jl: v0.8.3}, v0.8.3 Zenodo, \dodoi{10.5281/zenodo.8009875}

\bibitem[{D. {Mahdi} {et~al.}(2016){Mahdi}, {Soubiran}, {Blanco-Cuaresma}, \& {Chemin}}]{Mahdi2016}
{Mahdi}, D., {Soubiran}, C., {Blanco-Cuaresma}, S., \& {Chemin}, L. 2016, \bibinfo{title}{{Solar twins in the ELODIE archive},} \aap, 587, A131, \dodoi{10.1051/0004-6361/201527472}

\bibitem[{J. {Maldonado} {et~al.}(2010){Maldonado}, {Mart{\'\i}nez-Arn{\'a}iz}, {Eiroa}, {Montes}, \& {Montesinos}}]{Maldonado2010}
{Maldonado}, J., {Mart{\'\i}nez-Arn{\'a}iz}, R.~M., {Eiroa}, C., {Montes}, D., \& {Montesinos}, B. 2010, \bibinfo{title}{{A spectroscopy study of nearby late-type stars, possible members of stellar kinematic groups},} \aap, 521, A12, \dodoi{10.1051/0004-6361/201014948}

\bibitem[{E. {Mamajek} \& K. {Stapelfeldt}(2024){Mamajek} \& {Stapelfeldt}}]{Mamajek2024}
{Mamajek}, E., \& {Stapelfeldt}, K. 2024, \bibinfo{title}{{NASA Exoplanet Exploration Program (ExEP) Mission Star List for the Habitable Worlds Observatory (2023)},} arXiv e-prints, arXiv:2402.12414, \dodoi{10.48550/arXiv.2402.12414}

\bibitem[{E.~E. {Mamajek} \& L.~A. {Hillenbrand}(2008){Mamajek} \& {Hillenbrand}}]{Mamajek2008}
{Mamajek}, E.~E., \& {Hillenbrand}, L.~A. 2008, \bibinfo{title}{{Improved Age Estimation for Solar-Type Dwarfs Using Activity-Rotation Diagnostics},} \apj, 687, 1264, \dodoi{10.1086/591785}

\bibitem[{M. {Mayor} \& D. {Queloz}(1995){Mayor} \& {Queloz}}]{Mayor1995}
{Mayor}, M., \& {Queloz}, D. 1995, \bibinfo{title}{{A Jupiter-mass companion to a solar-type star},} \nat, 378, 355, \dodoi{10.1038/378355a0}

\bibitem[{J.~D. {McEwen} {et~al.}(2021){McEwen}, {Wallis}, {Price}, \& {Spurio Mancini}}]{McEwen2021}
{McEwen}, J.~D., {Wallis}, C. G.~R., {Price}, M.~A., \& {Spurio Mancini}, A. 2021, \bibinfo{title}{{Machine learning assisted Bayesian model comparison: learnt harmonic mean estimator},} arXiv e-prints, arXiv:2111.12720, \dodoi{10.48550/arXiv.2111.12720}

\bibitem[{T.~S. {Metcalfe} \& J. {van Saders}(2017){Metcalfe} \& {van Saders}}]{Metcalfe2017}
{Metcalfe}, T.~S., \& {van Saders}, J. 2017, \bibinfo{title}{{Magnetic Evolution and the Disappearance of Sun-Like Activity Cycles},} \solphys, 292, 126, \dodoi{10.1007/s11207-017-1157-5}

\bibitem[{M. {Mittag} {et~al.}(2023){Mittag}, {Schmitt}, \& {Schr{\"o}der}}]{Mittag2023}
{Mittag}, M., {Schmitt}, J.~H.~M.~M., \& {Schr{\"o}der}, K.-P. 2023, \bibinfo{title}{{Revisiting the cycle-rotation connection for late-type stars},} \aap, 674, A116, \dodoi{10.1051/0004-6361/202245060}

\bibitem[{N. {Nari} {et~al.}(2025){Nari}, {Dumusque}, {Hara}, {Su{\'a}rez Mascare{\~n}o}, {Cretignier}, {Gonz{\'a}lez Hern{\'a}ndez}, {Stefanov}, {Passegger}, {Rebolo}, {Pepe}, {Santos}, {Cristiani}, {Faria}, {Figueira}, {Sozzetti}, {Zapatero Osorio}, {Adibekyan}, {Alibert}, {Allende Prieto}, {Bouchy}, {Benatti}, {Castro-Gonz{\'a}lez}, {D'Odorico}, {Damasso}, {Delisle}, {Di Marcantonio}, {Ehrenreich}, {G{\'e}nova-Santos}, {Hobson}, {Lavie}, {Lillo-Box}, {Lo Curto}, {Lovis}, {Martins}, {Mehner}, {Micela}, {Molaro}, {Mordasini}, {Nunes}, {Palle}, {Quanz}, {S{\'e}gransan}, {Silva}, {Sousa}, {Udry}, {Unger}, \& {Venturini}}]{Nari2025}
{Nari}, N., {Dumusque}, X., {Hara}, N.~C., {et~al.} 2025, \bibinfo{title}{{Revisiting the multi-planetary system of the nearby star HD 20794: Confirmation of a low-mass planet in the habitable zone of a nearby G-dwarf},} \aap, 693, A297, \dodoi{10.1051/0004-6361/202451769}

\bibitem[{K. {Ol{\'a}h} {et~al.}(2016){Ol{\'a}h}, {K{\H{o}}v{\'a}ri}, {Petrovay}, {Soon}, {Baliunas}, {Koll{\'a}th}, \& {Vida}}]{Olah2016}
{Ol{\'a}h}, K., {K{\H{o}}v{\'a}ri}, Z., {Petrovay}, K., {et~al.} 2016, \bibinfo{title}{{Magnetic cycles at different ages of stars},} \aap, 590, A133, \dodoi{10.1051/0004-6361/201628479}

\bibitem[{K. {Ol{\'a}h} {et~al.}(2009){Ol{\'a}h}, {Koll{\'a}th}, {Granzer}, {Strassmeier}, {Lanza}, {J{\"a}rvinen}, {Korhonen}, {Baliunas}, {Soon}, {Messina}, \& {Cutispoto}}]{Olah2009}
{Ol{\'a}h}, K., {Koll{\'a}th}, Z., {Granzer}, T., {et~al.} 2009, \bibinfo{title}{{Multiple and changing cycles of active stars. II. Results},} \aap, 501, 703, \dodoi{10.1051/0004-6361/200811304}

\bibitem[{T.~E. {Oliphant}(2007){Oliphant}}]{Oliphant2007}
{Oliphant}, T.~E. 2007, \bibinfo{title}{{Python for Scientific Computing},} Computing in Science and Engineering, 9, 10, \dodoi{10.1109/MCSE.2007.58}

\bibitem[{N. {Olspert} {et~al.}(2018){Olspert}, {Lehtinen}, {K{\"a}pyl{\"a}}, {Pelt}, \& {Grigorievskiy}}]{Olspert2018}
{Olspert}, N., {Lehtinen}, J.~J., {K{\"a}pyl{\"a}}, M.~J., {Pelt}, J., \& {Grigorievskiy}, A. 2018, \bibinfo{title}{{Estimating activity cycles with probabilistic methods. II. The Mount Wilson Ca H\&K data},} \aap, 619, A6, \dodoi{10.1051/0004-6361/201732525}

\bibitem[{A. {Pietrinferni} {et~al.}(2004){Pietrinferni}, {Cassisi}, {Salaris}, \& {Castelli}}]{Pietrinferni2004}
{Pietrinferni}, A., {Cassisi}, S., {Salaris}, M., \& {Castelli}, F. 2004, \bibinfo{title}{{A Large Stellar Evolution Database for Population Synthesis Studies. I. Scaled Solar Models and Isochrones},} \apj, 612, 168, \dodoi{10.1086/422498}

\bibitem[{A. {Pietrinferni} {et~al.}(2006){Pietrinferni}, {Cassisi}, {Salaris}, \& {Castelli}}]{Pietrinferni2006}
{Pietrinferni}, A., {Cassisi}, S., {Salaris}, M., \& {Castelli}, F. 2006, \bibinfo{title}{{A Large Stellar Evolution Database for Population Synthesis Studies. II. Stellar Models and Isochrones for an {\ensuremath{\alpha}}-enhanced Metal Distribution},} \apj, 642, 797, \dodoi{10.1086/501344}

\bibitem[{A. {Pietrinferni} {et~al.}(2009){Pietrinferni}, {Cassisi}, {Salaris}, {Percival}, \& {Ferguson}}]{Pietrinferni2009}
{Pietrinferni}, A., {Cassisi}, S., {Salaris}, M., {Percival}, S., \& {Ferguson}, J.~W. 2009, \bibinfo{title}{{A Large Stellar Evolution Database for Population Synthesis Studies. V. Stellar Models and Isochrones with CNONa Abundance Anticorrelations},} \apj, 697, 275, \dodoi{10.1088/0004-637X/697/1/275}

\bibitem[{V. {Rajpaul} {et~al.}(2015){Rajpaul}, {Aigrain}, {Osborne}, {Reece}, \& {Roberts}}]{Rajpaul2015}
{Rajpaul}, V., {Aigrain}, S., {Osborne}, M.~A., {Reece}, S., \& {Roberts}, S. 2015, \bibinfo{title}{{A Gaussian process framework for modelling stellar activity signals in radial velocity data},} \mnras, 452, 2269, \dodoi{10.1093/mnras/stv1428}

\bibitem[{I. {Ram{\'\i}rez} {et~al.}(2012){Ram{\'\i}rez}, {Fish}, {Lambert}, \& {Allende Prieto}}]{Ramirez2012}
{Ram{\'\i}rez}, I., {Fish}, J.~R., {Lambert}, D.~L., \& {Allende Prieto}, C. 2012, \bibinfo{title}{{Lithium Abundances in nearby FGK Dwarf and Subgiant Stars: Internal Destruction, Galactic Chemical Evolution, and Exoplanets},} \apj, 756, 46, \dodoi{10.1088/0004-637X/756/1/46}

\bibitem[{P. {Robertson} {et~al.}(2019){Robertson}, {Anderson}, {Stefansson}, {Hearty}, {Monson}, {Mahadevan}, {Blakeslee}, {Bender}, {Ninan}, {Conran}, {Levi}, {Lubar}, {Cole}, {Dykhouse}, {Kanodia}, {Nitroy}, {Smolsky}, {Tuggle}, {Blank}, {Nelson}, {Blake}, {Halverson}, {Henderson}, {Kaplan}, {Li}, {Logsdon}, {McElwain}, {Rajagopal}, {Ramsey}, {Roy}, {Schwab}, {Terrien}, \& {Wright}}]{Robertson2019}
{Robertson}, P., {Anderson}, T., {Stefansson}, G., {et~al.} 2019, \bibinfo{title}{{Ultrastable environment control for the NEID spectrometer: design and performance demonstration},} Journal of Astronomical Telescopes, Instruments, and Systems, 5, 015003, \dodoi{10.1117/1.JATIS.5.1.015003}

\bibitem[{L.~J. {Rosenthal} {et~al.}(2021){Rosenthal}, {Fulton}, {Hirsch}, {Isaacson}, {Howard}, {Dedrick}, {Sherstyuk}, {Blunt}, {Petigura}, {Knutson}, {Behmard}, {Chontos}, {Crepp}, {Crossfield}, {Dalba}, {Fischer}, {Henry}, {Kane}, {Kosiarek}, {Marcy}, {Rubenzahl}, {Weiss}, \& {Wright}}]{Rosenthal2021}
{Rosenthal}, L.~J., {Fulton}, B.~J., {Hirsch}, L.~A., {et~al.} 2021, \bibinfo{title}{{The California Legacy Survey. I. A Catalog of 178 Planets from Precision Radial Velocity Monitoring of 719 Nearby Stars over Three Decades},} \apjs, 255, 8, \dodoi{10.3847/1538-4365/abe23c}

\bibitem[{J.~D. {Scargle}(1982){Scargle}}]{Scargle1982}
{Scargle}, J.~D. 1982, \bibinfo{title}{{Studies in astronomical time series analysis. II. Statistical aspects of spectral analysis of unevenly spaced data.},} \apj, 263, 835, \dodoi{10.1086/160554}

\bibitem[{J.~H.~M.~M. {Schmitt} {et~al.}(2026){Schmitt}, {Schr{\"o}der}, {Mittag}, {Hempelmann}, {Gonz{\'a}lez-P{\'e}rez}, \& {Jack}}]{Schmitt2026}
{Schmitt}, J.~H.~M.~M., {Schr{\"o}der}, K.-P., {Mittag}, M., {et~al.} 2026, \bibinfo{title}{{Six decades of TIGRE and Mount Wilson chromospheric monitoring in the H and K lines: The quest for an understanding of solar-type activity},} arXiv e-prints, arXiv:2606.26283, \dodoi{10.48550/arXiv.2606.26283}

\bibitem[{C. {Schwab} {et~al.}(2016){Schwab}, {Rakich}, {Gong}, {Mahadevan}, {Halverson}, {Roy}, {Terrien}, {Robertson}, {Hearty}, {Levi}, {Monson}, {Wright}, {McElwain}, {Bender}, {Blake}, {St{\"u}rmer}, {Gurevich}, {Chakraborty}, \& {Ramsey}}]{Schwab2016}
{Schwab}, C., {Rakich}, A., {Gong}, Q., {et~al.} 2016, \bibinfo{title}{{Design of NEID, an extreme precision Doppler spectrograph for WIYN},} in Society of Photo-Optical Instrumentation Engineers (SPIE) Conference Series, Vol. 9908, Ground-based and Airborne Instrumentation for Astronomy VI, ed. C.~J. {Evans}, L.~{Simard}, \& H.~{Takami}, 99087H, \dodoi{10.1117/12.2234411}

\bibitem[{C. {Soubiran} {et~al.}(2022){Soubiran}, {Brouillet}, \& {Casamiquela}}]{Soubiran2022}
{Soubiran}, C., {Brouillet}, N., \& {Casamiquela}, L. 2022, \bibinfo{title}{{Assessment of [Fe/H] determinations for FGK stars in spectroscopic surveys},} \aap, 663, A4, \dodoi{10.1051/0004-6361/202142409}

\bibitem[{C. {Soubiran} \& A. {Triaud}(2004){Soubiran} \& {Triaud}}]{Soubiran2004}
{Soubiran}, C., \& {Triaud}, A. 2004, \bibinfo{title}{{The Top Ten solar analogs in the ELODIE library},} \aap, 418, 1089, \dodoi{10.1051/0004-6361:20035708}

\bibitem[{F. {Spada} {et~al.}(2017){Spada}, {Demarque}, {Kim}, {Boyajian}, \& {Brewer}}]{Spada2017}
{Spada}, F., {Demarque}, P., {Kim}, Y.-C., {Boyajian}, T.~S., \& {Brewer}, J.~M. 2017, \bibinfo{title}{{The Yale-Potsdam Stellar Isochrones},} \apj, 838, 161, \dodoi{10.3847/1538-4357/aa661d}

\bibitem[{G. {Stef{\`a}nsson} {et~al.}(2022){Stef{\`a}nsson}, {Mahadevan}, {Petrovich}, {Winn}, {Kanodia}, {Millholland}, {Maney}, {Ca{\~n}as}, {Wisniewski}, {Robertson}, {Ninan}, {Ford}, {Bender}, {Blake}, {Cegla}, {Cochran}, {Diddams}, {Dong}, {Endl}, {Fredrick}, {Halverson}, {Hearty}, {Hebb}, {Hirano}, {Lin}, {Logsdon}, {Lubar}, {McElwain}, {Metcalf}, {Monson}, {Rajagopal}, {Ramsey}, {Roy}, {Schwab}, {Schweiker}, {Terrien}, \& {Wright}}]{Stefansson2022}
{Stef{\`a}nsson}, G., {Mahadevan}, S., {Petrovich}, C., {et~al.} 2022, \bibinfo{title}{{The Warm Neptune GJ 3470b Has a Polar Orbit},} \apjl, 931, L15, \dodoi{10.3847/2041-8213/ac6e3c}

\bibitem[{L. {Tal-Or} {et~al.}(2019){Tal-Or}, {Trifonov}, {Zucker}, {Mazeh}, \& {Zechmeister}}]{Tal-Or2019}
{Tal-Or}, L., {Trifonov}, T., {Zucker}, S., {Mazeh}, T., \& {Zechmeister}, M. 2019, \bibinfo{title}{{Correcting HIRES/Keck radial velocities for small systematic errors},} \mnras, 484, L8, \dodoi{10.1093/mnrasl/sly227}

\bibitem[{J.~T. {Teklu} {et~al.}(2025){Teklu}, {Perdelwitz}, {Butler}, {Trifonov}, {Vogt}, {Mukhija}, \& {Tal-Or}}]{Teklu2025}
{Teklu}, J.~T., {Perdelwitz}, V., {Butler}, R.~P., {et~al.} 2025, \bibinfo{title}{{An updated catalog of HIRES/Keck radial velocity measurements: Including Ca II H\&K measurements},} \aap, 702, A68, \dodoi{10.1051/0004-6361/202555034}

\bibitem[{Q.~H. {Tran} {et~al.}(2023){Tran}, {Bedell}, {Foreman-Mackey}, \& {Luger}}]{Tran2023}
{Tran}, Q.~H., {Bedell}, M., {Foreman-Mackey}, D., \& {Luger}, R. 2023, \bibinfo{title}{{Joint Modeling of Radial Velocities and Photometry with a Gaussian Process Framework},} \apj, 950, 162, \dodoi{10.3847/1538-4357/acd05c}

\bibitem[{S.~S. {Vogt} {et~al.}(1994){Vogt}, {Allen}, {Bigelow}, {Bresee}, {Brown}, {Cantrall}, {Conrad}, {Couture}, {Delaney}, {Epps}, {Hilyard}, {Hilyard}, {Horn}, {Jern}, {Kanto}, {Keane}, {Kibrick}, {Lewis}, {Osborne}, {Pardeilhan}, {Pfister}, {Ricketts}, {Robinson}, {Stover}, {Tucker}, {Ward}, \& {Wei}}]{Vogt1994}
{Vogt}, S.~S., {Allen}, S.~L., {Bigelow}, B.~C., {et~al.} 1994, \bibinfo{title}{{HIRES: the high-resolution echelle spectrometer on the Keck 10-m Telescope},} in Society of Photo-Optical Instrumentation Engineers (SPIE) Conference Series, Vol. 2198, Instrumentation in Astronomy VIII, ed. D.~L. {Crawford} \& E.~R. {Craine}, 362, \dodoi{10.1117/12.176725}

\bibitem[{S.~S. {Vogt} {et~al.}(2014){Vogt}, {Radovan}, {Kibrick}, {Butler}, {Alcott}, {Allen}, {Arriagada}, {Bolte}, {Burt}, {Cabak}, {Chloros}, {Cowley}, {Deich}, {Dupraw}, {Earthman}, {Epps}, {Faber}, {Fischer}, {Gates}, {Hilyard}, {Holden}, {Johnston}, {Keiser}, {Kanto}, {Katsuki}, {Laiterman}, {Lanclos}, {Laughlin}, {Lewis}, {Lockwood}, {Lynam}, {Marcy}, {McLean}, {Miller}, {Misch}, {Peck}, {Pfister}, {Phillips}, {Rivera}, {Sandford}, {Saylor}, {Stover}, {Thompson}, {Walp}, {Ward}, {Wareham}, {Wei}, \& {Wright}}]{Vogt2014}
{Vogt}, S.~S., {Radovan}, M., {Kibrick}, R., {et~al.} 2014, \bibinfo{title}{{APF{\textemdash}The Lick Observatory Automated Planet Finder},} \pasp, 126, 359, \dodoi{10.1086/676120}

\bibitem[{A. {Ware} {et~al.}(2026){Ware}, {Ruppert}, \& {Young}}]{Ware2026}
{Ware}, A., {Ruppert}, K., \& {Young}, P. 2026, \bibinfo{title}{{The HAges Catalog: Stellar Ages for High-priority HWO Target Stars},} \apj, 1003, 195, \dodoi{10.3847/1538-4357/ae69d7}

\bibitem[{O.~C. {Wilson}(1978){Wilson}}]{Wilson1978}
{Wilson}, O.~C. 1978, \bibinfo{title}{{Chromospheric variations in main-sequence stars.},} \apj, 226, 379, \dodoi{10.1086/156618}

\bibitem[{M. {Zechmeister} \& M. {K{\"u}rster}(2009){Zechmeister} \& {K{\"u}rster}}]{Zechmeister2009}
{Zechmeister}, M., \& {K{\"u}rster}, M. 2009, \bibinfo{title}{{The generalised Lomb-Scargle periodogram. A new formalism for the floating-mean and Keplerian periodograms},} \aap, 496, 577, \dodoi{10.1051/0004-6361:200811296}

\bibitem[{M. {Zechmeister} {et~al.}(2018){Zechmeister}, {Reiners}, {Amado}, {Azzaro}, {Bauer}, {B{\'e}jar}, {Caballero}, {Guenther}, {Hagen}, {Jeffers}, {Kaminski}, {K{\"u}rster}, {Launhardt}, {Montes}, {Morales}, {Quirrenbach}, {Reffert}, {Ribas}, {Seifert}, {Tal-Or}, \& {Wolthoff}}]{Zechmeister2018}
{Zechmeister}, M., {Reiners}, A., {Amado}, P.~J., {et~al.} 2018, \bibinfo{title}{{Spectrum radial velocity analyser (SERVAL). High-precision radial velocities and two alternative spectral indicators},} \aap, 609, A12, \dodoi{10.1051/0004-6361/201731483}

\end{thebibliography}
\bibliographystyle{aasjournalv7}

\end{document}